\documentclass{jfm}

\usepackage{amsmath,amssymb}
\usepackage{mathrsfs}
\usepackage{xcolor}
\usepackage{subfigure}
\usepackage{bbm}
\usepackage{bm}
\usepackage{multirow}
\usepackage{booktabs}
\usepackage{tabularx}
\usepackage{setspace}
\usepackage{textcomp}

\usepackage[colorlinks=true, linkcolor=blue, urlcolor=blue, citecolor=blue]{hyperref}
\usepackage{natbib}

\usepackage{multirow,bigdelim}
\usepackage{makecell}

\begin{document}

\newtheorem{lemma}{Lemma}
\newtheorem{corollary}{Corollary}

\title{Moment theories for a $d$-dimensional dilute granular gas of Maxwell molecules}

\shorttitle{Moment theories for a granular gas of Maxwell molecules} 
\shortauthor{Vinay Kumar Gupta} 

\author
 {
 {Vinay Kumar Gupta\aff{1,2,}
 \corresp{
 \email{vkg@iiti.ac.in}
 }}
 }

\affiliation
{
\aff{1}
Discipline of Mathematics, Indian Institute of Technology Indore, Indore 453552, India
\aff{2}
Mathematics Institute, University of Warwick, Coventry CV4 7AL, UK
}

\maketitle

\begin{abstract}
Various systems of moment equations---consisting of up to $(d+3)(d^2+6d+2)/6$ moments---in a general dimension $d$ for a dilute granular gas composed of Maxwell molecules are derived from the inelastic Boltzmann equation by employing the Grad moment method. 
The Navier--Stokes-level constitutive relations for the stress and heat flux appearing in the system of mass, momentum and energy balance equations are determined from the derived moment equations. 
It has been shown that the moment equations only for the hydrodynamic field variables (density, velocity and granular temperature), stress and heat flux---along with the time-independent value of the fourth cumulant---are sufficient for determining the Navier--Stokes-level constitutive relations in the case of inelastic Maxwell molecules, and that the other higher-order moment equations do not play any role in this case.
The homogeneous cooling state of a freely cooling granular gas is investigated with the system of the Grad $(d+3)(d^2+6d+2)/6$-moment equations and its various subsystems.
By performing a linear stability analysis in the vicinity of the homogeneous cooling state, the critical system size for the onset of instability is estimated through the considered Grad moment systems. 
The results on critical system size from the presented moment theories are found to be in reasonably good agreement with those from simulations.
%
\end{abstract}
\section{Introduction}\label{Sec:intro}
Under strong excitation, granular materials resemble ordinary (molecular) gases and are referred to as rapid granular flows or granular gases \citep{Campbell1990, Goldhirsch2003}.
The prototype model of a granular gas is a dilute system comprised of smooth (frictionless) identical hard spheres---with no interstitial fluid---colliding pairwise and inelastically with a constant coefficient of (normal) restitution $0 \leq e \leq 1$, with $e=0$ referring to perfectly sticky collisions and $e=1$ to perfectly elastic collisions \citep{Campbell1990, Goldhirsch2003, BP2004, Garzo2019}. 
In the dilute limit, this system can be described by a single-particle velocity distribution function, which is the fundamental quantity in kinetic theory and obeys the Boltzmann equation suitably modified to incorporate energy dissipation due to inelastic collisions. 
The resemblance of granular gases to ordinary gases has motivated the development of several kinetic theory based tools for granular gases by suitably modifying these tools to account for energy dissipation due to inelastic collisions in the last three decades, and it is still an active area of research; see, e.g., \cite{JR1985, JR1985PoF, SGJFM1998, BDKS1998, GS2003, Santos2003, BP2004, BST2004, GarzoSantos2011, KM2011, Garzo2013, KGS2014, KSG2014, GS2017, GST2018, Garzo2019}. 
Nevertheless, the non-conservation of energy in granular systems makes kinetic theory based tools much more involved and has profound consequences on their behaviour, leading to a raft of intriguing phenomena pertaining to granular matter. 

Kinetic theory of classical (monatomic) gases offers systematic ways of deriving the transport equations for the field variables. 
The two notable approaches in kinetic theory, around which various solution techniques and some other models have been developed, are the Chapman--Enskog (CE) expansion \citep{CC1970} and the Grad moment method \citep{Grad1949}. 
While these approaches have been instrumental in understanding several problems from a theoretical point of view, both have their own shortcomings. 
The former, which is adequate for flows close to 
equilibrium, 
considers the transport equations only for the hydrodynamic field variables (density, velocity and temperature) and provides the constitutive relations for additional unknowns, namely the stress and heat flux, in these equations. 
Despite being successful in deriving the Euler equations (at zeroth order of expansion) and the classical Navier--Stokes and Fourier (NSF) equations (at first order of expansion), the usefulness of models (the Burnett equations and beyond) resulting from the higher-order CE expansion remains scarce mainly due to inherent instabilities \citep{Bobylev1982}.
On the other hand, the Grad moment method \citep{Grad1949} furnishes the governing equations---referred to as the moment equations---for more field variables 
than the hydrodynamic ones and employs a Hermite polynomial expansion to close the system of moment equations. 
A set of moment equations emanating from the Grad moment method is always linearly stable but suffers from the loss of hyperbolicity \citep{MullerandRuggeri1998, CFL2014KRM}---an essential property for the well-posedness of a system of partial differential equations (PDEs). 
The loss of hyperbolicity renders a Grad moment system to show some unphysical behaviour, e.g.~unphysical sub-shocks within the shock profile above a critical Mach number.
Yet, the Grad moment method has a clear advantage of linearly stable equations and, hence, is preferred over the CE expansion for describing nonequilibrium flows of monatomic gases. 

To circumvent the problems associated with the Grad moment method, a number of moment methods have been proposed in the literature. 
\citet{Levermore1996} propounded the maximum-entropy approach for closing a moment system. 
Although the maximum-entropy approach of \citet{Levermore1996} produces hyperbolic systems of moment equations by construction, it is extremely difficult 
to obtain Levermore's moment equations in an explicit form beyond the 10-moment case (which does not include the heat flux) because the fluxes associated with higher moments cannot be expressed in a closed form. 
As Levermore's 10-moment system is not capable of describing heat conduction, it is not very useful for describing gaseous processes. 
In addition, larger moment systems resulting from the maximum-entropy approach are prone to serious mathematical issues \citep{Junk1998, JU2002}.   
To alleviate problems associated with the maximum-entropy approach, \citet{MT2013} proposed affordable numerical approximations to the maximum-entropy closures for problems involving heat transfer 
and presented a robust and affordable version of Levermore's 14-moment system (that includes the heat flux). 
Although the 14-moment system proposed by \citet{MT2013} is capable of predicting accurate and smooth shock structures even for relatively large Mach numbers, it is not globally hyperbolic.
In a completely different approach that focused on producing hyperbolic moment equations, \cite{Torrilhon2010} introduced a novel closure computed with multi-variate Pearson-IV-distributions for the 13-moment system; however the approach seems unlikely to work for higher-order moment systems.
In another approach, \citet{ST2003} introduced a model, termed as the regularised 13-moment (R13) equations, which regularises the original Grad 13-moment (G13) equations by employing a CE-like expansion around a pseudo-equilibrium state.  Subsequently, the R13 equations were also derived in an elegant and clean way by \citet{Struchtrup2005} via the order of magnitude approach. 
The R13 equations are linearly stable, predict smooth shock structure for all Mach numbers and can capture several rarefaction effects, such as Knudsen layers, with good accuracy for sufficiently small Knudsen numbers \citep{ST2003, TS2004}.
To cover more transition-flow regime, \cite{GuEmerson2009} employed the regularisation approach of \cite{ST2003} to derive the regularised 26-moment (R26) equations. 
It may be noted, however, that the R13 and R26 equations are also not  hyperbolic.
The Grad and regularised moment equations consisting of an arbitrary number of moments have also been implemented and solved numerically by \cite{CaiLi2010}.
In the last few years, various other regularisation techniques that yield globally hyperbolic moment equations have been introduced. 
\citet{CFL2013} proposed a regularisation of the Grad
moment equations in one space dimension based on investigating the properties of the Jacobian matrix of fluxes in the system and derived globally hyperbolic moment equations (HME) in one space dimension. 
Further, \citet{CFL2014} generalised the method to derive HME in multi-dimension. 
\cite{KST2014} employed quadrature-based projection methods, which alter the structure of a moment system in a desired way, to obtain hyperbolic systems of the so-called quadrature-based moment equations (QBME).
\cite{FKLLT2016} proposed a generalised framework, which is capable of deriving various existing as well as some new systems of regularised hyperbolic moment equations, based on the  so-called operator projection method. 
A remarkable drawback of HME and QBME is that they cannot be written in a conservative form \citep{KT2017}. 
Consequently, the standard finite volume schemes cannot be applied to solve systems of HME and QBME numerically. 
Recently, some non-conservative numerical schemes have been proposed by \citet{KT2017, KT2018} for the numerical solution of QBME in one and two dimensions.
While numerical methods for solving general three-dimensional unsteady flow problems with moment equations are still intractable, the method of fundamental solution (MFS) enables us to develop efficient meshfree numerical methods for solving three-dimensional steady flow problems with the linearised moment equations. 
Recently, the MFS has been developed  for the G13 and R13 equations in \cite{LC2016} and \cite{CSRSL2017}, respectively.

The system of the R13 equations (also the system of the R26 equations), despite being non-hyperbolic, may be regarded as the most promising continuum model for describing rarefied monatomic gas flows since it is accompanied with appropriate boundary conditions \citep{GE2007, TS2008}, and has already been successful in describing a number of canonical flows \cite[see][and references therein]{TorrilhonARFM}.
Motivated from the accomplishments of the moment method (in particular, the R13 equations) in the case of monatomic gases, the Grad and 
regularised moment equations have also been developed for monatomic gas mixtures \citep{GT2015PRSA, Gupta2015, GST2016}. 
It is important to note here that the derivation of the regularised moment equations requires higher-order Grad moment equations, for instance, the derivations of the R13 and R26 equations require the Grad 26-moment (G26) and Grad 45-moment equations, respectively, and that most of the aforementioned works on the moment method employ either some simplified kinetic models to replace the Boltzmann collision operator or the Maxwell potential for molecular interactions.
The latter, introduced by \citet{Maxwell1867}, is inversely proportional to the fourth power of the distance  
between the colliding molecules and makes the collision rate independent of the relative velocity between the colliding molecules, which greatly simplifies the original Boltzmann equation. 
Remarkably, for Maxwell molecules (i.e.~for molecules interacting with the Maxwell interaction potential), the collisional production terms---the terms emanating from the Boltzmann collision operator in the moment equations---can be computed without the knowledge of explicit form of the distribution function and, moreover, they turn out to be  bilinear combinations of moments of the same or lower order, resulting into a one-way coupling on the right-hand sides of a moment system.
This makes the moment equations for Maxwell molecules tractable. 
For more details on the moment method for Maxwell molecules, the reader is referred to a review paper by \citet{Santos2009}. 

The development of kinetic theory of granular gases started out with two seminal works by \cite{JS1983} and \cite{LSJC1984}, which introduced kinetic theory for smooth inelastic hard spheres (IHS), followed by the pioneering work of \cite{JR1985PoF} on kinetic theory for rough inelastic hard disks (IHD).
The aforementioned methods, namely the CE expansion and the Grad moment method, in kinetic theory of classical gases have also been extended to granular gases, with the main goal of determining the NSF-level transport coefficients appearing in the expressions for the stress and heat flux, since the hydrodynamic equations closed with the NSF-level constitutive relations are sufficient to describe flows involving small spatial gradients.
The CE expansion to zeroth order was first employed by \citet{GS1995} to obtain the Euler-like hydrodynamic equations for rough granular flows.
Subsequently, \cite{BDKS1998} and \cite{GD1999} determined the NSF-level transport coefficients for dilute and dense granular gases of IHS, respectively, by means of the first-order CE expansion in powers of a uniformity parameter that estimates the strength of spatial gradients of the hydrodynamic field variables.
The derivation of Burnett equations (i.e.~the second-order CE expansion) even for the prototype model of a granular gas is an arduous task. Yet, by performing a generalised CE expansion in powers of two small parameters, namely the Knudsen number and the degree of inelasticity, 
\citet{SGJFM1998} 
determined the constitutive relations for the stress and heat flux up to Burnett order for a smooth granular gas of IHS. 
The requirement of the degree of inelasticity being small for performing asymptotic expansion limits the validity of Burnett equations derived by \cite{SGJFM1998} to nearly elastic granular gases.
\citet{Lutsko2005} further extended the CE expansion to dense granular fluids with arbitrary energy loss models and determined the NSF-level constitutive relations. Not only did his work consider arbitrary inelasticity but also a velocity-dependent coefficient of restitution,  providing the NSF-level constitutive relations for more realistic granular fluids. 

Granular flows of interest often fall beyond the regime covered by Newtonian hydrodynamics since the strength of spatial gradients in flows of practical interest is not small due to the inherent coupling between the spatial gradients and inelasticity \citep{Goldhirsch2003}. 
Consequently, for such flows, the granular NSF equations obtained from the first-order CE expansion are not adequate and the Burnett equations for IHS are not meaningful due to their validity being restricted to nearly elastic granular gases besides Bobylev's instability.
Such granular flows can alternatively be modelled by the Grad moment method. 
The method was extended to granular fluids first by Jenkins \& Richman who derived the G13 equations for a dense and smooth granular gas of IHS \citep{JR1985} and the Grad 16-moment equations for a dense and rough granular gas of IHD \citep{JR1985PoF}.  
It is well-established that the fourth cumulant (scalar fourth moment of the velocity distribution function) ought to be included in the list of the field variables for appropriate description of processes in granular gases; for instance, a theoretical description of the recently observed Mpemba effect in granular fluids requires the fourth cumulant as a field variable \citep{Mpemba_granular_PRL2017}. 
Keeping that in mind, \citet{RC2002} included the fourth cumulant in the list of the field variables to derive the Grad 9-moment equations for a bidimensional granular gas, and utilised them to investigate the homogeneous cooling state (HCS) and the steadily heated state of a bidimensional granular gas.
\citet{BST2004} attempted to extend the Grad moment method to one-dimensional dilute granular flows of viscoelastic hard spheres. 
It may be noted that all the aforementioned works on moment method for granular fluids are also  restricted to nearly elastic particles.
The Grad 14-moment (G14) equations for a dilute granular gas of IHS were introduced by \citet{KM2011} wherein the authors exploited the G14 equations to obtain the NSF-level constitutive relations for the stress and heat flux via the Maxwell iteration procedure and to investigate the linear stability of the HCS. 
Although the G14 equations introduced by \citet{KM2011} were not restricted to nearly elastic particles, their procedure to obtain the constitutive relations did not incorporate the effect of the collisional dissipation. Consequently, the constitutive relations determined by them are valid only for nearly elastic granular gases.
The issue was resolved by \citet{Garzo2013} who proposed a procedure to determine the NSF-level constitutive relations incorporating the contributions through the collisional dissipation as well. 
Although the work of \cite{Garzo2013} yielded the accurate NSF-level constitutive relations for moderately dense granular gases in a general dimension, it only computed the collisional contributions to stress and heat flux exploiting the G14 distribution function but did not provide the G14 equations explicitly.
Very recently, \cite{GST2018} derived the fully nonlinear G26 equations for dilute granular gases of IHS. 
Following the approach of \cite{Garzo2013}, they determined the NSF-level constitutive relations for the stress and heat flux through the G26 equations. 
The coefficient of the shear viscosity found by them through the G26 equations turned out to be the best among those obtained via any other theory so far. 
Notwithstanding, the other transport coefficients related to the heat flux obtained through the G26 equations in \cite{GST2018} were exactly the same as those obtained via the CE expansion at the first Sonine approximation \citep{BDKS1998} or via the G14 distribution function \citep{Garzo2013}, and the authors adduced that the Grad 29-moment (G29) theory, which includes the flux of the fourth cumulant as field variable, would be able to improve the transport coefficients related to the heat flux. 

Despite these ever-improving developments, the fact is that the Boltzmann equation for IHS, and hence the models stemming from the Boltzmann equation for IHS, remains difficult to deal with.
To circumvent the difficulties pertaining to models for IHS, a model of inelastic Maxwell molecules (IMM) was proposed at the beginning of this century \citep{B-NK2000, CCG2000, ErnstBrito2002}.
Similarly to the model of Maxwell molecules for monatomic gases, the IMM model also makes the collision rate of the inelastic Boltzmann equation independent of the relative velocity of the colliding molecules and thereby simplifies the inelastic Boltzmann equation greatly. 
In the past few years, the IMM model has received tremendous attention as the simple structure of the Boltzmann collision operator for IMM enables us to describe many properties of granular gases analytically, such as the high-velocity tails \citep{B-NK2002, ErnstBrito2002} and the fourth cumulant \citep{ErnstBrito2002, Santos2003} in the HCS, and the NSF-level transport coefficients \citep{Santos2003}.
Moreover, the experimental results on the velocity distribution in driven granular gases composed of magnetic grains are well-described by the IMM model \citep{KSSAOB-N2005}. 
The paper by \cite{GarzoSantos2011} presents a comprehensive review of the IMM model.
The two relevant works here are by \citet{Santos2003} and \citet{KGS2014}, which respectively derive the NSF- and Burnett-level transport coefficients for a $d$-dimensional dilute granular gas of IMM by means of the CE expansion.
It is worthwhile to note that 
the work of \citet{KGS2014}, in contrast to that of \cite{SGJFM1998}, contains only one smallness parameter, proportional to the spatial gradient of a hydrodynamic field,  
for performing the CE expansion but is not restricted to nearly elastic granular gases.
Nevertheless, as also pointed out in \citet{KGS2014} as a cautionary note, a regularisation of Burnett equations for IMM is apparently necessary to extricate Bobylev's instability.
Furthermore, as mentioned above, for a proper description of many processes in granular fluids, it is imperative to include the scalar fourth moment as a field variable.
Therefore, a moment-based modelling of granular gases seems to be necessary for proper description of processes involving large spatial gradients.

Aiming to the long-term perspective of establishing a complete set of predictive moment equations---for which appropriate boundary conditions, the MFS for steady flow problems and a general numerical framework for unsteady flow problems
can be developed---for granular gases, the main objective of this paper is to derive the Grad moment equations---comprising of up to $(d+3)(d^2+6d+2)/6$ moments---for a $d$-dimensional dilute (unforced) granular gas of IMM. 
Here, $d=2$ refers to planar disk flows and $d=3$ to three-dimensional sphere flows. 
Following the procedure due to \citet{Garzo2013}, the NSF-level transport coefficients for a dilute granular gas of IMM are determined from the derived Grad moment equations for IMM. 
The Grad $(d+3)(d^2+6d+2)/6$-moment equations are then utilised to study the HCS of a freely cooling granular gas of IMM.  
As it is well-known that the HCS of a granular gas is unstable but the instabilities are confined to large systems \citep[see, e.g.,][and references therein]{BP2004}, the linear stability of the HCS is investigated with the considered Grad moment systems and the results are employed to estimate the critical system size for the onset of instability.

It is worthwhile to note that a Grad moment system for a dilute granular gas differs from that for a rarefied monatomic gas only on the right-hand sides by virtue of different Boltzmann collision operators, therefore it is expected that a Grad moment system for dilute granular gases, similarly to a Grad moment system for monatomic gases, will also suffer from the loss of hyperbolicity. 
A detailed investigation of the hyperbolicity of the Grad moment systems derived in this paper and their regularisations will, however, be considered elsewhere in the future. 
From an application point of view, the Grad moment equations derived in the present work have limited applications at present due to unavailability of the associated boundary conditions, which are beyond the scope of the present paper and will also be considered elsewhere in the future. 
Nonetheless, the Grad moment equations developed in this paper can be utilised to investigate problems that do not require boundary conditions, e.g.~the shock-tube problem, by employing numerical techniques specialised to moment equations developed, for example, in \cite{Torrilhon2006, CaiLi2010, KT2017}.

The layout of the paper is as follows. 
The Boltzmann equation for IMM and the basic transport equations (i.e.~mass, momentum and energy balance equations)  for granular gases of IMM are presented in \S\,\ref{Sec:reviewKT}. 
The considered Grad moment systems are presented in \S\,\ref{Sec:GradMethod}.
The NSF transport coefficients for a dilute granular gas of IMM are determined from the Grad moment equations in \S\,\ref{Sec:NSF_rels}. 
The HCS of a freely cooling granular gas is explored through the Grad moment equations in \S\,\ref{Sec:HCS}. 
The linear stability analysis of the HCS is performed 
in \S\,\ref{Sec:Stability}. 
The paper ends with a short summary and conclusion in \S\,\ref{Sec:Conclusion}.%
\section{The Boltzmann equation and the hydrodynamic equations for IMM} \label{Sec:reviewKT}
We consider a dilute granular gas composed of smooth-identical-inelastic $d$-dimensional spherical particles of mass $m$ and diameter $\mathbbm{d}$. 
%
The state of such a gas can be fully described by a single-particle velocity distribution function $f\equiv f(t, \bm{x}, \bm{c})$---where $t$, $\bm{x}$ and $\bm{c}$ denote the time, position and instantaneous velocity of a particle, respectively---that obeys the inelastic Boltzmann equation \citep{BP2004}
\begin{align}
\label{BE}
\frac{\partial f}{\partial t}+c_i\,\frac{\partial f}{\partial x_i}+F_i\,\frac{\partial f}{\partial c_i}=J[\bm{c}|f,f],
\end{align} 
where $\bm{F}$ is the external force per unit mass that does not usually depend on $\bm{c}$, $J[\bm{c}|f,f]$ is the (inelastic) Boltzmann collision operator and the Einstein summation applies over repeated indices throughout the paper (unless mentioned otherwise). 
For $d$-dimensional IMM, the Boltzmann collision operator has a simplified form given by \citep{B-NK2002, ErnstBrito2002, GarzoSantos2007, Garzo2019}
\begin{align}
\label{CollOpMM}
J[\bm{c}|f,f]
=\frac{\mathring{\nu}}{n \Omega_d}\int_{\mathbb{R}^d} \!\int_{S^{d-1}} \!\left[\frac{1}{e} f(t, \bm{x},\bm{c}^{\prime\prime}) \, f(t, \bm{x},\bm{c}_1^{\prime\prime})
-f(t, \bm{x},\bm{c}) \, f(t, \bm{x},\bm{c}_1)\right]
\,\mathrm{d}\hat{\bm{k}}\,\mathrm{d}\bm{c}_1.
\end{align}
In the Boltzmann collision operator for IMM \eqref{CollOpMM}, $e$ is the (constant) coefficient of restitution and 
$\mathring{\nu} \equiv \mathring{\nu}(e)$---a free parameter in the  model---is an effective collision frequency that is typically chosen in such a way that the results from the Boltzmann equations for IHS and IMM agree in an optimal way \citep{GarzoSantos2007}. In particular, the agreement of cooling rates from the Boltzmann equations for IHS and IMM leads to \citep{GarzoSantos2007, KGS2014}
\begin{subequations}
\label{effeciveFreq}
\begin{align}
\mathring{\nu} = \frac{d+2}{2} \nu, \qquad \textrm{where} \qquad 
\nu = \frac{4 \, \Omega_d}{\sqrt{\pi} (d+2)} n \,\mathbbm{d}^{d-1} \sqrt{\frac{T}{m}}
\tag{\theequation $a$,$b$}
\end{align}
\end{subequations}
is the collision frequency associated with the Navier--Stokes shear viscosity of an elastic (monatomic) gas with $\Omega_d=2\pi^{d/2} / \Gamma(d/2)$ 
being the total solid angle in $d$ dimensions, $n\equiv n(t, \bm{x})$ the number density and $T\equiv T(t, \bm{x})$ the granular temperature, which is a measure of the fluctuating kinetic energy. 
The velocities $\bm{c}^{\prime\prime}$ and $\bm{c}_1^{\prime\prime}$ in \eqref{CollOpMM} are the pre-collisional velocities of the colliding molecules that transform to the post-collisional velocities $\bm{c}$ and $\bm{c}_1$ in an inverse collision following the relations \citep{SGJFM1998, BP2004}: 
\begin{align}
\label{VelTranInverse}
\bm{c}^{\prime\prime}=\bm{c}-\frac{1+e}{2e}(\hat{\bm{k}}\cdot\bm{g})\hat{\bm{k}}
\qquad\textrm{and}\qquad
\bm{c}_1^{\prime\prime}=\bm{c}_1+\frac{1+e}{2e}(\hat{\bm{k}}\cdot\bm{g})\hat{\bm{k}},
\end{align}
where $\bm{g}=\bm{c}-\bm{c}_1$ is the relative velocity of the colliding molecules, $\hat{\bm{k}}$ is the unit vector joining the centres of the colliding molecules at the time of collision.
The integration limits of $\hat{\bm{k}}$ in \eqref{CollOpMM} extend over the $d$-dimensional unit sphere $S^{d-1}$. 
Although the limits of integration will be dropped henceforth for the sake of succinctness, an integration over any velocity space will stand for the volume integral over $\mathbb{R}^d$ and that over $\hat{\bm{k}}$ will stand for the volume integral over the $d$-dimensional unit sphere $S^{d-1}$. 

The hydrodynamic variables---number density $n\equiv n(t, \bm{x})$, macroscopic velocity $\bm{v}\equiv \bm{v}(t, \bm{x})$ and granular temperature $T\equiv T(t, \bm{x})$---relate to the velocity distribution function via
\begingroup
\allowdisplaybreaks
\begin{align}
n(t, \bm{x})&=\int\! f(t, \bm{x}, \bm{c})\,\mathrm{d}\bm{c},
\\
n(t, \bm{x})\, \bm{v}(t, \bm{x})&=\int\! \bm{c}\,f(t, \bm{x}, \bm{c})\,\mathrm{d}\bm{c},
\\
\label{granTemp}
\frac{d}{2}n(t, \bm{x})\, T(t, \bm{x})&=\frac{1}{2}m\int\! C^2\,f(t, \bm{x}, \bm{c})\,\mathrm{d}\bm{c},
\end{align}
\endgroup
where $\bm{C}(t, \bm{x}, \bm{c})=\bm{c}-\bm{v}(t, \bm{x})$ is the peculiar velocity. 
The governing equations for the hydrodynamic variables---namely, the mass, momentum and energy balance equations---can be derived from the Boltzmann equation \eqref{BE} by multiplying it with $1$, $c_i$ and $\frac{1}{d}m\, C^2$, and integrating each of the resulting equations over the velocity space successively. 
The mass, momentum and energy balance equations, respectively, read 
\begingroup
\allowdisplaybreaks
\begin{align}
\label{massBal}
\frac{\mathrm{D} n}{\mathrm{D}t}+ n \frac{\partial v_i}{\partial x_i} &= 0,
\\
\label{momBal}
\frac{\mathrm{D} v_i} {\mathrm{D}t} + \frac{1}{m\,n} \left[\frac{\partial \sigma_{ij}}{\partial x_j} + \frac{\partial (n T)}{\partial x_i}\right] - F_i &= 0,
\\
\label{energyBal}
\frac{\mathrm{D} T} {\mathrm{D}t} + \frac{2}{d}\frac{1}{n} \left( \frac{\partial q_i}{\partial x_i} +\sigma_{ij}\frac{\partial v_i}{\partial x_j}+n T\frac{\partial v_i}{\partial x_i}\right) &= - \zeta\,T,
\end{align}
\endgroup
%
%
where $\mathrm{D}/\mathrm{D}t \equiv \partial / \partial t + v_k \, \partial / \partial x_k$ is the material derivative. The right-hand sides of the mass and momentum balance equations \eqref{massBal} and \eqref{momBal} vanish due to the conservation of mass and momentum. 
However, owing to dissipative collisions among grains, the energy is not conserved, yielding a nonzero right-hand side in the energy balance equation \eqref{energyBal} with the nonzero cooling rate $\zeta$ being given by
\begin{align}
\label{CoolingRate}
\zeta=-\frac{m}{d\,n\,T} \int \! C^2 \, J[\bm{c}|f,f] \, \mathrm{d}\bm{c}.
\end{align}
Furthermore, $\sigma_{ij}\equiv \sigma_{ij}(t, \bm{x})$ and $q_i\equiv q_i(t, \bm{x})$ in \eqref{momBal} and \eqref{energyBal} are the stress tensor and heat flux, respectively, and are given by
\begin{align}
\label{stressHFdef}
\sigma_{ij}=m\int\!\, C_{\langle i}C_{j\rangle} f\,\mathrm{d}\bm{c} \quad\textrm{and}\quad q_i=\frac{1}{2}m\int\!\, C^2 C_i f\,\mathrm{d}\bm{c},
\end{align}
where the angle brackets around the indices denote the symmetric and traceless part of the corresponding tensor; see appendix \ref{app:traceless} for its definition.

Needless to say, the system of mass, momentum and energy balance equations \eqref{massBal}--\eqref{energyBal} for the hydrodynamic variables $n$, $v_i$ and $T$ is not closed since it encompasses the additional unknowns $\sigma_{ij}$, $q_i$ and $\zeta$, and in order to deal with this system any further, it is indispensable to close it. Typically, the closure for system \eqref{massBal}--\eqref{energyBal} is obtained by means of the CE expansion, which yields the constitutive relations for $\sigma_{ij}$, $q_i$ and $\zeta$ to various orders of approximation \cite[see, e.g.,][]{BDKS1998, SGJFM1998, GD1999, Gupta2011, KGS2014}.
However, as also stated in 
\S\,\ref{Sec:intro}, system \eqref{massBal}--\eqref{energyBal} closed with the constitutive relations obtained at the zeroth and first orders of the CE expansion is not adequate for describing processes involving large spatial gradients while system \eqref{massBal}--\eqref{energyBal} closed with the constitutive relations obtained at the second and higher orders of the CE expansion suffers from Bobylev's instability. 
On the other hand, the Grad moment method is capable of yielding more accurate models that do not suffer from Bobylev's instability and are expected to be valid for  processes involving large spatial gradients.
Therefore, in what follows, the Grad moment method will be employed for deriving a few closed sets of macroscopic transport equations for a $d$-dimensional dilute granular gas of IMM.
\section{Grad moment method} \label{Sec:GradMethod}
The central goal of the moment method is to have reduced complexity while allowing for more accurate models for rarefied gases.  
It is well-known that the direct solutions of the Boltzmann equation are computationally expensive since the Boltzmann equation is solved for the velocity distribution function, which depends on total $2d+1$ variables (1 for time, $d$ for space and $d$ for velocity). 
The idea of moment method is to consider a finite number of equations for moments, instead of the Boltzmann equation, that depend only on $d+1$ variables (1 for time and $d$ for space); and the hope is that a sufficient number of moment equations would recover the solution from the Boltzmann equation (to a certain extent).
The details of the Grad moment method are skipped here for the sake of brevity but they---for monatomic gases---can be found in \cite{Grad1949} and in standard textbooks, e.g.~\cite{Struchtrup2005,Kremer2010}, and---for granular gases of three-dimensional hard spheres---
in \cite{GST2018}.%

Inclusion of the governing equations for the stress ($\sigma_{ij}$) and heat flux ($q_i$) along with the system of mass, momentum and energy balance equations \eqref{massBal}--\eqref{energyBal} leads to the well-known system of the 13-moment equations in three dimensions. 
In this paper, some Grad moment systems consisting of higher-order moments will also be derived and investigated. 
%
To this end, it is convenient to introduce a general symmetric-traceless moment
\begin{align}
\label{genmoment}
u_{i_1 i_2 \dots i_r}^{a} :=m\int\!C^{2a} C_{\langle i_1}C_{i_2}\dots C_{i_r\rangle}\, f\,\mathrm{d}\bm{c}, \quad a,r\in\mathbb{N}_0
\end{align} 
and its associated collisional production term (or collisional moment)
\begin{align}
\label{genProdTermMM}
\mathcal{P}_{i_1 i_2 \dots i_r}^a
:= m \int C^{2a} C_{\langle i_1}C_{i_2}\dots C_{i_r\rangle} \, J[\bm{c}|f,f] \, \mathrm{d}\bm{c},
\end{align}
where the angle brackets around the indices again denote the symmetric and traceless part of the corresponding quantity; see appendix~\ref{app:traceless} for its definition.
From definitions \eqref{genmoment} and \eqref{genProdTermMM}, it is straightforward to verify that $u^{0}=m\, n=\rho$, $u_i^{0}=0$, $u^{1}=d \, n\,T = d \, \rho \, \theta$, $u_{ij}^{0}=\sigma_{ij}$, $u_i^{1}=2\,q_i$, $\mathcal{P}^0=\mathcal{P}_i^0=0$ and $\mathcal{P}^1=-d\,n\,T\,\zeta$. 
Here $\rho=m\, n$ is the mass density and $\theta = T/m$. 
\subsection{Counting moments in \texorpdfstring{$d$}{} dimensions}
Before deriving the various moment systems, it is worthwhile to know how many moments a Grad moment system contains in a general dimension $d$. 
As it is more convenient to work with symmetric-traceless moments, the number of moments in a Grad moment system in a general dimension can be determined by knowing the number of independent components in a symmetric $r$-rank tensor and the number of traces in this tensor. 
Indeed, the number of independent components of a fully symmetric $r$-rank tensor in $d$ dimensions is $\binom{d+r-1}{r} = \frac{(d+r-1)!}{r! \, (d-1)!}$, and the number of traces in this tensor is $0$ for $r\in\{0,1\}$ while $\binom{d+r-3}{r-2} = \frac{(d+r-3)!}{(r-2)! \, (d-1)!}$ for 
$r\in \mathbb{N} \setminus \{1\}$.
Consequently, the number of independent components of a fully symmetric-traceless $r$-rank tensor ($r\in \mathbb{N} \setminus \{1\}$)
in $d$ dimensions is 
\begin{align*}
\binom{d+r-1}{r} - \binom{d+r-3}{r-2} 
=\frac{d + 2 r - 2}{d+r-2} \binom{d+r-2}{r}.
\end{align*}
Notably, any symmetric-traceless $r$-rank tensor ($r\in\mathbb{N}$) in two dimensions has only two independent components while any symmetric-traceless $r$-rank tensor ($r\in\mathbb{N}$) in three dimensions has $2r+1$ independent components. The counting of number of moments in some of the Grad moment systems considered in this paper is illustrated in table~\ref{table:counting}. 
Notwithstanding, any Grad moment system considered in this paper henceforth will be referred by its number of moments in three dimensions since Grad moment systems with the number of moments in three dimensions are more familiar to us \citep[see, e.g.,][]{JR1985, Levermore1996, Struchtrup2005, KM2011}. For instance, the Grad 
$\frac{d^2+5d+2}{2!}$-, $\frac{(d+1)(d^2+8d+6)}{3!}$- or $\frac{(d+3)(d^2+6d+2)}{3!}$-moment systems will simply be referred to as the Grad $13$-, $26$- or $29$-moment systems, respectively, which we are more acquainted with. 
%
%
%
{
\renewcommand{\arraystretch}{1.5}
\begin{table}
\setlength{\tabcolsep}{1em}
  \begin{center}
\def~{\hphantom{0}}
     \begin{tabular}{>{$}c<{$} >{$}c<{$}@{\hskip 1cm} c c c c}
        \text{Field variables} & \text{Unknowns in $d$ dimensions} 
        & \makecell{Unknowns in \\$3$ dimensions} & \makecell{Unknowns in \\$2$ dimensions} 
        \\
        \rho & \hspace*{2mm} 1 \rdelim\}{5}{-5mm}[\hspace*{0.1mm} $\frac{d^2+5d+2}{2!}$]\hspace*{1.6mm}
        \rdelim\}{8}{-5mm}[$\frac{(d+1)(d^2+8d+6)}{3!}$] 
        & \hspace*{0.5mm} 1 \hspace*{-2.5mm} \rdelim\}{5}{0pt}[\hspace*{0.1mm} 13] \hspace*{0.5mm}\rdelim\}{8}{0mm}[26] 
        & \hspace*{0.5mm} 1 \hspace*{-2.5mm} \rdelim\}{5}{0pt}[\hspace*{0.1mm} 8] \hspace*{0.5mm}\rdelim\}{8}{0mm}[13] 
        \\                
        v_i & d & 3 & 2                       \\   
        \theta & 1 & 1 & 1 \\
        \sigma_{ij} & \hspace*{-10mm} \frac{(d+2)(d-1)}{2!} & 5 & 2 \\
        q_i & d & 3 & 2 \\
        u_{ijk}^0 & \hspace*{-12mm} \frac{(d+4)d(d-1)}{3!} & 7 & 2 \\
        u^2 & 1 & 1 & 1 \\
        u_{ij}^1 & \hspace*{-10mm} \frac{(d+2)(d-1)}{2!} & 5 & 2
        \\
        u_i^2 & d & 3 & 2
        \\
        \hline 
        \text{Total} & \frac{(d+3)(d^2+6d+2)}{3!} & 29 & 15 \\
        \hline
      \end{tabular}
  \caption{Number of unknown field variables in Grad moment systems in $d$ dimensions}
  \label{table:counting}
  \end{center}
\end{table}
}
\subsection{The system of the 29-moment equations}
The system of the 29-moment equations includes the governing equations for the third rank tensor, for one- and full-traces of the fourth rank tensor and for full-trace of the fifth rank tensor along with the governing equations for the well-known 13 moments. 
In other words, the system of the 29-moment equations consists of the governing equations for the moments $n$, $v_i$, $T$, $\sigma_{ij}$, $q_i$, $u_{ijk}^0$, $u^2$, $u_{ij}^1$, $u_i^2$, and is obtained by multiplying the Boltzmann equation~\eqref{BE} with
$1$, $c_i$, $\frac{1}{d}m\, C^2$, $m\,C_{\langle i} C_{j\rangle}$, $\frac{1}{2}m\,C^2C_i$, $m\,C_{\langle i} C_j C_{k\rangle}$, $m\,C^4$, $m\,C^2 C_{\langle i} C_{j\rangle}$ and $m\,C^4 C_i$, and integrating each of the resulting equations over the velocity space successively. 
The detailed derivation of the 29-moment equations is provided as supplementary material. Here, they are presented directly.  
The system of the 29-moment equations consists of the mass, momentum and energy balance equations \eqref{massBal}--\eqref{energyBal} and other higher-order moment equations, which on using the abbreviations
\begin{align}
\label{abbreviations}
\left.
\begin{aligned}
m_{ijk}&:=u_{ijk}^0,
\qquad & \Delta &:= \frac{u^2}{d(d+2)\rho\theta^2} - 1,
\\
R_{ij}&:=u_{ij}^1-(d+4)\theta\sigma_{ij},
\qquad 
&\varphi_i &:= u_i^2 - 4 (d+4) \theta q_i
\end{aligned}
\right\}
\end{align}
read
\begingroup
\allowdisplaybreaks
\begin{align}
\label{eqn:stress}
\frac{\mathrm{D}\sigma_{ij}}{\mathrm{D}t}+\frac{\partial m_{ijk}}{\partial x_k}+\frac{4}{d+2}\frac{\partial q_{\langle i}}{\partial x_{j\rangle}}+\sigma_{ij}\frac{\partial v_k}{\partial x_k}+2\sigma_{k\langle i}\frac{\partial v_{j\rangle}}{\partial x_k}+2\rho\theta\frac{\partial v_{\langle i}}{\partial x_{j\rangle}}=\mathcal{P}_{ij}^0,
\end{align}
%
%
\begin{align}
\label{eqn:HF}
\frac{\mathrm{D}q_i}{\mathrm{D} t} 
&+ \frac{1}{2} \frac{\partial R_{ij}}{\partial x_j} 
+ \frac{d+2}{2} \left[ \rho\theta^2 \frac{\partial \Delta}{\partial x_i} 
+ \Delta \theta^2 \frac{\partial \rho}{\partial x_i}
+ (1 + 2 \Delta) \rho\theta \frac{\partial \theta}{\partial x_i} 
+ \sigma_{ij} \frac{\partial \theta}{\partial x_j}\right]
\nonumber\\
&+ \theta \frac{\partial \sigma_{ij}}{\partial x_j} 
- \frac{\sigma_{ij}}{\rho} \left(\frac{\partial \sigma_{jk}}{\partial x_k} 
- \theta \frac{\partial \rho}{\partial x_j} \right)
+ m_{ijk} \frac{\partial v_j}{\partial x_k}
\nonumber\\
&+ \frac{d+4}{d+2} q_i \frac{\partial v_j}{\partial x_j} 
+ \frac{d+4}{d+2} q_j \frac{\partial v_i}{\partial x_j} 
+ \frac{2}{d+2} q_j \frac{\partial v_j}{\partial x_i} 
=\frac{1}{2} \mathcal{P}_i^1,
\end{align} 
\begin{align}
\label{eqn:mijk}
\frac{\mathrm{D}m_{ijk}}{\mathrm{D} t} 
&+ \frac{\partial u_{ijkl}^0}{\partial x_l} 
+ \frac{3}{d+4} \frac{\partial R_{\langle ij}}{\partial x_{k \rangle}} 
+ 3 \theta \frac{\partial \sigma_{\langle ij}}{\partial x_{k \rangle}} 
-3 \frac{\sigma_{\langle ij}}{\rho} \left(\frac{\partial \sigma_{k\rangle l}}{\partial x_l}+\theta\frac{\partial \rho}{\partial x_{k\rangle}}\right)
\nonumber\\
&+ m_{ijk}\frac{\partial v_l}{\partial x_l} 
+ 3 m_{l \langle ij} \frac{\partial v_{k \rangle}}{\partial x_l} + \frac{12}{d+2} q_{\langle i} \frac{\partial v_j}{\partial x_{k \rangle}}
=\mathcal{P}_{ijk}^0,
\end{align} 
%
%
%
\begin{align}
\label{eqn:Delta}
\frac{\mathrm{D} \Delta}{\mathrm{D} t}
&+ \frac{8}{d(d+2)} \frac{1}{\rho\theta} \left(1 - \frac{d+2}{2}\Delta\right) \left(\frac{\partial q_i}{\partial x_i} + \sigma_{ij} \frac{\partial v_i}{\partial x_j} \right)
\nonumber\\
&+ \frac{1}{d(d+2)} \frac{1}{\rho\theta^2} \left[
\frac{\partial \varphi_i}{\partial x_i}
+ 4 (d+2) q_i \frac{\partial \theta}{\partial x_i} 
- 8 \frac{q_i}{\rho} \left( \frac{\partial \sigma_{ij}}{\partial x_j} + \theta \frac{\partial \rho}{\partial x_i} \right)
+4 R_{ij} \frac{\partial v_i}{\partial x_j}
\right]
\nonumber\\
&= \frac{1}{d(d+2)} \frac{1}{\rho\theta^2} 
\Big[\mathcal{P}^2 - 2 (d+2) (1+\Delta) \theta \, \mathcal{P}^1\Big],
\end{align}
\begin{align}
\label{eqn:Rij}
\frac{\mathrm{D} R_{ij}}{\mathrm{D} t} 
&+ \frac{2}{d+2} \frac{\partial \varphi_{\langle i}}{\partial x_{j \rangle}} 
+ \frac{4(d+4)}{d+2} \left( \theta \frac{\partial q_{\langle i}}{\partial x_{j \rangle}} 
+ q_{\langle i} \frac{\partial \theta}{\partial x_{j \rangle}} 
- \frac{q_{\langle i}}{\rho} \frac{\partial \sigma_{j \rangle k}}{\partial x_k}
- \frac{\theta}{\rho} q_{\langle i} \frac{\partial \rho}{\partial x_{j \rangle}}
\right)
\nonumber\\
&
+ 4 \theta \sigma_{k\langle i} \frac{\partial v_k}{\partial x_{j \rangle}}
+ 4 \theta \sigma_{k\langle i} \frac{\partial v_{j \rangle}}{\partial x_k}
- \frac{8}{d} \theta\sigma_{ij}\frac{\partial v_k}{\partial x_k}
- \frac{2(d+4)}{d} \frac{\sigma_{ij}}{\rho}\left( \frac{\partial q_k}{\partial x_k} 
+\sigma_{kl}\frac{\partial v_k}{\partial x_l}
\right)
\nonumber\\
&+ \frac{\partial u_{ijk}^1}{\partial x_k} 
- (d+4) \theta \frac{\partial m_{ijk}}{\partial x_k}
- 2 \frac{m_{ijk}}{\rho} \left(\frac{\partial \sigma_{kl}}{\partial x_l} 
+ \rho \frac{\partial \theta}{\partial x_k} + \theta \frac{\partial \rho}{\partial x_k}\right)
+2 u_{ijkl}^0 \frac{\partial v_k}{\partial x_l} 
\nonumber\\
&+ \frac{d+6}{d+4} \left( R_{ij} \frac{\partial v_k}{\partial x_k} 
+ 2 R_{k\langle i} \frac{\partial v_{j \rangle}}{\partial x_k} 
\right)
+ \frac{4}{d+4} R_{k\langle i} \frac{\partial v_k}{\partial x_{j \rangle}} 
+ 2(d+4) \Delta \rho \theta^2  \frac{\partial v_{\langle i}}{\partial x_{j \rangle}}
\nonumber\\
&=\mathcal{P}_{ij}^1 - (d+4) \theta \mathcal{P}_{ij}^0 - \frac{d+4}{d} \frac{\sigma_{ij}}{\rho} \mathcal{P}^1,
\end{align}
\begin{align}
\label{eqn:phii}
\frac{\mathrm{D} \varphi_i}{\mathrm{D} t} 
&- \frac{8(d+4)}{d}\frac{q_i}{\rho} \left( \frac{\partial q_j}{\partial x_j} +\sigma_{jk}\frac{\partial v_j}{\partial x_k}+\rho\theta\frac{\partial v_j}{\partial x_j}\right) 
+ \frac{\partial u_{ij}^2}{\partial x_j} 
+ \frac{1}{d} \frac{\partial u^3}{\partial x_i} 
- 2 (d+4) \theta \frac{\partial R_{ij}}{\partial x_j} 
\nonumber\\
&- 4 R_{ij} \frac{\partial \theta}{\partial x_j}
- (d+4)\big[ (d+6) + (d+2) \Delta \big] \theta^2 \frac{\partial \sigma_{ij}}{\partial x_j} 
- 2 (d+4)^2  \theta \sigma_{ij} \frac{\partial \theta}{\partial x_j}
\nonumber\\
&- (d+2) (d+4) \left[ 2 \rho\theta^3 \frac{\partial \Delta}{\partial x_i} 
+(1+3\Delta)  \theta^3 \frac{\partial \rho}{\partial x_i}
+(3 + 5 \Delta) \rho\theta^2 \frac{\partial \theta}{\partial x_i}
\right]
\nonumber\\
&- 4 \frac{R_{ij}}{\rho} \left(\frac{\partial \sigma_{jk}}{\partial x_k} + \theta \frac{\partial \rho}{\partial x_j}\right)  
+ 4 u_{ijk}^1 \frac{\partial v_j}{\partial x_k} 
- 4 (d+4) \theta m_{ijk} \frac{\partial v_j}{\partial x_k}
\nonumber\\
&+ \frac{8 (d+4)}{d+2} \theta \left( 
 q_i \frac{\partial v_j}{\partial x_j} 
+ q_j \frac{\partial v_i}{\partial x_j} 
+ q_j \frac{\partial v_j}{\partial x_i}
\right)
+ \frac{d+6}{d+2} \varphi_i \frac{\partial v_j}{\partial x_j} 
+ \frac{d+6}{d+2} \varphi_j \frac{\partial v_i}{\partial x_j} 
\nonumber\\
&+ \frac{4}{d+2} \varphi_j \frac{\partial v_j}{\partial x_i}
=\mathcal{P}_i^2 
- 2 (d+4) \theta \mathcal{P}_i^1
- \frac{4 (d+4)}{d} \frac{q_i }{\rho} \mathcal{P}^1.
\end{align}
\endgroup
The abbreviations \eqref{abbreviations} are introduced in such a way that 
$m_{ijk}$, $\Delta$, $R_{ij}$ and $\varphi_i$ vanish if computed with the well-known G13 distribution function 
\begin{align}
\label{G13disfun}
f_{|\mathrm{G13}}
&=f_M\left[1 + \frac{1}{2} \frac{\sigma_{ij} C_i C_j}{\rho\theta^2} + \frac{q_i C_i}{\rho\theta^2}\left( \frac{1}{d+2}\frac{C^2}{\theta}-1\right)  \right],
\end{align}
where 
\begin{align}
\label{Maxwellian}
f_M\equiv f_M(t, \bm{x}, \bm{c})=n\left(\frac{1}{2\,\pi\, \theta}\right)^{\!d/2}\exp{\left(-\frac{C^2}{2\,\theta}\right)}
\end{align}
is the Maxwellian distribution function~\citep{Garzo2013}. 
%
%
In general, the computation of the collisional production terms $\mathcal{P}_{i_1 i_2 \dots i_r}^a$ requires the knowledge of the distribution function and is not easy for particles interacting with a general interaction potential.  
Nevertheless, for IMM (considered in this work), the collisional production terms can be evaluated easily---indeed, without the knowledge of the explicit form of the distribution function. 
A strategy for computing them for IMM in an automated way using the computer algebra software {\textsc{Mathematica}}\textsuperscript{\textregistered} is demonstrated in appendix~\ref{app:CompProd}. 
Using this strategy, the production terms associated with the G29 equations for $d$-dimensional IMM have been computed. They turn out to be
\begingroup
\allowdisplaybreaks
\begin{align}
\label{P1}
\mathcal{P}^1=& - \zeta_0^\ast \, \nu\,d \,\rho\,\theta
\\
\label{Pij0}
\mathcal{P}_{ij}^0=& - \nu_\sigma^\ast \, \nu\,\sigma_{ij},
\\
\label{Pi1}
\mathcal{P}_i^1=& - 2 \, \nu_q^\ast \, \nu\,q_i,
\\
\label{Pijk0}
\mathcal{P}_{ijk}^0=& - \nu_m^\ast \, \nu\,m_{ijk},
\\
\label{P2}
\mathcal{P}^2=& - \nu \left[\big(\alpha_0 + \alpha_1 \Delta \big) \rho\,\theta^2 + d(d+2) \varsigma_0 \frac{\sigma_{ij}\sigma_{ij}}{\rho}\right],
\\
\label{Pij1}
\mathcal{P}_{ij}^1=& - \nu \left[\nu_R^\ast \, R_{ij} + \alpha_2 \theta \sigma_{ij} + \varsigma_1 \frac{\sigma_{k\langle i}\sigma_{j\rangle k}}{\rho}\right],
\\
\label{Pi2}
\mathcal{P}_i^2=& - \nu \left[\nu_\varphi^\ast \, \varphi_i + \alpha_3 \theta q_i + \varsigma_2 \frac{\sigma_{ij} q_j}{\rho} 
+ \varsigma_3 \frac{m_{ijk} \sigma_{jk}}{\rho}
\right],
\end{align}
\endgroup
where the coefficients $\zeta_0^\ast$, $\nu_\sigma^\ast$, $\nu_q^\ast$, $\nu_m^\ast$, $\nu_R^\ast$, $\nu_\varphi^\ast $, $\alpha_0$, $\alpha_1$, $\alpha_2$, $\alpha_3$, $\varsigma_0$, $\varsigma_1$, $\varsigma_2$ and $\varsigma_3$ depend only on the dimension $d$ and coefficient of restitution $e$, and are relegated to appendix~\ref{app:coeffprodterms} for better readability. 
Collisional production terms \eqref{P1}--\eqref{Pij1} for IMM agree with those obtained in \cite{GarzoSantos2007}, wherein they have been computed till fourth order. 
Moreover, the coefficients $\zeta_0^\ast$, $\nu_\sigma^\ast$, $\nu_q^\ast$, $\nu_R^\ast$, $\nu_\varphi^\ast $, and $\alpha_1$ relate to the collisional rate $\nu_{2r|s}$---associated with the Ikenberry polynomial $Y_{2r|i_1 i_2 \dots i_s}(\bm{C})$---given in \cite{SG2012} for $s\in\{0,1,2\}$ via 
$\zeta_0^\ast \, \nu = \nu_{2|0}$, 
$\nu_\sigma^\ast \, \nu = \nu_{0|2}$, 
$\nu_q^\ast \, \nu = \nu_{2|1}$, 
$\nu_R^\ast \, \nu = \nu_{2|2}$, 
$\nu_\varphi^\ast \, \nu = \nu_{4|1}$ and 
$\alpha_1 \, \nu = d (d+2) \nu_{4|0}$. 
I could not find the full expression for the collisional production term \eqref{Pi2} for granular gases in the existing literature. Nonetheless, for monatomic gases (i.e.~for $d=3$ and $e=1$), it can be found, for instance, in \cite{GuEmerson2009}---although not explicitly.
The source code for computing the above collisional production terms is provided as supplementary material with the present paper.
The collisional production terms associated with the G26 equations for three-dimensional IHS can be found in \cite{GT2012, GST2018}.

The relation $\mathcal{P}^1=-d\,n\,T\,\zeta$ on exploiting \eqref{P1} gives the cooling rate for IMM:
\begin{align}
\label{coolingRateG29MM}
\zeta = \zeta_0^\ast \, \nu, 
\end{align}
where $\zeta_0^\ast = (d+2)(1-e^2) / (4d)$ (see \eqref{zeta0ast}). 
The cooling rate~\eqref{coolingRateG29MM} is the same as that obtained in \cite{Santos2003, GarzoSantos2011, KGS2014}, and vanishes identically for monatomic gases (i.e.~for $e=1$), guaranteeing the conservation of energy for them. It is important to note from \eqref{coolingRateG29MM} that the cooling rate for IMM neither depends on the gradients of any field nor on any higher-order moment (in contrast to the cooling rate for IHS that also depends on the scalar fourth moment $\Delta$; see \cite{GST2018}). 
%
%
\subsection{Grad 29-moment closure}
The system of the 29-moment equations for IMM (eqs.~\eqref{massBal}--\eqref{energyBal} and \eqref{eqn:stress}--\eqref{eqn:phii} along with collisional production terms \eqref{P1}--\eqref{Pi2}) is still not closed as it possesses the additional unknown moments $u_{ijkl}^0$, $u_{ijk}^1$, $u_{ij}^2$, $u^3$. 
The system is closed with the Grad distribution function based on the considered 29 moments, which is referred to as the G29 distribution function. 
The ($d$-dimensional) G29 distribution function $f_{|\mathrm{G29}}$ reads
\begin{align}
\label{G29disfun}
f_{|\mathrm{G29}}
&=f_M\left[1 + \frac{1}{2} \frac{\sigma_{ij} C_i C_j}{\rho\theta^2} + \frac{q_i C_i}{\rho\theta^2}\left( \frac{1}{d+2}\frac{C^2}{\theta}-1\right) 
 + \frac{1}{6}\frac{m_{ijk} C_i C_j C_k}{\rho\theta^3} 
 \right.\nonumber\\
&\left.\quad + \frac{d(d+2) \Delta}{8} \left(1 - \frac{2}{d} \frac{C^2}{\theta} + \frac{1}{d(d+2)}\frac{C^4}{\theta^2} \right)
+ \frac{1}{4}  \frac{R_{ij} C_i C_j}{\rho\theta^3} \left(\frac{1}{d+4}\frac{C^2}{\theta}-1\right)
\right.\nonumber\\
&\left.\quad + \frac{1}{8}
\frac{\varphi_i C_i}{\rho\theta^3}  \left(1- \frac{2}{d+2} \frac{C^2}{\theta} + \frac{1}{(d+2)(d+4)} \frac{C^4}{\theta^2}\right)
\right].
\end{align}
%
The details of computing the G29 distribution function \eqref{G29disfun} can be found in appendix~\ref{app:G29}. 
Insertion of the G29 distribution function \eqref{G29disfun} into the definitions of unknown moments $u_{ijkl}^0$, $u_{ijk}^1$, $u_{ij}^2$ and $u^3$ expresses them in terms of the considered 29 moments:
\begin{subequations}
\allowdisplaybreaks
\label{closure}
\begin{align}
u_{ijkl|\mathrm{G29}}^0&=0,
\\
u_{ijk|\mathrm{G29}}^1 &= (d+6)\,\theta\, m_{ijk},
\\
u_{ij|\mathrm{G29}}^2 &= (d+6)\,\theta\big[2 R_{ij} + (d+4) \theta \sigma_{ij} \big],
\\
u_{|\mathrm{G29}}^3 &= d(d+2)(d+4)(1+3\Delta)\rho \theta^3,
\end{align}
\end{subequations}
where the subscript ``$|\mathrm{G29}$" denotes that these moments are computed with the G29 distribution function \eqref{G29disfun}. 
%
\subsection{The G29 system for IMM}
Equations \eqref{massBal}--\eqref{energyBal} and \eqref{eqn:stress}--\eqref{eqn:phii} closed with \eqref{closure} and \eqref{P1}--\eqref{Pi2} form the system of the G29 equations for $d$-dimensional IMM. Combining all of them, the system of the G29 equations for $d$-dimensional IMM reads
\begin{align}
\label{massBalG29}
\frac{\mathrm{D} n}{\mathrm{D}t}+ n \frac{\partial v_i}{\partial x_i}=0,
\end{align}
%
\begin{align}
\label{momBalG29}
\frac{\mathrm{D} v_i} {\mathrm{D}t} + \frac{1}{m\,n} \left[\frac{\partial \sigma_{ij}}{\partial x_j} + \frac{\partial (n\, T)}{\partial x_i}\right] - F_i=0,
\end{align}
%
\begin{align}
\label{energyBalG29}
\frac{\mathrm{D} T} {\mathrm{D}t} + \frac{2}{d}\frac{1}{n} \left[ \frac{\partial q_i}{\partial x_i} +\sigma_{ij}\frac{\partial v_i}{\partial x_j}+n \,T\frac{\partial v_i}{\partial x_i}\right] 
= - \zeta_0^\ast \, \nu \, T,
\end{align}
\begin{align}
\label{eqn:stressG29}
\frac{\mathrm{D}\sigma_{ij}}{\mathrm{D}t}+\frac{\partial m_{ijk}}{\partial x_k}+\frac{4}{d+2}\frac{\partial q_{\langle i}}{\partial x_{j\rangle}}+\sigma_{ij}\frac{\partial v_k}{\partial x_k}+2\sigma_{k\langle i}\frac{\partial v_{j\rangle}}{\partial x_k}+2\rho\theta\frac{\partial v_{\langle i}}{\partial x_{j\rangle}} = - \nu_\sigma^\ast \, \nu\,\sigma_{ij},
\end{align}
\begin{align}
\label{eqn:HFG29}
\frac{\mathrm{D}q_i}{\mathrm{D} t} 
&+ \frac{1}{2} \frac{\partial R_{ij}}{\partial x_j} 
+ \frac{d+2}{2} \left[ \rho\theta^2 \frac{\partial \Delta}{\partial x_i} 
+ \Delta \theta^2 \frac{\partial \rho}{\partial x_i}
+ (1 + 2 \Delta) \rho\theta \frac{\partial \theta}{\partial x_i} 
+ \sigma_{ij} \frac{\partial \theta}{\partial x_j}\right]
\nonumber\\
&+ \theta \frac{\partial \sigma_{ij}}{\partial x_j} 
- \frac{\sigma_{ij}}{\rho} \left(\frac{\partial \sigma_{jk}}{\partial x_k} 
- \theta \frac{\partial \rho}{\partial x_j} \right)
+ m_{ijk} \frac{\partial v_j}{\partial x_k}
\nonumber\\
&+ \frac{d+4}{d+2} q_i \frac{\partial v_j}{\partial x_j} 
+ \frac{d+4}{d+2} q_j \frac{\partial v_i}{\partial x_j} 
+ \frac{2}{d+2} q_j \frac{\partial v_j}{\partial x_i} 
= - \nu_q^\ast \, \nu\,q_i,
\end{align} 
\begin{align}
\label{eqn:mijkG29}
\frac{\mathrm{D}m_{ijk}}{\mathrm{D} t} 
&+ \frac{3}{d+4} \frac{\partial R_{\langle ij}}{\partial x_{k \rangle}} 
+ 3 \theta \frac{\partial \sigma_{\langle ij}}{\partial x_{k \rangle}} 
-3 \frac{\sigma_{\langle ij}}{\rho} \left(\frac{\partial \sigma_{k\rangle l}}{\partial x_l}+\theta\frac{\partial \rho}{\partial x_{k\rangle}}\right)
\nonumber\\
&+ m_{ijk}\frac{\partial v_l}{\partial x_l} 
+ 3 m_{l \langle ij} \frac{\partial v_{k \rangle}}{\partial x_l} + \frac{12}{d+2} q_{\langle i} \frac{\partial v_j}{\partial x_{k \rangle}}
=- \nu_m^\ast \, \nu\,m_{ijk},
\end{align} 
\begin{align}
\label{eqn:DeltaG29}
\frac{\mathrm{D} \Delta}{\mathrm{D} t}
&+ \frac{8}{d(d+2)} \frac{1}{\rho\theta} \left(1 - \frac{d+2}{2}\Delta\right) \left(\frac{\partial q_i}{\partial x_i} + \sigma_{ij} \frac{\partial v_i}{\partial x_j} \right)
\nonumber\\
&+ \frac{1}{d(d+2)} \frac{1}{\rho\theta^2} \left[
\frac{\partial \varphi_i}{\partial x_i}
+ 4 (d+2) q_i \frac{\partial \theta}{\partial x_i} 
- 8 \frac{q_i}{\rho} \left( \frac{\partial \sigma_{ij}}{\partial x_j} + \theta \frac{\partial \rho}{\partial x_i} \right)
+4 R_{ij} \frac{\partial v_i}{\partial x_j}
\right]
\nonumber\\
&= - \nu \left[ \nu_{\!\Delta}^\ast \left\{ \Delta - \frac{6(1-e)^2}{4d-7+6e-3e^2} \right\} +  \underline{
\varsigma_0 \frac{\sigma_{ij}\sigma_{ij}}{\rho^2\theta^2}
}
\right],
\end{align}
\begin{align}
\label{eqn:RijG29}
\frac{\mathrm{D} R_{ij}}{\mathrm{D} t} 
&+ \frac{2}{d+2} \frac{\partial \varphi_{\langle i}}{\partial x_{j \rangle}} 
+ \frac{4(d+4)}{d+2} \left( \theta \frac{\partial q_{\langle i}}{\partial x_{j \rangle}} 
+ q_{\langle i} \frac{\partial \theta}{\partial x_{j \rangle}} 
- \frac{q_{\langle i}}{\rho} \frac{\partial \sigma_{j \rangle k}}{\partial x_k}
- \frac{\theta}{\rho} q_{\langle i} \frac{\partial \rho}{\partial x_{j \rangle}}
\right)
\nonumber\\
&
+ 4 \theta \sigma_{k\langle i} \frac{\partial v_k}{\partial x_{j \rangle}}
+ 4 \theta \sigma_{k\langle i} \frac{\partial v_{j \rangle}}{\partial x_k}
- \frac{8}{d} \theta\sigma_{ij}\frac{\partial v_k}{\partial x_k}
- \frac{2(d+4)}{d} \frac{\sigma_{ij}}{\rho}\left( \frac{\partial q_k}{\partial x_k} 
+\sigma_{kl}\frac{\partial v_k}{\partial x_l}
\right)
\nonumber\\
&+ 2\theta \frac{\partial m_{ijk}}{\partial x_k}
+ (d+4) m_{ijk} \frac{\partial \theta}{\partial x_k}
- 2 \frac{m_{ijk}}{\rho} \left(\frac{\partial \sigma_{kl}}{\partial x_l} + \theta \frac{\partial \rho}{\partial x_k}\right)
\nonumber\\
&+ \frac{d+6}{d+4} \left( R_{ij} \frac{\partial v_k}{\partial x_k} 
+ 2 R_{k\langle i} \frac{\partial v_{j \rangle}}{\partial x_k} 
\right)
+ \frac{4}{d+4} R_{k\langle i} \frac{\partial v_k}{\partial x_{j \rangle}} 
+ 2(d+4) \Delta \rho \theta^2  \frac{\partial v_{\langle i}}{\partial x_{j \rangle}}
\nonumber\\
&=-\nu \left(\nu_R^\ast R_{ij} - \nu_{R\sigma}^\ast \theta \sigma_{ij} + \underline{
\varsigma_1 \frac{\sigma_{k\langle i}\sigma_{j\rangle k}}{\rho}
}
\right),
\end{align}
\begin{align}
\label{eqn:phiiG29}
\frac{\mathrm{D} \varphi_i}{\mathrm{D} t} 
&- \frac{8(d+4)}{d}\frac{q_i}{\rho} \left( \frac{\partial q_j}{\partial x_j} +\sigma_{jk}\frac{\partial v_j}{\partial x_k}+\rho\theta\frac{\partial v_j}{\partial x_j}\right) 
+ 4 \theta \frac{\partial R_{ij}}{\partial x_j} 
\nonumber\\
&+ (d+2) (d+4) \theta^2 \left[ \rho\theta \frac{\partial \Delta}{\partial x_i} 
+ 4 \Delta \rho \frac{\partial \theta}{\partial x_i}
- \Delta \frac{\partial \sigma_{ij}}{\partial x_j}
\right]
\nonumber\\
&- 4 \frac{R_{ij}}{\rho} \left(\frac{\partial \sigma_{jk}}{\partial x_k} + \theta \frac{\partial \rho}{\partial x_j}\right)  
+ 2(d+4) R_{ij} \frac{\partial \theta}{\partial x_j}
+4 (d+4) \theta \sigma_{ij} \frac{\partial \theta}{\partial x_j}
+ 8 \theta m_{ijk} \frac{\partial v_j}{\partial x_k}
\nonumber\\
&+ \frac{8 (d+4)}{d+2} \theta \left( 
 q_i \frac{\partial v_j}{\partial x_j} 
+ q_j \frac{\partial v_i}{\partial x_j} 
+ q_j \frac{\partial v_j}{\partial x_i}
\right)
+ \frac{d+6}{d+2} \varphi_i \frac{\partial v_j}{\partial x_j} 
+ \frac{d+6}{d+2} \varphi_j \frac{\partial v_i}{\partial x_j} 
\nonumber\\
&+ \frac{4}{d+2} \varphi_j \frac{\partial v_j}{\partial x_i}
= - \nu \left(
\nu_\varphi^\ast \, \varphi_i - \nu_{\varphi q}^\ast \theta q_i 
+ \underline{
\varsigma_2 \frac{\sigma_{ij}q_j}{\rho} + \varsigma_3 \frac{m_{ijk} \sigma_{jk}}{\rho}
}
\right),
\end{align}
where
\begin{align}
\left.
\begin{aligned}
\nu_{\!\Delta}^\ast &= \frac{(1+e)^2(4d-7+6e-3e^2)}{16d},
\\
\nu_{R\sigma}^\ast&= \frac{3(1+e)^2 (1-e) (d+2-2e)}{4d},
\\
\nu_{\varphi q}^\ast & = \frac{3(1+e)^2 (1-e) [5(d+2)-(d+14) e]}{4d},
\end{aligned}
\right\}
\end{align}
and the other coefficients 
$\zeta_0^\ast$, 
$\nu_\sigma^\ast$, $\nu_q^\ast$, $\nu_m^\ast$, 
$\nu_R^\ast$, $\nu_\varphi^\ast$,  
$\varsigma_0$, $\varsigma_1$, $\varsigma_2$ and $\varsigma_3$ appearing on the right-hand sides of the G29 equations \eqref{massBalG29}--\eqref{eqn:phiiG29} depend only on the dimension $d$ and coefficient of restitution $e$ (see 
appendix~\ref{app:coeffprodterms} for their expressions).
%
For $d=3$ and $e=1$, these coefficients become $\nu_\sigma^\ast = 1$, $\nu_q^\ast = 2/3$, $\nu_m^\ast = 3/2$, $\nu_{\!\Delta}^\ast = 2/3$, $\nu_R^\ast = 7/6$, $\nu_\varphi^\ast = 1$, $\nu_{R\sigma}^\ast = 0$, $\nu_{\varphi q}^\ast = 0$, $\varsigma_0 = 2/45$, $\varsigma_1 = 2/3$, $\varsigma_2 = 28/15$ and $\varsigma_3 = 2/3$, which are the same as the respective coefficients for monatomic gases of Maxwell molecules; see, e.g.,~\cite{Struchtrup2005} and \cite{GuEmerson2009}. 
In particular, the vanishing coefficients $\nu_{R\sigma}^\ast = 0$ and  $\nu_{\varphi q}^\ast = 0$ make the right-hand sides of the linearised G29 equations for monatomic gases of Maxwell molecules completely decoupled, which is not the case for granular gases.
Furthermore, the underlined nonlinear terms in \eqref{eqn:DeltaG29}--\eqref{eqn:phiiG29} will be discarded for simplicity while investigating the HCS of a granular gas in \S\,\ref{Sec:HCS}.
%
\subsection{Various Grad moment systems}\label{Subsec:various}
%
The abbreviations \eqref{abbreviations} have been introduced in such a way that the smaller systems of the Grad moment equations can be obtained directly from the G29 system \eqref{massBalG29}--\eqref{eqn:phiiG29}. 
The other Grad moment systems considered in this paper are as follows.
\begin{enumerate}\itemsep1.5ex
\item 
\emph{The G13 system:}
The system of the 13-moment equations contains the governing equations for variables $n$, $v_i$, $T$, $\sigma_{ij}$ and $q_i$, i.e.~it consists of equations \eqref{massBalG29}--\eqref{eqn:HFG29}. 
However, equations \eqref{massBalG29}--\eqref{eqn:HFG29} contain additional unknowns $m_{ijk}$, $\Delta$ and $R_{ij}$ that vanish on being computed with the G13 distribution function \eqref{G13disfun}.
%
Thus, the G13 system for $d$-dimensional IMM consists of equations \eqref{massBalG29}--\eqref{eqn:HFG29} with $m_{ijk}=\Delta=R_{ij}=0$.
%
\item 
\emph{The G14 system:}
The system of the 14-moment equations contains the governing equations for variables $n$, $v_i$, $T$, $\sigma_{ij}$, $q_i$ and $\Delta$, i.e.~it consists of equations \eqref{massBalG29}--\eqref{eqn:HFG29} and \eqref{eqn:DeltaG29}. 
However, equations \eqref{massBalG29}--\eqref{eqn:HFG29} and \eqref{eqn:DeltaG29} contain additional unknowns $m_{ijk}$, $R_{ij}$ and $\varphi_i$ that also vanish on being computed with the G14 distribution function 
\begin{align}
\label{G14disfun}
f_{|\mathrm{G14}}
&=f_M\left[1 + \frac{1}{2} \frac{\sigma_{ij} C_i C_j}{\rho\theta^2} + \frac{q_i C_i}{\rho\theta^2}\left( \frac{1}{d+2}\frac{C^2}{\theta}-1\right) 
\right.\nonumber\\
&\left.\quad 
+ \frac{d(d+2) \Delta}{8} \left(1 - \frac{2}{d} \frac{C^2}{\theta} + \frac{1}{d(d+2)}\frac{C^4}{\theta^2} \right)
\right],
\end{align}
which can be obtained easily by following a similar procedure presented in appendix~\ref{app:G29}.
%
%
Thus, the G14 system for $d$-dimensional IMM consists of equations \eqref{massBalG29}--\eqref{eqn:HFG29} and \eqref{eqn:DeltaG29} with $m_{ijk}=R_{ij}=\varphi_i=0$.
\item 
\emph{The G26 system:}
The system of the 26-moment equations contains the governing equations for variables $n$, $v_i$, $T$, $\sigma_{ij}$, $q_i$, $m_{ijk}$, $\Delta$ and $R_{ij}$, i.e.~it consists of equations \eqref{massBalG29}--\eqref{eqn:HFG29} and \eqref{eqn:mijk}--\eqref{eqn:Rij} with the right-hand sides computed using the collisional production terms \eqref{P1}--\eqref{Pi2}. 
However, equations \eqref{eqn:mijk}--\eqref{eqn:Rij} contain additional unknowns $u_{ijkl}^0$,  $\varphi_i$ and $u_{ijk}^1$ that are computed with the G26 distribution function 
\begin{align}
\label{G26disfun}
f_{|\mathrm{G26}}
&=f_M\left[1 + \frac{1}{2} \frac{\sigma_{ij} C_i C_j}{\rho\theta^2} + \frac{q_i C_i}{\rho\theta^2}\left( \frac{1}{d+2}\frac{C^2}{\theta}-1\right) 
 + \frac{1}{6}\frac{m_{ijk} C_i C_j C_k}{\rho\theta^3} 
 \right.\nonumber\\
&\left.\quad + \frac{d(d+2) \Delta}{8} \left(1 - \frac{2}{d} \frac{C^2}{\theta} + \frac{1}{d(d+2)}\frac{C^4}{\theta^2} \right)
+ \frac{1}{4}  \frac{R_{ij} C_i C_j}{\rho\theta^3} \left(\frac{1}{d+4}\frac{C^2}{\theta}-1\right)
\right],
\end{align}
which can also be obtained easily by following a similar procedure presented in appendix~\ref{app:G29}.
With the G26 distribution function \eqref{G26disfun}, $u_{ijkl}^0$ and $\varphi_i$ vanish and $u_{ijk}^1$ turns out to be $u_{ijk|\mathrm{G26}}^1 = (d+6)\,\theta\, m_{ijk}$, which is exactly the same as the value of $u_{ijk}^1$ obtained with the G29 distribution function \eqref{G29disfun}. 
Therefore, inserting the G26 closure (i.e.~$u_{ijkl}^0 = \varphi_i = 0$ and $u_{ijk}^1 = (d+6)\,\theta\, m_{ijk}$), equations \eqref{eqn:mijk}--\eqref{eqn:Rij} turn to \eqref{eqn:mijkG29}--\eqref{eqn:RijG29} in which $\varphi_i=0$. 
%
Thus, the G26 system for $d$-dimensional IMM consists of equations \eqref{massBalG29}--\eqref{eqn:RijG29} with $\varphi_i=0$.
\end{enumerate}

It is worthwhile to note that the G13 and G26 theories belong to the category of \emph{ordered moment theories}, which always include the neglected fluxes of a moment theory at the previous level \citep{Torrilhon2015}. 
Also, there are other moment theories, which consider complete (traces and traceless) moments of a given order; such moment theories are referred to as \emph{full moment theories} \citep{Torrilhon2015}. 
The first few examples of full moment theories are the Grad 10-, 20- and 35-moment theories (in three dimensions). 
In this sense, the G14 and G29 theories considered in the present work neither belong to the category of ordered moment theories nor to that of full moment theories.
%
%
\section{Transport coefficients in the NSF laws}
\label{Sec:NSF_rels}
Recall that system \eqref{massBal}--\eqref{energyBal} of the mass, momentum and energy balance equations was not closed due to the presence of additional unknowns: the stress $\sigma_{ij}$, heat flux $q_i$ and cooling rate $\zeta$. 
One of the major goals of kinetic theory is to furnish a closure for the system of the mass, momentum and energy balance equations in the form of constitutive relations. 
Traditionally, these constitutive relations are derived by performing the CE expansion on the Boltzmann equation. 
An alternative, but relatively much easier, way to determine the constitutive relations is by means of a CE-like expansion---in powers of a small parameter (usually, the Knudsen number)---performed on the Grad moment system.
For monatomic gases of Maxwell molecules, it can be shown via the order of magnitude approach that a CE-like expansion on the G13 equations yields the Euler, NSF and Burnett constitutive relations at the zeroth, first and second orders of expansion, respectively \citep{Struchtrup2005}. 
Thus, for monatomic gases of Maxwell molecules, the G13 equations already contain the Burnett equations. 
Such a CE-like expansion procedure of \citet{Struchtrup2005} on the Grad moment equations for IMM is much more involved due to non-conservation of energy, and---at its present understanding---does not yield the correct transport coefficients appearing in the constitutive relations. 
I still believe that a formal CE-like expansion procedure based on the order of magnitude of moments, which would yield the correct transport coefficients for granular gases (in particular, for IMM) can be devised; although it will be a topic for future research.
%
Here, I follow the approach of \cite{Garzo2013} to determine the transport coefficients in the NSF laws for a dilute granular gas of IMM through the Grad moment equations developed above.

The cooling rate in the energy balance equation \eqref{energyBal} for IMM is given by \eqref{coolingRateG29MM} while the constitutive relations for the stress and heat flux for closing the system of the mass, momentum and energy balance equations \eqref{massBal}--\eqref{energyBal}---to the linear approximation in spatial gradients---read \citep{JR1985, JR1985PoF, GD1999, Garzo2013}%
\begin{subequations}
\label{NSFconstitutiveRels}
\begin{align}
\label{NSFconstitutiveRelsStress}
\sigma_{ij} &= -2 \eta \frac{\partial v_{\langle i}}{\partial x_{j\rangle}},
\\
\label{NSFconstitutiveRelsHF}
q_i &= - \kappa \frac{\partial T}{\partial x_i} - \lambda \frac{\partial n}{\partial x_i},
\end{align}
\end{subequations}
where $\eta$, $\kappa$ and $\lambda$ are the transport coefficients. The coefficient $\eta$ is referred to as the \emph{shear viscosity} and $\kappa$ as the  \emph{thermal conductivity}; the coefficient $\lambda$ is a Dufour-like coefficient \citep{ACG2009, KSG2014, SBG2019} that vanishes identically for monatomic gases.
Equations \eqref{NSFconstitutiveRelsStress} and \eqref{NSFconstitutiveRelsHF} are the Navier--Stokes' law and the Fourier's law for granular gases, respectively. 
Equations \eqref{NSFconstitutiveRelsStress} and \eqref{NSFconstitutiveRelsHF} together are referred to as the NSF laws for granular gases.
\subsection{Zeroth-order contributions in spatial gradients}
To zeroth order in the spatial gradients, the mass, momentum and energy balance equations \eqref{massBalG29}--\eqref{energyBalG29} reduce to
\begin{align}
\label{zerothorderConservationlaws}
\frac{\partial n}{\partial t} = 0, \qquad \frac{\partial v_i}{\partial t} = 0 \qquad \textrm{and} \qquad \frac{\partial T}{\partial t} = - \zeta_0^\ast \, \nu \, T
\end{align}
while the balance equations for the other higher moments (eqs.~\eqref{eqn:stressG29}--\eqref{eqn:phiiG29}) reduce to 
\begin{align}
\label{zerothorderEq}
\left.
\begin{gathered}
\sigma_{ij} = 0, \qquad q_i = 0, 
\qquad 
m_{ijk} = 0, 
\qquad 
\Delta - \frac{6(1-e)^2}{4d-7+6e-3e^2} 
= 0, 
\\ 
\nu_R^\ast R_{ij} - \nu_{R\sigma}^\ast \theta \sigma_{ij} = 0, 
\qquad 
\nu_\varphi^\ast \, \varphi_i - \nu_{\varphi q}^\ast \theta q_i  = 0.
\end{gathered}
\right\}
\end{align}
Equations in \eqref{zerothorderEq} readily imply that
\begin{align}
\label{zerothorderMom}
\sigma_{ij} = q_i = m_{ijk} = R_{ij} = \varphi_i = 0 \quad\textrm{and}\quad 
\Delta = a_2.
\end{align}
where
\begin{align}
\label{a2MM}
a_2 = \frac{6(1-e)^2}{4d-7+6e-3e^2}
\end{align}
is the same as the value of the fourth cumulant for IMM reported in previous studies \citep{Santos2003, KGS2014}.
Thus, to zeroth order in spatial gradients, $\sigma_{ij}$, $q_i$, $m_{ijk}$, $R_{ij}$ and $\varphi_i$ are zero while $\Delta = a_2$. 
%
\subsection{First-order contributions in spatial gradients}
%
Now, the terms having first-order spatial derivatives are also retained in the moment equations.
To first order in spatial gradients, moment equations \eqref{massBalG29}--\eqref{eqn:HFG29}, read
\begingroup
\allowdisplaybreaks
\begin{align}
\label{FirstOrderMassEq}
\frac{\partial n}{\partial t} &= - v_i \frac{\partial n}{\partial x_i} - n \frac{\partial v_i}{\partial x_i}, 
\\
\label{FirstOrderMomEq}
\frac{\partial v_i}{\partial t} &= - v_j \frac{\partial v_i}{\partial x_j} - \frac{1}{m\,n} \frac{\partial (n T)}{\partial x_i},
\\ 
\label{FirstOrderEnergyEq}
\frac{\partial T}{\partial t} &= - v_i \frac{\partial T}{\partial x_i} - \frac{2}{d} T\frac{\partial v_i}{\partial x_i} - \zeta \, T,
\\
\label{FirstOrderSigmaEq}
\frac{\partial \sigma_{ij}}{\partial t} &= - 2\rho\theta\frac{\partial v_{\langle i}}{\partial x_{j\rangle}}
- \nu_\sigma^\ast \, \nu\,\sigma_{ij},
\\
\label{FirstOrderHFEq}
\frac{\partial q_i}{\partial t} &= - \frac{d+2}{2} \left[a_2 \theta^2\frac{\partial \rho}{\partial x_i}+ (1 + 2 a_2) \rho\theta \frac{\partial \theta}{\partial x_i}\right]
- \nu_q^\ast \, \nu\,q_i.
\end{align}
\endgroup
%
%
%
Notice that, unlike IHS (see~\cite{GST2018}), here 
none of the balance equations \eqref{eqn:mijkG29}--\eqref{eqn:phiiG29} is required for determining the transport coefficients for IMM up to first-order accuracy in spatial gradients, since the stress and heat flux balance equations (eqs.~\eqref{FirstOrderSigmaEq} and \eqref{FirstOrderHFEq}) have no coupling with the higher moments. 
The balance equations \eqref{eqn:mijkG29}--\eqref{eqn:phiiG29} will only be needed for computing the transport coefficients beyond the first-order accuracy in spatial gradients, which is not the focus of the present work. 

The time derivatives of the stress and heat flux in \eqref{FirstOrderSigmaEq} and \eqref{FirstOrderHFEq} are computed using dimensional analysis and using the zeroth-order accurate mass, momentum and energy balance equations \eqref{zerothorderConservationlaws}. They turn out to be  \citep{Garzo2013, GST2018}
\begin{align}
\label{timeDerStressHF}
\frac{\partial \sigma_{ij}}{\partial t} = \eta \, \zeta \frac{\partial v_{\langle i}}{\partial x_{j\rangle}}
\qquad\textrm{and}\qquad
\frac{\partial q_i}{\partial t} = 2 \kappa \, \zeta \frac{\partial T}{\partial x_i} + \left(\kappa \frac{T}{n} + \frac{3}{2} \lambda \right) \zeta \frac{\partial n}{\partial x_i}.
\end{align}
Now, in the first-order accurate stress and heat flux balance equations (\eqref{FirstOrderSigmaEq} and \eqref{FirstOrderHFEq}), one replaces $\sigma_{ij}$ and $q_i$ using \eqref{NSFconstitutiveRels} and their time derivatives using \eqref{timeDerStressHF}. 
Subsequent comparison of the coefficients of each hydrodynamic gradient in both the resulting equations leads to the transport coefficients in the NSF laws \eqref{NSFconstitutiveRels}:
\begin{align} 
\label{transCoeffNSF}
\eta = \eta_0 \, \eta^\ast, \qquad 
\kappa = \kappa_0 \, \kappa^\ast \qquad \textrm{and} \qquad 
\lambda = \frac{\kappa_0 \, T}{n}\, \lambda^\ast
\end{align}
%
where
\begin{align}
\label{eta0kappa0}
\eta_0 = \frac{n\,T}{\nu} 
\qquad \textrm{and} \qquad 
\kappa_0 = \frac{d(d+2)}{2(d-1) m} \eta_0 
\end{align}
are the shear viscosity and thermal conductivity, respectively, in the elastic limit; 
and $\eta^\ast$, $\kappa^\ast$ and $\lambda^\ast$ are the reduced shear viscosity, reduced thermal conductivity and reduced Dufour-like  coefficient, respectively. These reduced transport coefficients are given by
\begin{subequations}
\label{ReducedTransCoeff}
\begin{align}
\label{ReducedTransCoeffeta}
\eta^\ast &= \frac{1}{\nu_\sigma^\ast - \frac{1}{2} \zeta_0^\ast}, 
\\
\label{ReducedTransCoeffkappa}
\kappa^\ast &= \frac{d-1}{d} \frac{1 + 2\,a_2}{\nu_q^\ast - 2 \zeta_0^\ast},
\\
\label{ReducedTransCoefflambda}
\lambda^\ast &= \frac{\kappa^\ast \zeta_0^\ast + \frac{d-1}{d} a_2}{\nu_q^\ast - \frac{3}{2} \zeta_0^\ast} = \frac{\kappa^\ast}{1+2 a_2} \frac{\zeta_0^\ast + \nu_q^\ast a_2}{\nu_q^\ast - \frac{3}{2} \zeta_0^\ast}.
\end{align} 
\end{subequations}
Expressions~\eqref{ReducedTransCoeff} for the reduced transport coefficients agree with those obtained at first order of the CE expansion for IMM, e.g.~in \cite{Santos2003, KGS2014, GarzoSantos2011}. 
Indeed, the structure of these transport coefficients is very similar to those for IHS \citep{BDKS1998, Garzo2013} 
except for the fact that $a_2$, $\zeta_0^\ast$, $\nu_\sigma^\ast$ and $\nu_q^\ast$ for IMM and IHS  are different. 
Despite the structural similarity, the transport coefficients $\kappa^\ast$ and $\lambda^\ast$ for IMM 
(\eqref{ReducedTransCoeffkappa} and
\eqref{ReducedTransCoefflambda}) diverge at a certain value of the coefficient of restitution and do not yield meaningful values below it---in contrast to the transport coefficients for IHS which are meaningful for all values of the coefficient of restitution and are in reasonably good agreement with the simulations. 
This issue pertaining to IMM can readily be appreciated by inspecting the explicit dependence of the reduced transport coefficients on the coefficient of restitution and dimension as follows. Inserting $a_2$, $\zeta_0^\ast$, $\nu_\sigma^\ast$ and $\nu_q^\ast$ from \eqref{a2MM}, \eqref{zeta0ast}, 
\eqref{nusigma} and \eqref{nuq} in the reduced transport coefficients \eqref{ReducedTransCoeff}, they are expressed as a function of the coefficient of restitution and dimension \citep{GarzoSantos2011}:
\begin{subequations}
\label{ReducedTransCoeff_in_de}
\begin{align}
\eta^\ast &= \frac{8d}{(1+e) [3d+2+(d-2) e]}, 
\\
\kappa^\ast &= \frac{8(d-1)(4 d+5-18 e+9 e^2)}{(1+e) (d-4+3 d e) (4 d-7+6 e-3 e^2)},
\\
\lambda^\ast &= \frac{16 (1-e) [2 d^2+8d-1-6 (d+2)e+9 e^2]}{(1+e)^2 (d-4+3 d e) (4 d-7+6 e-3 e^2)}.
\end{align} 
\end{subequations}
Clearly, both $\kappa^\ast$ and $\lambda^\ast$ have singularities at $e = (4-d)/(3d)$, for which the denominators of both of them vanish. 
In particular, the denominators of both $\kappa^\ast$ and $\lambda^\ast$ vanish at $e=1/3$ for $d=2$ and at $e=1/9$ for $d=3$. 
Moreover, below the singularities, i.e. for $e < (4-d)/(3d)$, both $\kappa^\ast$ and $\lambda^\ast$ are negative, which is unphysical. 
The existence of these singularities is apparently attributed to the breakdown of hydrodynamics in granular gases of IMM for $e \leq (4-d)/(3d)$ due to the lack of time scale separation between the kinetic and hydrodynamic parts of the distribution function \citep{BGM2010}. 
%
Owing to these singularities, it is customary to write the heat flux as a linear combination of the gradients of $T$ and $n\sqrt{T}$ instead of its usual representation \eqref{NSFconstitutiveRelsHF} \citep[see, e.g.,][]{GSM2007, GarzoSantos2011}. 
The heat flux in the new form reads
\begin{align}
\label{HFnewform}
q_i = - \kappa^\prime \frac{\partial T}{\partial x_i} - \frac{\lambda}{\sqrt{T}} \frac{\partial (n\sqrt{T})}{\partial x_i},
\qquad\textrm{where}\qquad 
\kappa^\prime = \kappa - \lambda \frac{n}{2T}
\end{align}
is referred to as the modified thermal conductivity \citep{GSM2007}. 
The reduced modified thermal conductivity $\kappa^{\prime\ast} = \kappa^\prime/\kappa_0$---using \eqref{ReducedTransCoeff}--\eqref{HFnewform}---is given by
\begin{align}
\label{Reducedkappaprime}
\kappa^{\prime\ast} 
= \frac{d-1}{d} \frac{1 + \frac{3}{2}  a_2}{\nu_q^\ast - \frac{3}{2} \zeta_0^\ast} 
= \frac{8 (2 d+1-6 e+3 e^2)}{(1+e)^2 (4 d-7+6 e-3 e^2)}.
\end{align}
The reduced modified thermal conductivity $\kappa^{\prime\ast}$ does not possess the above singularity and hence is finite for all $0 \leq e \leq 1$ and for $d=2,3$. 
%
\subsection{Comparison with existing theories and computer simulations}
I have not found any simulation data on the transport coefficients for IMM.
Therefore, in this subsection, I compare the reduced transport coefficients for IMM obtained above with those for IHD ($d=2$) and IHS ($d=3$) obtained through various theoretical and simulation methods. 

\begin{figure}
\centering
\includegraphics[width = 0.47\textwidth]{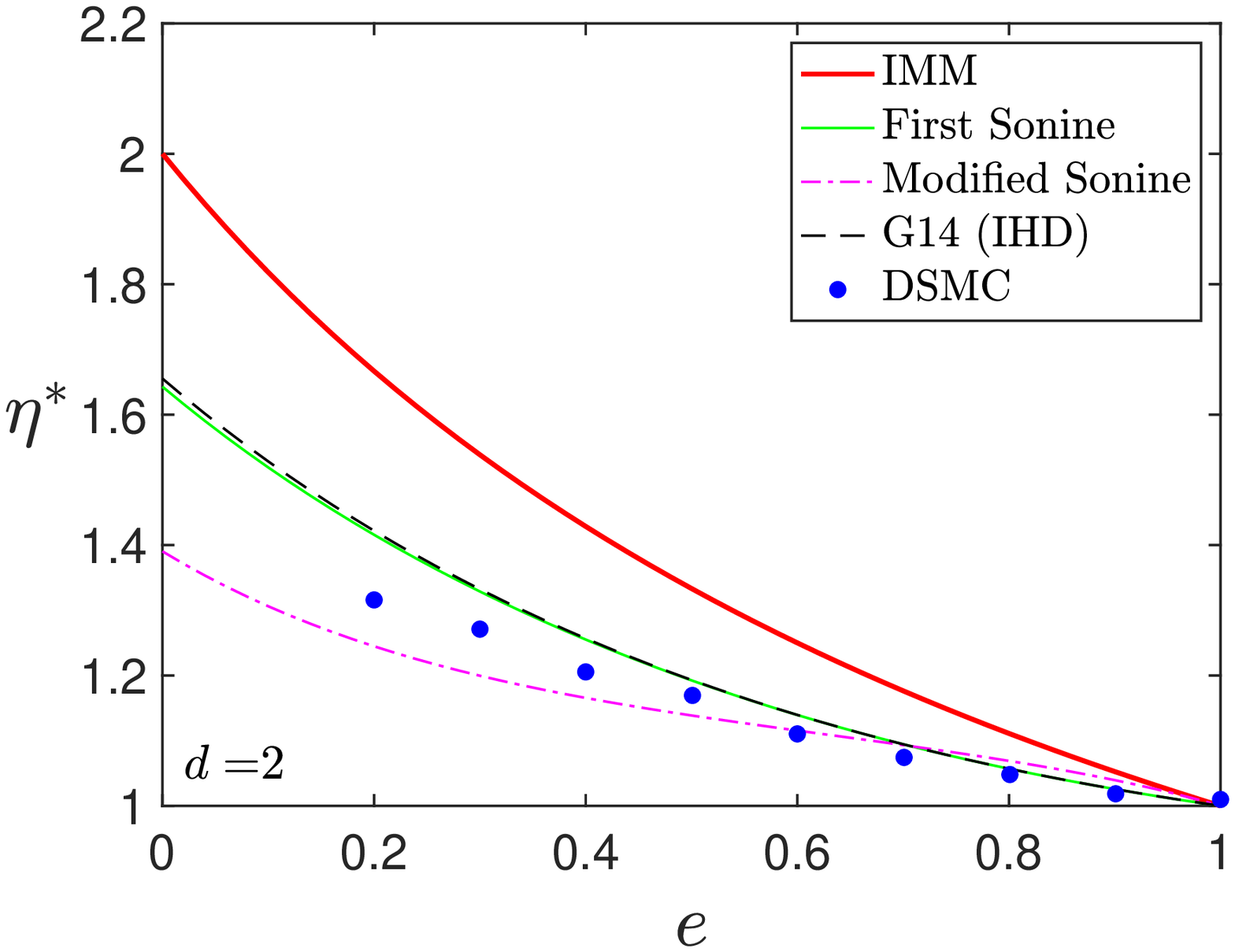} 
\hfill
\includegraphics[width = 0.47\textwidth]{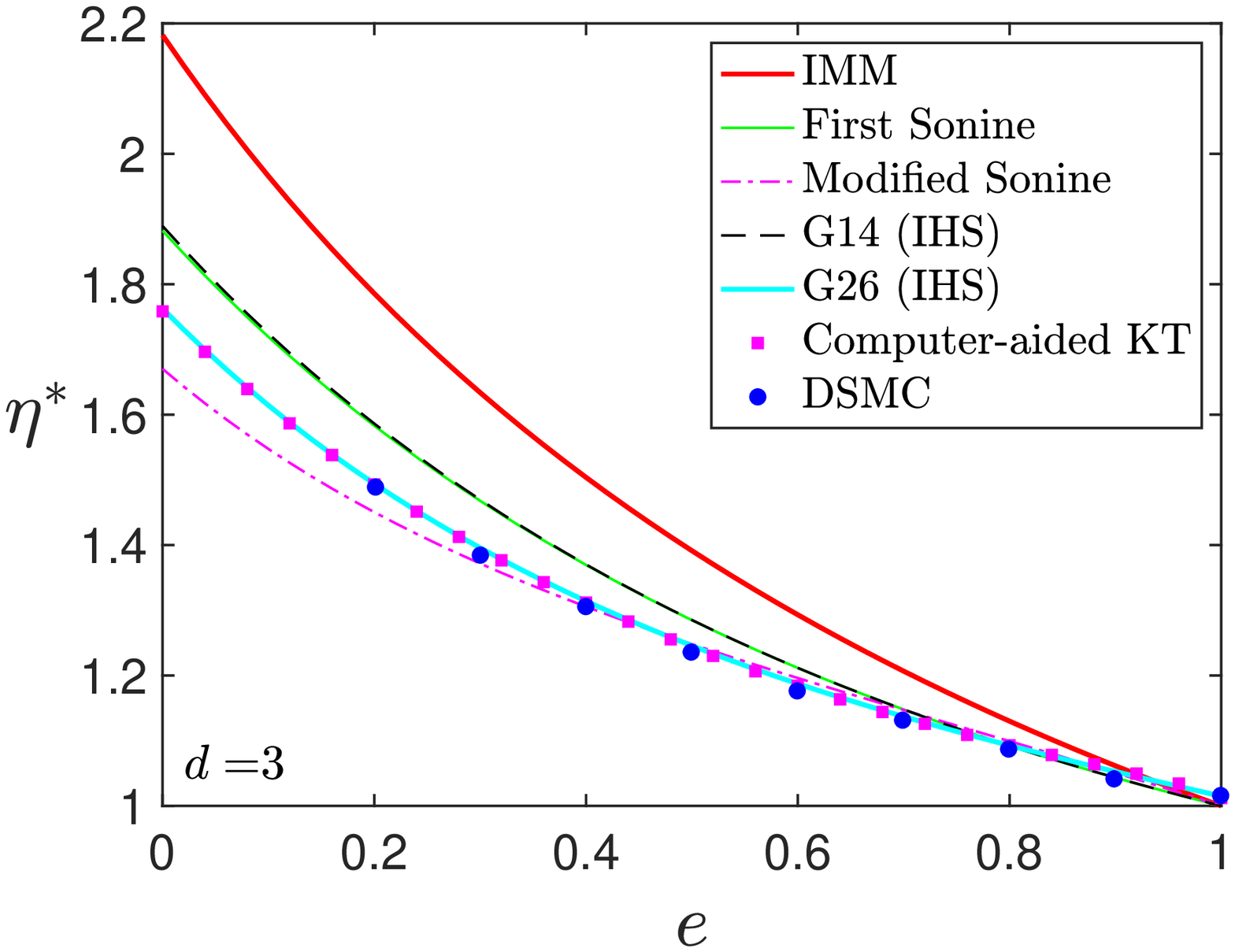} 
\caption{Reduced shear viscosity $\eta^\ast$ plotted over the coefficient of restitution $e$ for (left) two and (right) three dimensions. The thick solid (red) line represents the result for IMM. 
All other lines or symbols are the results for IHD ($d=2$) or IHS ($d=3$).
The thin solid (green) and dash-dotted (magenta) lines delineate the results from the first Sonine and modified first Sonine approximations, respectively \citep{GSM2007}.
The dashed (black) line depicts the result obtained with the G14 distribution function \citep{Garzo2013} while the solid cyan line depicts that obtained with the G26 equations \citep{GST2018}.
The squares are the results from the theoretical expressions obtained via the computer-aided method devised by \cite{NBSG2007}. The circles are the DSMC results from \cite{MSG2005} and  \cite{GSM2007}. 
}
\label{fig:viscosity}
\end{figure}

\begin{figure}
\centering
\includegraphics[width = 0.47\textwidth]{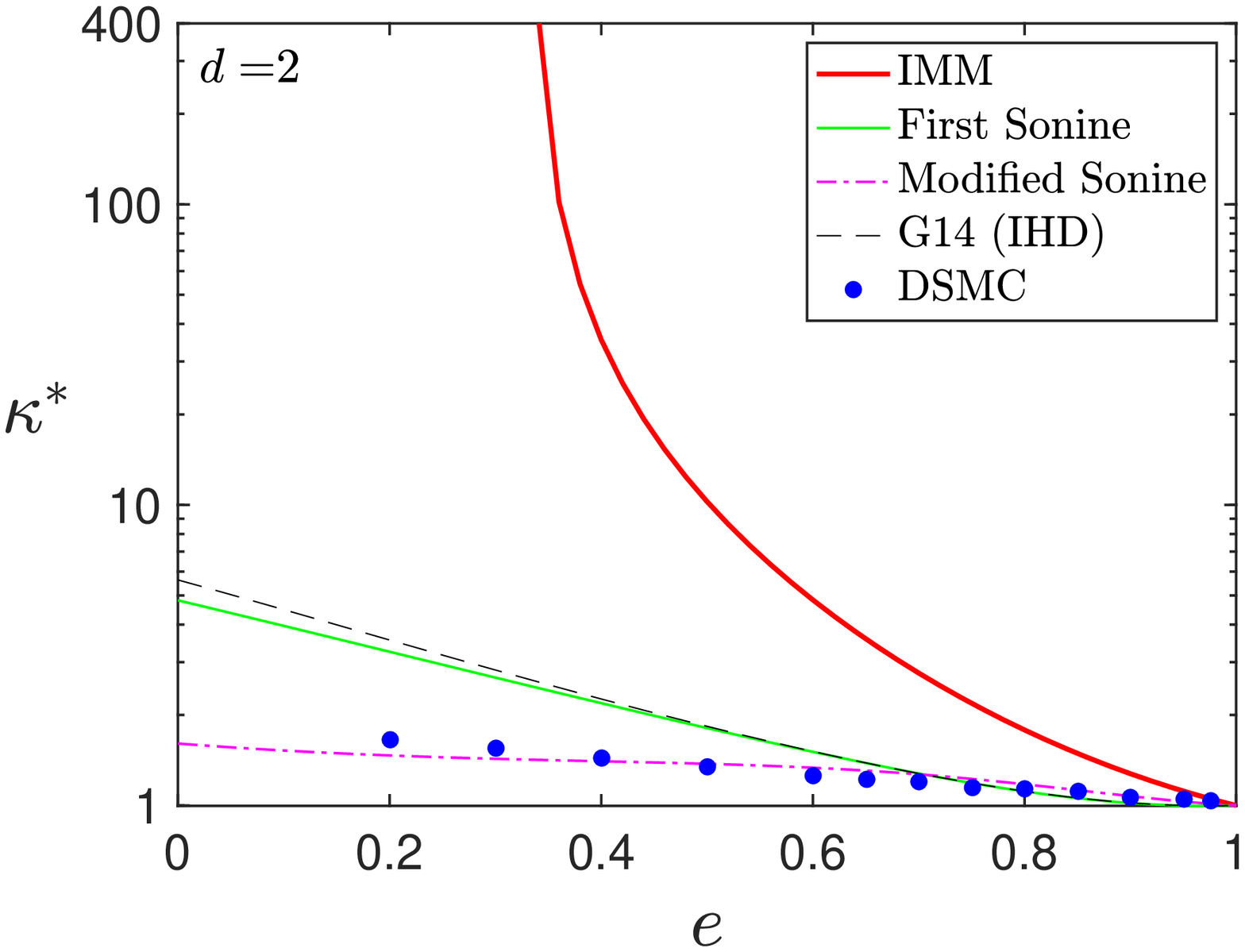} 
\hfill
\includegraphics[width = 0.47\textwidth]{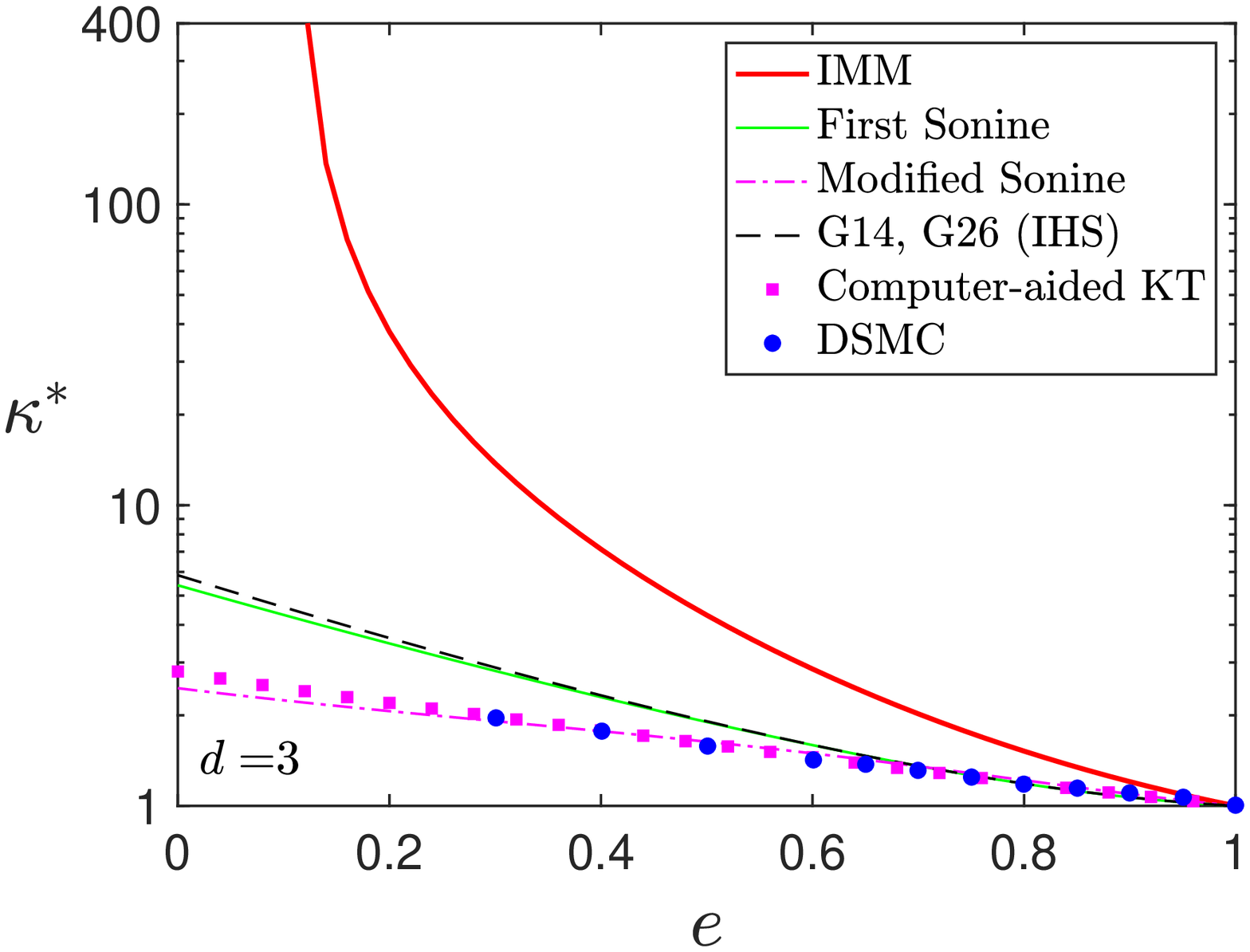} 
\caption{Reduced thermal conductivity $\kappa^\ast$ plotted over the coefficient of restitution $e$ for (left) two and (right) three dimensions. 
The circles are the DSMC results from \cite{BR-M2004} for $d=2$ and in \cite{BRMGdS2005} for $d=3$.
The lines and squares are the same as described in figure~\ref{fig:viscosity}.
}
\label{fig:kappa}
\end{figure}

\begin{figure}
\centering
\includegraphics[width = 0.47\textwidth]{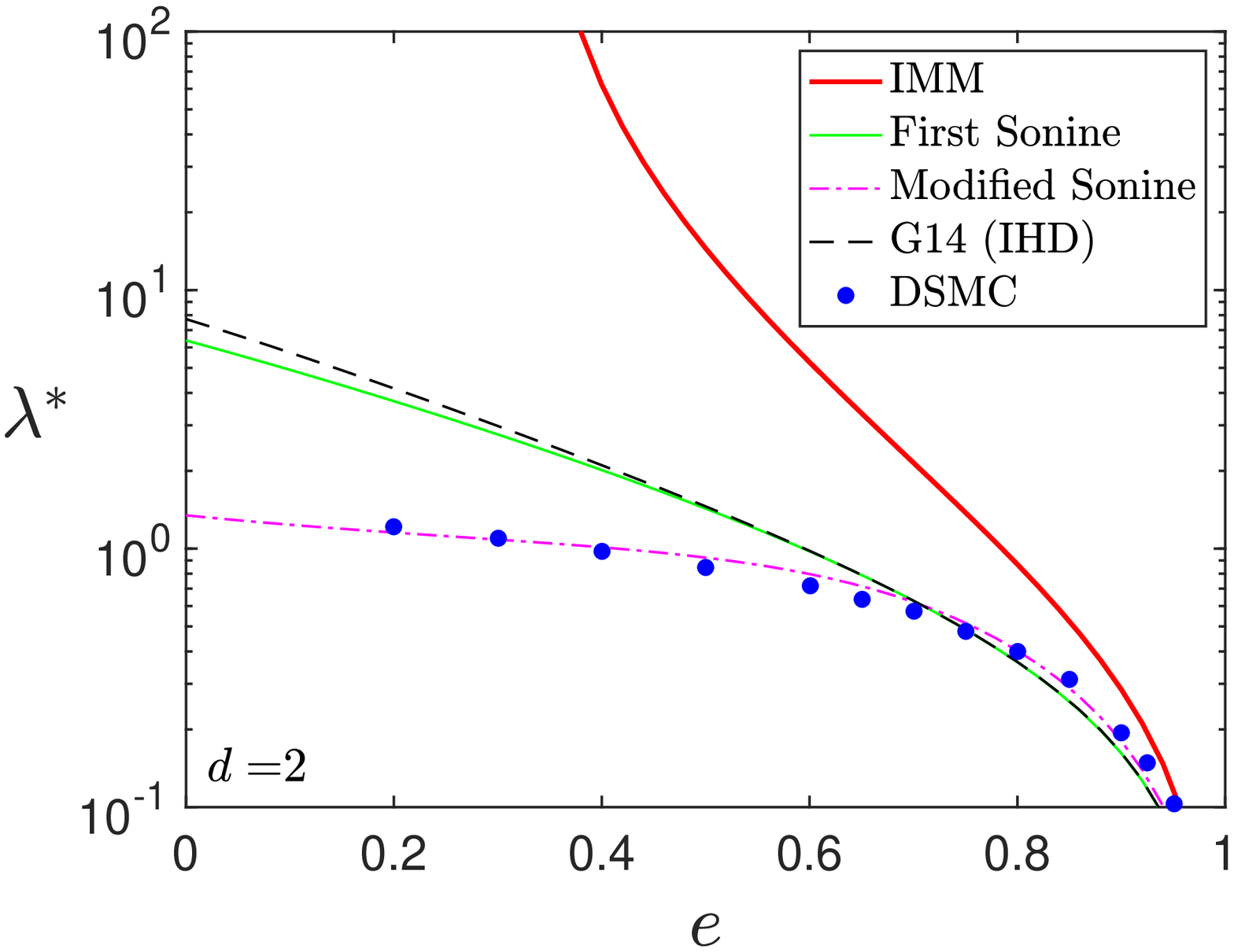} 
\hfill
\includegraphics[width = 0.47\textwidth]{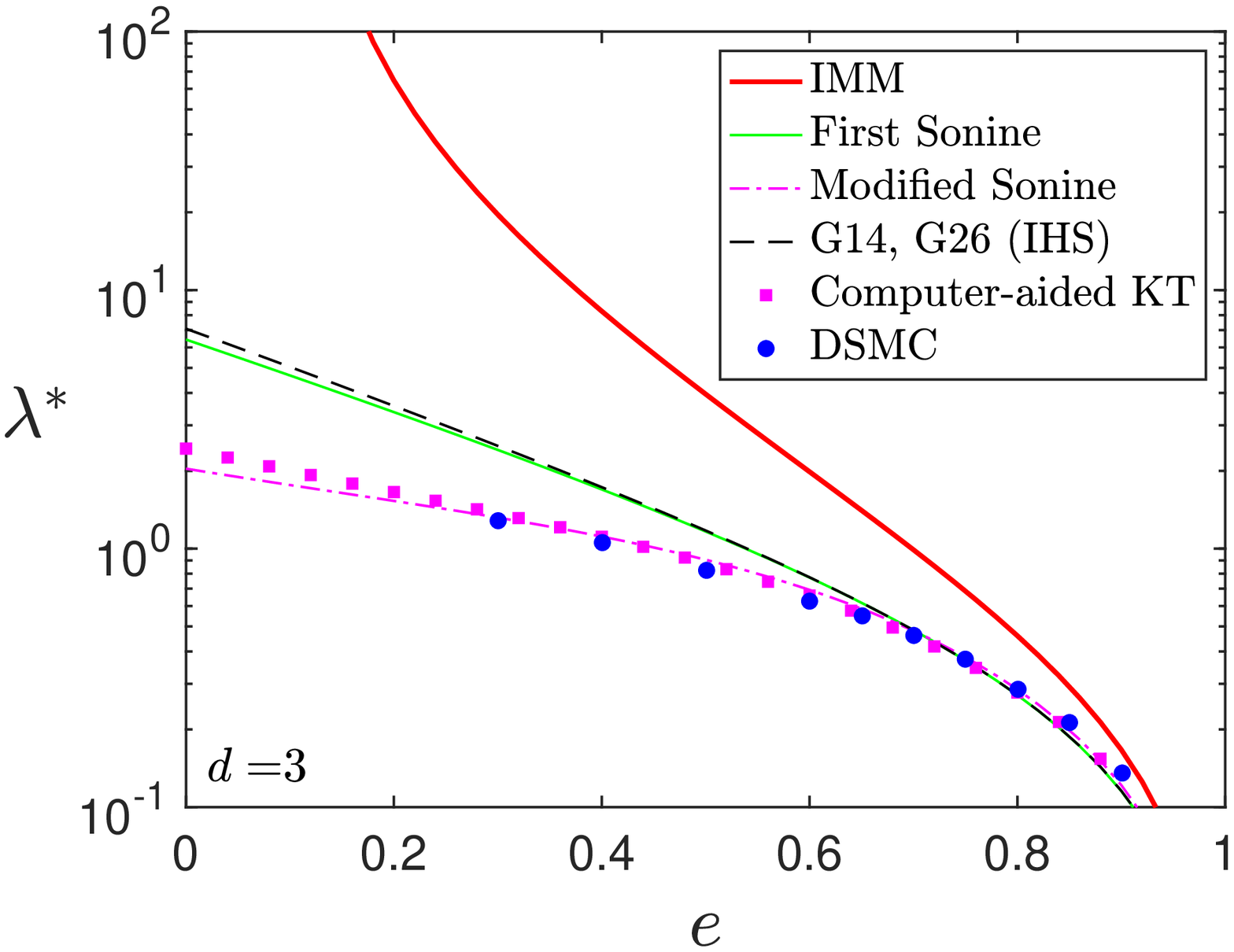} 
\caption{Reduced coefficient $\lambda^\ast$ plotted over the coefficient of restitution $e$ for (left) two and (right) three dimensions. 
The circles are the DSMC results from \cite{BR-M2004} for $d=2$ and in \cite{BRMGdS2005} for $d=3$.
The lines and squares are the same as described in figure~\ref{fig:viscosity}.
}
\label{fig:lambda}
\end{figure}

\begin{figure}
\centering
\includegraphics[width = 0.47\textwidth]{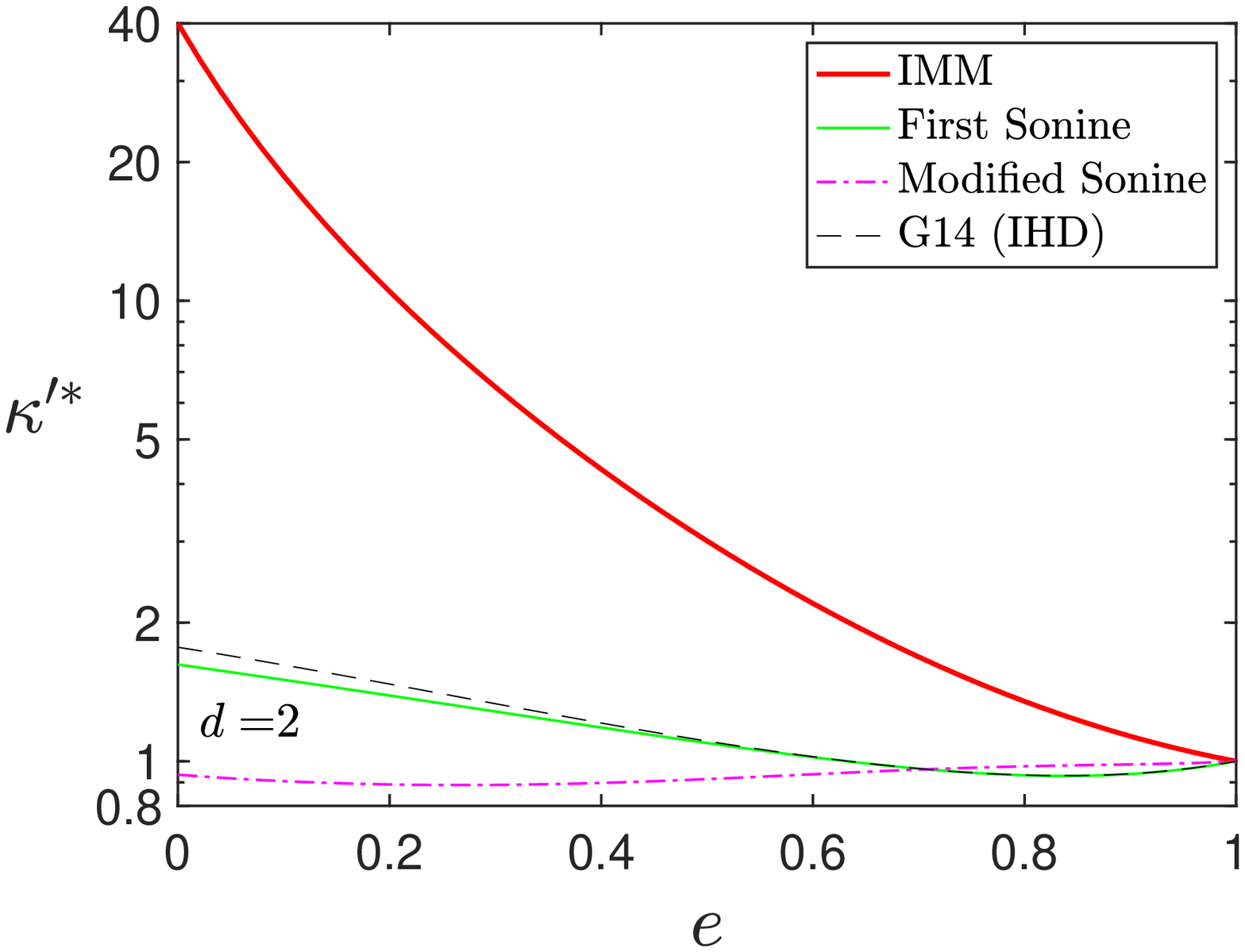} 
\hfill
\includegraphics[width = 0.47\textwidth]{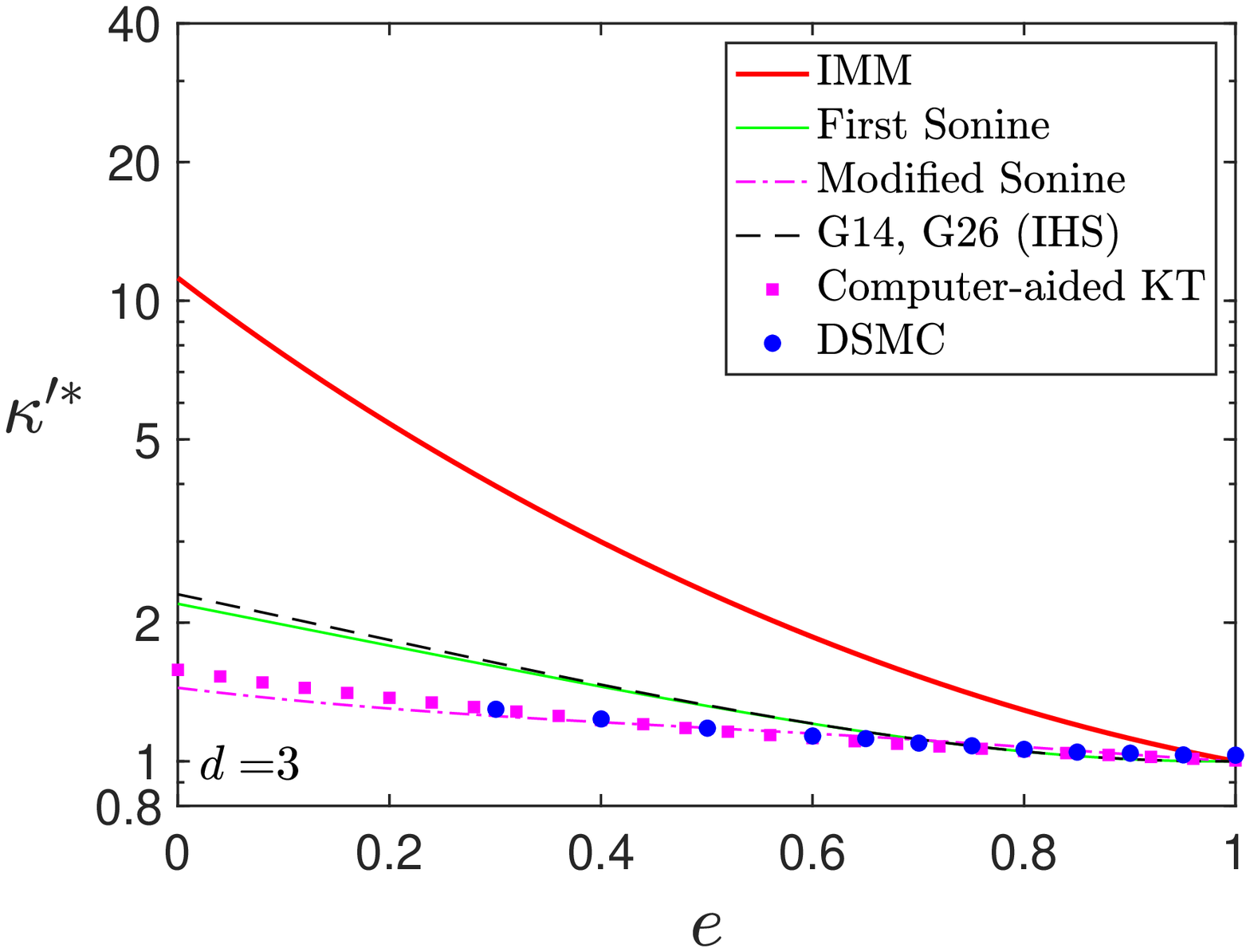} 
\caption{Reduced modified thermal conductivity $\kappa^{\prime\ast}$ plotted over the coefficient of restitution $e$ for (left) two and (right) three dimensions. 
The circles in the right panel are the DSMC results from \cite{MSG2007}. 
The lines and symbols are the same as described in figure~\ref{fig:viscosity}.
}
\label{fig:kappaprime}
\end{figure}

The reduced transport coefficients $\eta^\ast$, $\kappa^\ast$, $\lambda^\ast$ and $\kappa^{\prime\ast}$ for a dilute granular gas are plotted over the coefficient of restitution $e$ in figures~\ref{fig:viscosity}--\ref{fig:kappaprime}, respectively.
The left and right panels in each figure exhibit the results for $d=2$ and $d=3$, respectively. 
The thick solid (red) line in each figure denotes the result for IMM obtained from \eqref{ReducedTransCoeff} or \eqref{ReducedTransCoeff_in_de} and \eqref{Reducedkappaprime}, which have been obtained in this paper through the Grad moment equations. 
Recall that the reduced transport coefficients for IMM obtained through the moment method above are exactly the same as those obtained at first order of the CE expansion \cite[see][]{Santos2003, GarzoSantos2011}. 
Therefore, the  thick solid (red) line in each figure also represents the results for IMM from the first-order CE expansion. 
The remaining lines and symbols in figures~\ref{fig:viscosity}--\ref{fig:kappaprime} depict the results for IHD (in case of $d=2$) or for IHS (in case of $d=3$).
The thin solid (green) and dash-dotted (magenta) lines are the plots for the reduced transport coefficients from \cite{GSM2007} obtained at the first Sonine and modified first Sonine approximations, respectively, in the CE expansion. The dashed (black) lines depict the reduced transport coefficients obtained through the G14 distribution function for $d$-dimensional IHS in \cite{Garzo2013}. 
%
The right panel of figure~\ref{fig:viscosity} also displays a solid cyan line, which is not present in the other figures. This solid cyan line is the result for the reduced shear viscosity obtained with the G26 equations for IHS very recently by \citet{GST2018}. 
Indeed, the other transport coefficients from the G14 or G26 equations remain the same; consequently, the solid cyan line in the right panels of each of figures~\ref{fig:kappa}--\ref{fig:kappaprime} coincides with the dashed black line, and hence has not been shown separately. 
The squares are the results from the theoretical expressions derived via the so-called \emph{computer-aided} method devised by \cite{NBSG2007} and the circles are the numerical solutions of the Boltzmann equation obtained via the direct simulation Monte Carlo (DSMC) method in \cite{BR-M2004, MSG2005, BRMGdS2005, MSG2007}.
The paper by \cite{GSM2007} summarises and presents the DSMC results in the aforementioned references that are computed with two approaches: (i) through Green--Kubo relations in \cite{BR-M2004} (for $d=2$) and \cite{BRMGdS2005} (for $d=3$), and (ii) by implementing an external force in \cite{MSG2005} and \cite{MSG2007}. 
The external force in the latter compensates for collisional cooling and yields somewhat better results. 
Therefore, the DSMC results in figures~\ref{fig:viscosity}--\ref{fig:kappaprime} are shown with the latter for whichever coefficient they are available else they are shown with the former---figure~\ref{fig:viscosity} and the right panel of figure~\ref{fig:kappaprime} display the DSMC results with the latter while figures~\ref{fig:kappa} and \ref{fig:lambda} show the DSMC results with the former. 
The DSMC results for the reduced shear viscosity from the latter approach  were obtained by \cite{MSG2005} for $e = 0.6, 0.7, 0.8, 0.9, 1$ in the case of $d=3$ while that for $e = 0.2, 0.3, 0.4, 0.5$ in the case of $d=3$ and that for all $e$ in the case of $d=2$ were obtained by \cite{GSM2007}.
The DSMC data for $\kappa^{\prime\ast}$ in two dimensions (left panel of figure~\ref{fig:kappaprime}) are apparently unavailable.


It is clear from figures~\ref{fig:viscosity}--\ref{fig:kappaprime} that the IMM model overpredicts all the transport coefficients significantly in comparison to the IHS model, despite the transport coefficients for IMM computed from the Grad moment method in the present work being exactly the same as those obtained through CE expansion for IMM, for example, in \cite{Santos2003, GarzoSantos2011}.
%
%
The discrepancies between the results for IMM and IHS are apparently linked to the choice of the effective collision frequency $\mathring{\nu}$ (see (\ref{effeciveFreq}$a$)) in the IMM model \citep{Santos2003}, which is chosen in such a way that the cooling rates from the Boltzmann equation for IHS and IMM remain exactly the same. 
%
Furthermore, the reduced transport coefficients $\kappa^\ast$ and $\lambda^\ast$ for IMM (shown by the thick solid red lines in figures~\ref{fig:kappa} and \ref{fig:lambda}) diverge at $e=1/3$ for $d=2$ (left panels of figures~\ref{fig:kappa} and \ref{fig:lambda}) and at $e=1/9$ for $d=3$ (right panels of figures~\ref{fig:kappa} and \ref{fig:lambda}), and remain unphysical below these values of the coefficient of restitution. 
On the other hand, the reduced modified thermal conductivity $\kappa^{\prime\ast}$ for IMM remains positive for all values of the coefficient of restitution in both dimensions (see figure~\ref{fig:kappaprime}). Nevertheless, $\kappa^{\prime\ast}$ for IMM is also much higher than that for IHS. 
In the case of $d=2$ (left panel of figure~\ref{fig:kappaprime}), $\kappa^{\prime\ast}$ for IHS from any theory first decreases then increases on increasing the coefficient of restitution (although the profiles of $\kappa^{\prime\ast}$ from the modified first Sonine approximation and first Sonine approximation/G14 theory differ significantly) whereas that for IMM decreases monotonically on increasing the coefficient of restitution. 
However, as the DSMC data are not available in this case, it is difficult to discern which theory for IHS yields better results in this case.

Among fully theoretical methods, the modified version of the first Sonine approximation (dash-dotted magenta lines) proposed by \citet{GSM2007} for IHS seems to be the best model, which captures all the transport coefficient very well, although the G26 model of \citet{GST2018} was able to capture the coefficient of the reduced shear viscosity (but not the other transport coefficients) better than the modified first Sonine approximation.
%
\section{The HCS of a freely cooling granular gas of IMM}
\label{Sec:HCS}
The state of a force-free granular gas when its granular temperature decays constantly while its spatial homogeneity is maintained is termed as the HCS  
\citep{BP2004}. 
For studying the HCS, one considers a force-free (i.e.~$\bm{F}=\bm{0}$) granular gas having an initial number density as $n(0,\bm{x})=n_0$ and initial granular temperature $T(0,\bm{x})=T_0$ at time $t=0$ when the gas is left to cool down freely due to inelastic collisions while maintaining the spatial homogeneity (i.e.~$\partial(\cdot) / \partial x_i=0$). 

In this section, I investigate the HCS of a $d$-dimensional granular gas of IMM with the Grad moment equations \eqref{massBalG29}--\eqref{eqn:phiiG29} presented above. 
The nonlinear (underlined) contributions on the right-hand sides of the Grad moment equations \eqref{massBalG29}--\eqref{eqn:phiiG29} are discarded in this section for simplicity. This means that our focus is on the early evolution stage of homogeneously cooling granular gas. 
Hence, the possibility of increase in the granular temperature in a cooling granular gas (reported recently for granular gases of aggregating particles by \cite{BFP2018}) is disregarded, which possibly occurs at large times.

It is convenient to study the HCS with dimensionless variables obtained by introducing the following scaling:
\begin{align}
\label{scaling}
\left.
\begin{gathered}
n_{\ast}=\frac{n}{n_0},\quad v_i^{\ast}=\frac{v_i}{\sqrt{\theta_0}},\quad T_{\ast}=\frac{T}{T_0},\quad \sigma_{ij}^{\ast}=\frac{\sigma_{ij}}{n_0 T_0},\quad q_i^{\ast}=\frac{q_i}{n_0 T_0\sqrt{\theta_0}}, 
\\
m_{ijk}^{\ast}=\frac{m_{ijk}}{n_0 T_0\sqrt{\theta_0}}, \quad R_{ij}^{\ast}=\frac{R_{ij}}{n_0 T_0\theta_0},\quad \varphi_i^{\ast} = \frac{\varphi_i}{n_0 T_0\theta_0\sqrt{\theta_0}},\quad t_{\ast}=\nu_0 t,
\end{gathered}
\right\}
\end{align}
where 
$\theta_0 = T_0 / m$ and $\nu_0 = \nu(t=0) = 4 \, \Omega_d \, n_0 \,\mathbbm{d}^{d-1} \sqrt{T_0 / m} / [\sqrt{\pi} (d+2)]$.
%
With scaling \eqref{scaling}, 
the G29 equations \eqref{massBalG29}--\eqref{eqn:phiiG29}---without the underlined terms---in the HCS (i.e.~with $\partial(\cdot)/\partial x_i = 0$, $\bm{F}=\bm{0}$) reduce to
\begingroup
\allowdisplaybreaks
\begin{align}
\label{HCS:massBal}
\frac{\mathrm{d} n_{\ast}}{\mathrm{d}t_{\ast}}&=0,
\\
\label{HCS:momBal}
\frac{\mathrm{d} v_i^{\ast}} {\mathrm{d}t_{\ast}}&=0,
\\
\label{HCS:energyBal}
\frac{\mathrm{d} T_{\ast}} {\mathrm{d}t_{\ast}} 
&= - \zeta_0^\ast 
\, n_{\ast}T_{\ast}^{3/2},
\\
\label{HCS:stress}
\frac{\mathrm{d}\sigma_{ij}^{\ast}}{\mathrm{d}t_{\ast}} 
&= - \nu_\sigma^\ast \, n_{\ast}\sqrt{T_{\ast}} \, \sigma_{ij}^{\ast},
\\
\label{HCS:HF}
\frac{\mathrm{d} q_i^{\ast}}{\mathrm{d} t_{\ast}} 
&=-\nu_q^\ast \, n_{\ast}\sqrt{T_{\ast}}\,q_i^{\ast},
\\
\label{HCS:mijk}
\frac{\mathrm{d}m_{ijk}^{\ast}}{\mathrm{d}t_{\ast}}
&=-\nu_m^\ast \, n_{\ast}\sqrt{T_{\ast}}\,m_{ijk}^{\ast},
\\
\label{HCS:DeltaBal}
\frac{\mathrm{d}\Delta}{\mathrm{d}t_{\ast}}
&=- \nu_{\!\Delta}^\ast \, n_{\ast} \sqrt{T_{\ast}} \, (\Delta - a_2),
\\
\label{HCS:Rij}
\frac{\mathrm{d} R_{ij}^{\ast}}{\mathrm{d} t_{\ast}} 
&=- n_{\ast}\sqrt{T_{\ast}} \left(\nu_R^\ast R_{ij}^{\ast} - \nu_{R\sigma}^\ast T_{\ast}\sigma_{ij}^{\ast}\right),
\\
\label{HCS:phii}
\frac{\mathrm{d} \varphi_i^{\ast}}{\mathrm{d} t_{\ast}} 
&=- n_{\ast}\sqrt{T_{\ast}} \left(\nu_\varphi^\ast \varphi_i^{\ast} - \nu_{\varphi q}^\ast T_{\ast}q_i^{\ast}\right).
\end{align}
\endgroup
\subsection{Haff's law}\label{Subsec:HaffLaw}

Equations \eqref{HCS:massBal} and \eqref{HCS:momBal} with the initial conditions of the HCS imply $n_\ast(t_\ast) = 1$ and $v_i^{\ast}(t_\ast) = 0$. 
Therefore, equation~\eqref{HCS:energyBal} using the initial conditions of the HCS yields Haff's law \citep{Haff1983} for the evolution of the granular temperature:
\begin{subequations}
\label{eq:HaffLaw}
\begin{align}
T_\ast (t_\ast) = \frac{1}{(1+t_\ast/\tau_\ast)^2}
\qquad \textrm{or} \qquad
T(t) = \frac{T_0}{(1+t/\tau_0)^2},
\tag{\theequation $a$,$b$}
\end{align}
\end{subequations}
where 
\begin{subequations}
\label{tauMM}
\begin{align}
\tau_\ast = \frac{2}{\zeta_0^\ast}
\qquad \textrm{and} \qquad
\tau_0 = \frac{\tau_\ast}{\nu_0} = \frac{2}{\zeta_0^\ast \nu_0}.
\tag{\theequation $a$,$b$}
\end{align}
\end{subequations}
Here, $\tau_0$ is the time scale in Haff's law for IMM and $\tau_\ast$ is the corresponding dimensionless time scale. 
Haff's law \eqref{eq:HaffLaw} with  time scale \eqref{tauMM} is exactly the same as that obtained in \cite{GarzoSantos2011} for IMM.
It is worthwhile to note that, unlike the energy balance equation in the case of IHS that also contains the scalar fourth moment $\Delta$ \cite[see, e.g.,][]{KM2011, GST2018},
equation \eqref{HCS:energyBal} does not contain any other moment except $n_\ast$ and $T_\ast$. Consequently, Haff's law for IMM does not depend on higher moments; or in other words, Haff's law remains unchanged for IMM, no matter how large a moment system it is determined from. 

Note that the dimensionless time scale $\tau_\ast$ in Haff's law \eqref{eq:HaffLaw} for $d$-dimensional IHS obtained at first approximation of the Sonine expansion is given by \citep{vNE1998}
\begin{align}
\label{tauHS}
\tau_\ast^{\mathrm{(IHS)}} = \frac{8d}{(d+2) (1-e^2)} 
\left(1+\frac{3}{16} a_2^{\mathrm{(IHS)}}\right)^{\!-1}
\end{align}
with an excellent estimate for the fourth cumulant \citep{vNE1998}
\begin{align}
\label{a2HS}
a_2^{\mathrm{(IHS)}} = \frac{16(1-e)(1-2e^2)}{24d+9+e(8d-41)+30e^2 (1-e)}.
\end{align}

\begin{figure}
\centering
\includegraphics[width=0.47\textwidth]{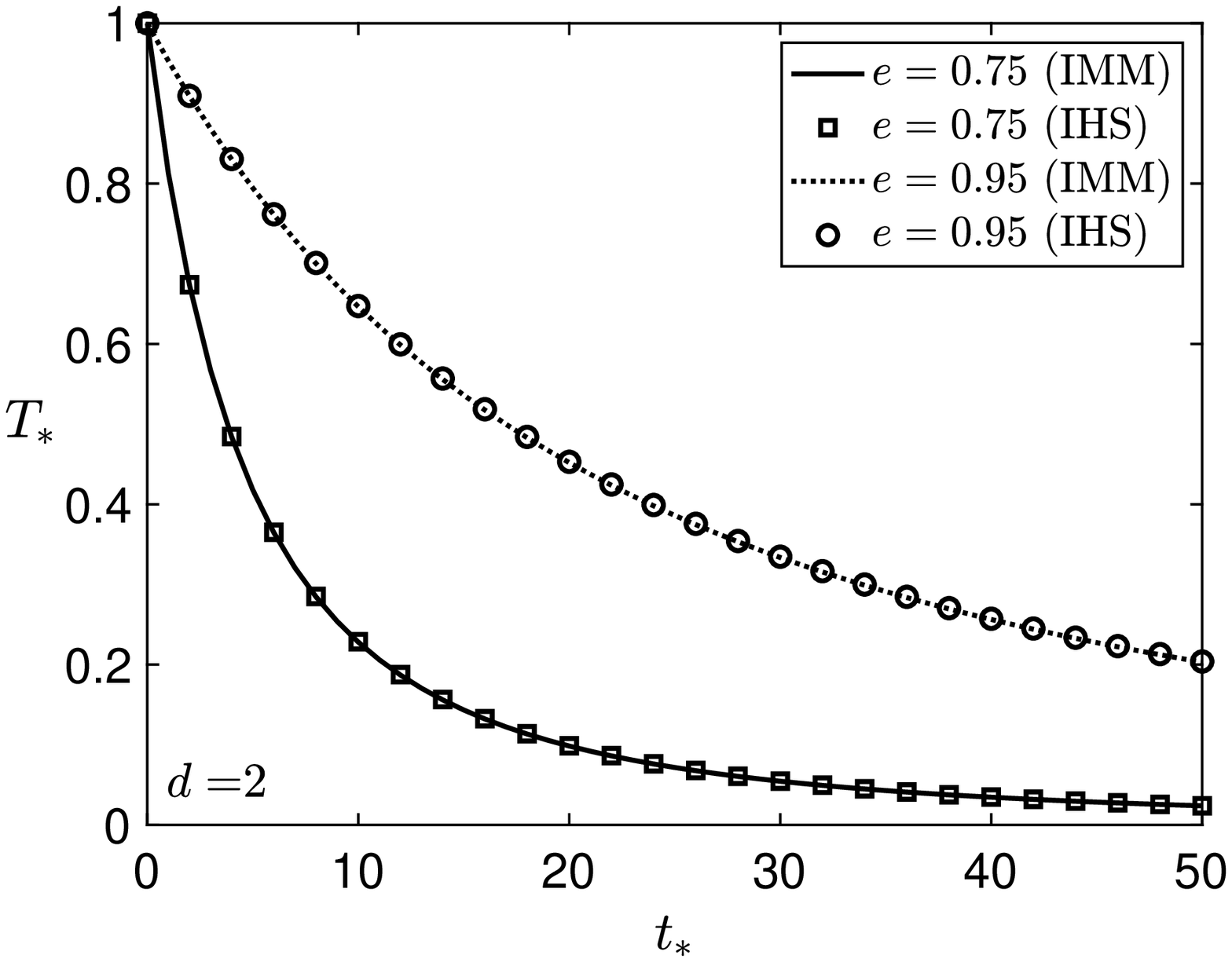}
\qquad
\includegraphics[width=0.47\textwidth]{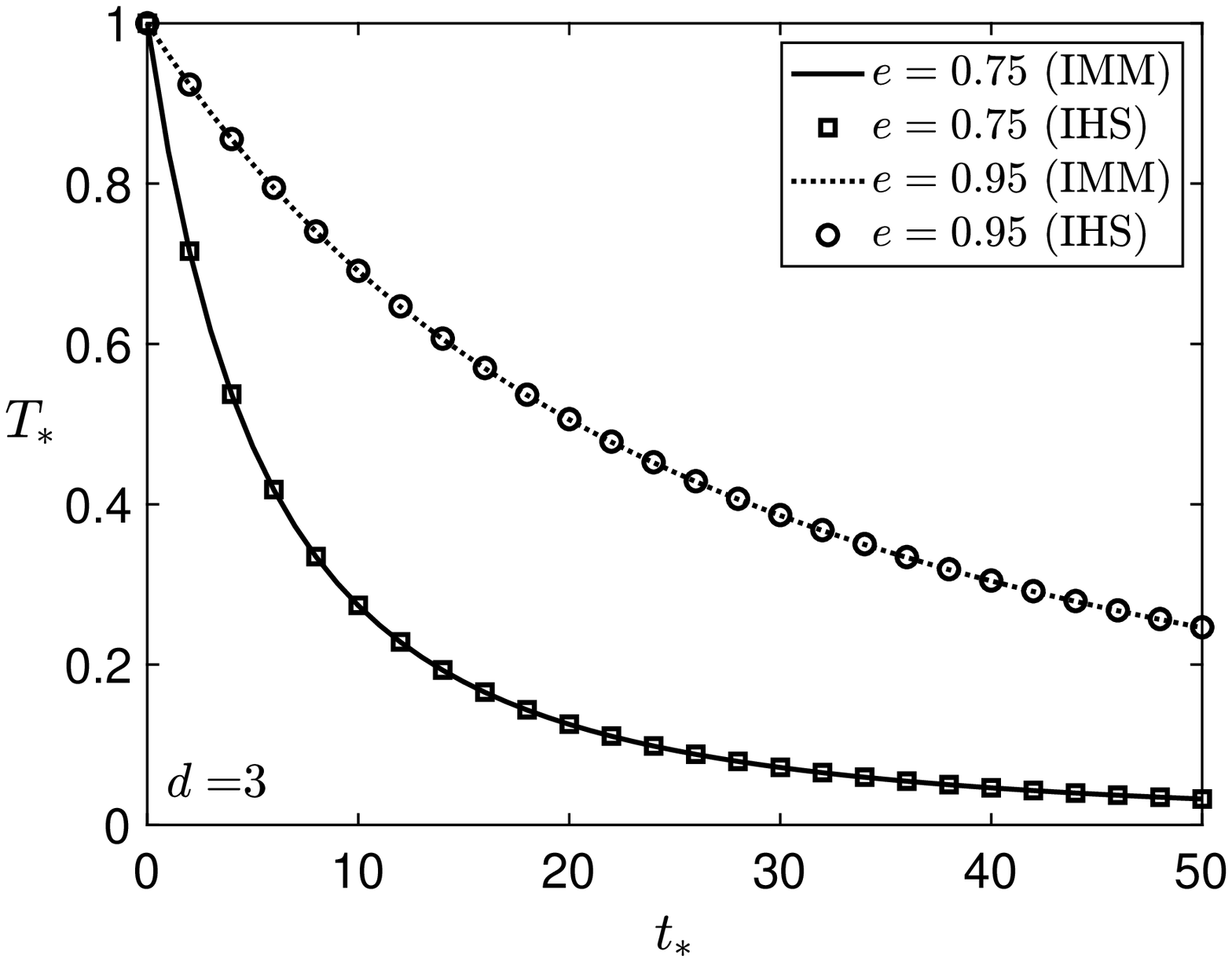}
\caption{\label{fig:TempHCS} Relaxation of the granular temperature $T_\ast$ in the HCS via Haff's law (\ref{eq:HaffLaw}$a$) for $d=2$ (left) and $d=3$ (right) with the initial conditions $n_\ast(0)=T_\ast(0)=1$. The lines depict granular temperature profiles for IMM, i.e. with $\tau_\ast$ as in (\ref{tauMM}$a$), while the symbols denote those for IHS, i.e. with $\tau_\ast$ as in \eqref{tauHS}.
}
\end{figure}

Figure~\ref{fig:TempHCS} illustrates the decay of the dimensionless granular temperature in the HCS via Haff's law (\ref{eq:HaffLaw}$a$) for the coefficients of restitution $e=0.75$ (depicted by solid lines and squares) and $e=0.95$ (depicted by dotted lines and circles). 
The lines denote Haff's law for IMM, where the associated (dimensionless) time scale $\tau_\ast$ for the decay is given by (\ref{tauMM}$a$), while the symbols denote that for IHS at first approximation of the Sonine expansion, where the associated (dimensionless) time scale $\tau_\ast$ is given by \eqref{tauHS}. 
Although the time scale $\tau_\ast$ for IMM does not contain the fourth cumulant $a_2$ while that for IHS does contain it, Haff's law from IMM (lines) is still in very good agreement with that from IHS (symbols) in both two and three dimensions. 
The granular temperature relaxes faster with increasing inelasticity 
due to the fact that more inelastic particles dissipate more energy during a collision in comparison to the less inelastic ones.

\subsection{Relaxation of other moments in the HCS}
\label{Relax}
For monatomic gases, it can be shown through the order of magnitude method devised by \cite{Struchtrup2004} that all the nonequilibrium moments ($\sigma_{ij}$, $q_i$, $m_{ijk}$, $\Delta$, $R_{ij}$ and beyond) are at least of first order in spatial gradients \cite[see][]{Struchtrup2004, Struchtrup2005}. 
In contrast, the order of magnitude method in the case of granular gases is not straightforward due to non-conservation of energy and, to the best of my knowledge, has never been attempted so far. 
Notwithstanding, I would expect that all the higher vectorial and tensorial moments ($\sigma_{ij}$, $q_i$, $m_{ijk}$, $R_{ij}$, $\varphi_i$ and beyond) for granular gases are also at least of first order in spatial gradients; this conjecture is well known for $\sigma_{ij}$ and $q_i$ in the case of granular gases as well. 
This means that all the vectorial and tensorial moments remain zero in the HCS because spatial gradients are zero in this state. 
Nonetheless, it is interesting to analyse how these higher-order moments relax in the HCS if started with non-vanishing initial conditions. 

Equations \eqref{HCS:stress}--\eqref{HCS:phii} are coupled with \eqref{HCS:massBal} and \eqref{HCS:energyBal}, but can be solved analytically. 
In order to compare the decay rates of the moments, the initial conditions for all the variables in \eqref{HCS:stress}--\eqref{HCS:phii} are taken as unity. 
With these initial conditions, equations~\eqref{HCS:stress}--\eqref{HCS:phii} yield the following solution for the other variables:
\begingroup
\allowdisplaybreaks
\begin{subequations}
\label{eq:OtherMomRelax}
\begin{align}
\sigma_{ij}^{\ast} (t_\ast) 
&= \left(1+\frac{t_\ast}{\tau_\ast}\right)^{\!-\nu_\sigma^\ast \tau_\ast},
\\
q_i^{\ast} (t_\ast) 
&= \left(1+\frac{t_\ast}{\tau_\ast}\right)^{\!-\nu_q^\ast \tau_\ast},
\\
m_{ijk}^{\ast} (t_\ast) 
&= \left(1+\frac{t_\ast}{\tau_\ast}\right)^{\!-\nu_m^\ast \tau_\ast},
\\
\Delta (t_\ast) &= a_2 + (1-a_2) \left(1+\frac{t_\ast}{\tau_\ast}\right)^{\!-\nu_{\!\Delta}^\ast \tau_\ast},
\\
R_{ij}^{\ast} (t_\ast) & = \left( 1 - \varkappa_R \right) \left(1+\frac{t_\ast}{\tau_\ast}\right)^{\!-\nu_R^\ast \tau_\ast} 
+ \varkappa_R \left(1+\frac{t_\ast}{\tau_\ast}\right)^{\!-\nu_\sigma^\ast \tau_\ast-2},
\\
\varphi_i^{\ast} (t_\ast) & = \left( 1 - \varkappa_\varphi \right) \left(1+\frac{t_\ast}{\tau_\ast}\right)^{\!-\nu_\varphi^\ast \tau_\ast}
+ \varkappa_\varphi \left(1+\frac{t_\ast}{\tau_\ast}\right)^{\!-\nu_q^\ast \tau_\ast-2},
\end{align}
\end{subequations}
\endgroup
where $\varkappa_R = \nu_{R\sigma}^\ast \tau_\ast / [(\nu_R^\ast - \nu_\sigma^\ast) \tau_\ast -2]$ and $\varkappa_\varphi = \nu_{\varphi q}^\ast \tau_\ast / [(\nu_\varphi^\ast - \nu_q^\ast) \tau_\ast -2]$. 
%
It is important to note that for dilute monatomic gases (i.e.~in the case of $d=3$ and $e=1$) of Maxwell molecules, equations \eqref{HCS:massBal}--\eqref{HCS:phii} with the same initial conditions yield the solution
\begin{align}
\label{OtherMomRelaxMonatomic}
\left.
\begin{aligned}
\sigma_{ij}^{\ast} (t_\ast) 
&= \mathrm{e}^{-t_\ast}, 
&\quad
q_i^{\ast}(t_\ast) 
&= \mathrm{e}^{-\frac{2}{3} t_\ast},
&\quad
m_{ijk}^{\ast}(t_\ast) &= \mathrm{e}^{-\frac{3}{2} t_\ast},
\\
\Delta (t_\ast) 
&= \mathrm{e}^{-\frac{2}{3} t_\ast},
&\quad
R_{ij}^{\ast} (t_\ast) 
&= \mathrm{e}^{-\frac{7}{6} t_\ast},
&\quad
\varphi_i^{\ast} (t_\ast) 
&= \mathrm{e}^{-t_\ast}.
\end{aligned}
\right\}
\end{align}
From solution \eqref{OtherMomRelaxMonatomic}, it is clear that, for dilute monatomic gases of Maxwell molecules, the third-order moment $m_{ijk}^{\ast}$ relaxes faster than all other higher-order moments considered in the present work; $R_{ij}^{\ast}$ relaxes slower than $m_{ijk}^{\ast}$ but faster than the remaining moments ($\sigma_{ij}^{\ast}$, $q_i^{\ast}$, $\Delta$ and $\varphi_i^{\ast}$);
$\sigma_{ij}^{\ast}$ and $\varphi_i^{\ast}$ relax with equal relaxation rates but faster than $q_i^{\ast}$ and $\Delta$, which also relax with equal relaxation rates.

\begin{figure}
\centering
\includegraphics[width=0.47\textwidth]{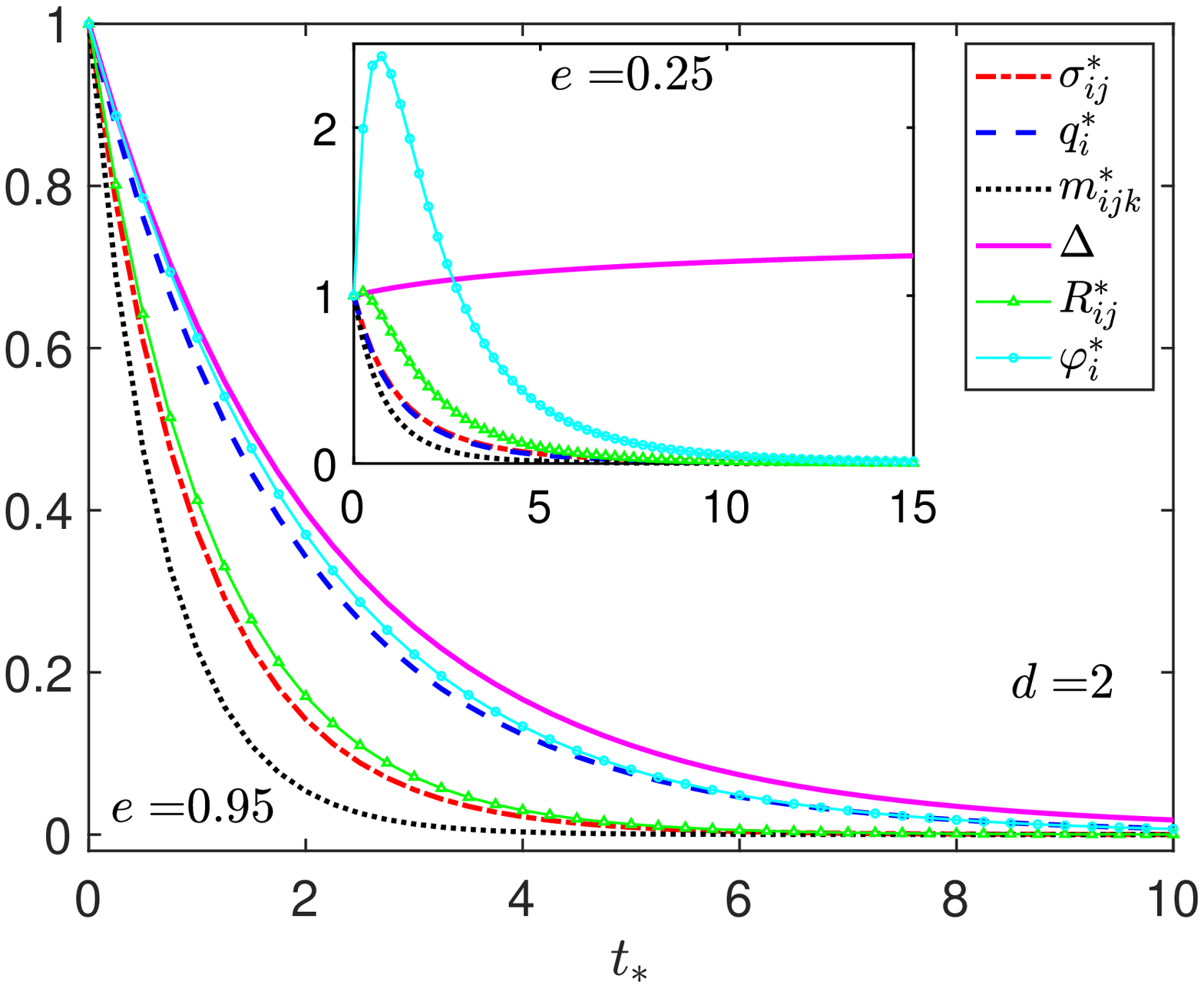}
\qquad
\includegraphics[width=0.47\textwidth]{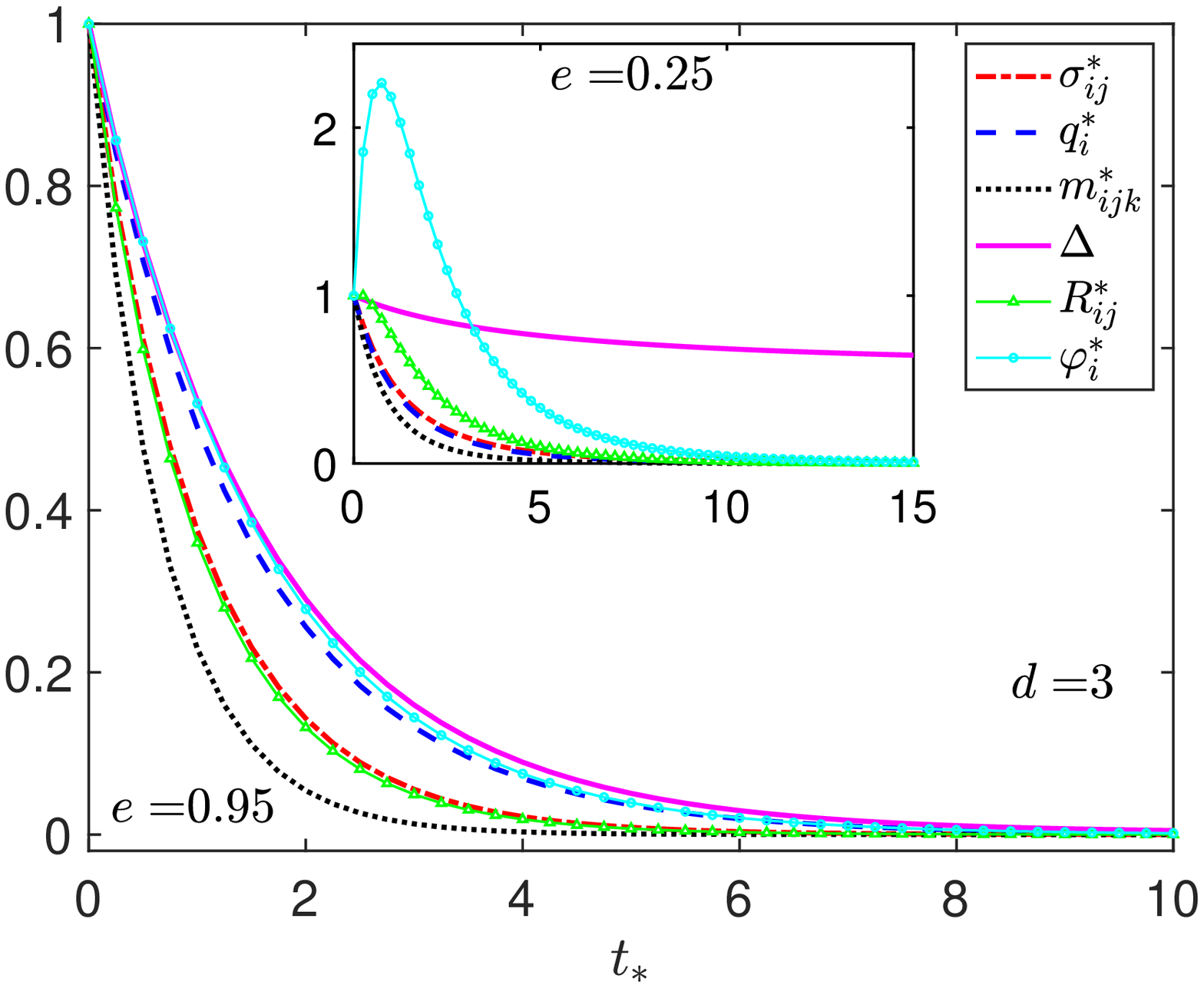}
\caption{\label{fig:othermomHCS} Relaxation of the other moments---$\sigma_{ij}^\ast$, $q_i^\ast$, $m_{ijk}^\ast$, $\Delta$, $R_{ij}^\ast$ and $\varphi_i^\ast$---for IMM in the HCS for $d=2$ (left) and $d=3$ (right) evaluated using analytical solutions \eqref{eq:OtherMomRelax} and $\tau_\ast$ from (\ref{tauMM}$a$). The coefficient of restitution is taken as $e=0.95$ (main panels) and $e=0.25$ (insets). 
Initial conditions are taken as $n_\ast(0)=T_\ast(0)=\sigma_{ij}^{\ast}(0)=q_i^{\ast}(0)=m_{ijk}^{\ast}(0)=\Delta(0) = R_{ij}^{\ast}(0) = \varphi_i^\ast(0)=1$.
}
\end{figure}

Figure~\ref{fig:othermomHCS} illustrates the relaxation of the other (dimensionless) moments---$\sigma_{ij}^\ast$, $q_i^\ast$, $m_{ijk}^\ast$, $\Delta$, $R_{ij}^\ast$ and $\varphi_i^\ast$ over the (dimensionless) time $t_\ast$ (via analytical solutions \eqref{eq:OtherMomRelax}) in two and three dimensions for granular gases (i.e.~for $e \neq 1$): $e=0.95$ (main panels) and $e=0.25$ (insets).
It turns out that all these moments relax with time much faster than the granular temperature. 
It is interesting to note that all the vectorial and tensorial moments relax to zero whereas the scalar moment $\Delta$ relaxes to a nonzero value $a_2$ for all $e\neq1$. These can also be verified from analytical solutions \eqref{eq:OtherMomRelax} in the limit $t_\ast \to \infty$, since all the exponents in \eqref{eq:OtherMomRelax} are negative for all $e \neq 1$ (note that $\tau_\ast$,  
$\nu_\sigma^\ast$, $\nu_q^\ast$, $\nu_m^\ast$, $\nu_\Delta^\ast$, $\nu_R^\ast$, 
$\nu_{R\sigma}^\ast$, $\nu_\varphi^\ast$, $\nu_{\varphi q}^\ast$, $\alpha_0$, $\alpha_1$, $\alpha_2$ and $\alpha_3$ are positive for $e \neq 1$).
For large values of the coefficient of restitution (main panels), all these moments decay monotonically and their decay rates are, apparently, proportional to their tensorial orders, i.e. the third-order moment ($m_{ijk}^\ast$) decays faster than the second-order moments ($\sigma_{ij}^\ast$ and $R_{ij}^\ast$), which decay faster than the vectorial moment ($q_i^\ast$ and $\varphi_i^\ast$), which decay faster than the scalar moment ($\Delta$). 
However, for sufficiently small values of the coefficient of restitution (insets), the higher-order moments ($R_{ij}^\ast$ and $\varphi_i^\ast$) do not decay monotonically. 
This is attributed to the fact that higher-order (sixth-order and beyond) moments for IMM are prone to diverge for sufficiently small values of the coefficient of restitution  in the HCS \citep{SG2012} and it is manifested already through the non-monotonic relaxation of $R_{ij}^\ast$ and $\varphi_i^\ast$ for small coefficients of restitution (insets), although they themselves do not diverge.
\section{Linear stability analysis of the HCS}
\label{Sec:Stability}
In this section, the temporal stability of the HCS of a freely cooling granular gas of IMM due to small perturbations will be analysed through the G29 and other Grad moment theories described in \S\,\ref{Subsec:various} with $F_i=0$. The amplitudes of these perturbations are assumed to be sufficiently small in order to ensure the validity of the linear analysis.%

%
For the linear stability analysis, all the field variables in \eqref{massBalG29}--\eqref{eqn:phiiG29} are decomposed into their reference state values (i.e.~their solutions in the HCS) plus perturbations from their respective reference state values. In other words, the field variables in the G29 system~\eqref{massBalG29}--\eqref{eqn:phiiG29} are written as 
\begingroup
\allowdisplaybreaks
\begin{subequations}
\label{perturbations}
\begin{align}
n(t,\bm{x}) &= n_0 \big[1+\tilde{n}(t,\bm{x})\big],
\\
v_i(t,\bm{x}) &= v_H(t) \, \tilde{v}_i(t,\bm{x}),
\\
T(t,\bm{x}) &= T_H(t) \big[1+\tilde{T}(t,\bm{x})\big],
\\
\sigma_{ij}(t,\bm{x}) &= n_0 \,T_H(t)\, \tilde{\sigma}_{ij}(t,\bm{x}),
\\
q_i(t,\bm{x}) &= n_0\, T_H(t)\,v_H(t)\, \tilde{q}_i(t,\bm{x}),
\\
m_{ijk}(t,\bm{x}) &= n_0 \,T_H(t)\,v_H(t)\, \tilde{m}_{ijk}(t,\bm{x}),
\\
\Delta(t,\bm{x}) &= a_2 + \tilde{\Delta}(t,\bm{x}),
\\
R_{ij}(t,\bm{x}) &= n_0 \,T_H(t)\,v_H(t)^2\, \tilde{R}_{ij}(t,\bm{x}),
\\
\varphi_i(t,\bm{x}) &= n_0 \,T_H(t)\,v_H(t)^3\, \tilde{\varphi}_i(t,\bm{x}),
\end{align}
\end{subequations}
\endgroup
where $T_H(t)$ is the granular temperature in the HCS and $v_H(t)=\sqrt{T_H(t)/m}$ is a reference speed proportional to the adiabatic sound speed in the HCS; and the quantities with tilde denote the dimensionless perturbations in the field variables from their respective solutions in the HCS. 

Inserting expressions \eqref{perturbations} for the field variables into 
\eqref{massBalG29}--\eqref{eqn:phiiG29} and discarding all the nonlinear terms of the perturbations, one obtains the system of linear PDEs in (dimensionless) perturbed field variables with time-dependent coefficients. 
This system of PDEs is transformed to a new system of PDEs having time-independent coefficients by introducing a length scale 
\begin{align}
\label{ell}
\ell := \frac{v_H(t)}{\nu_H(t)}, \quad\textrm{where}\quad
\nu_H(t) = \frac{4 \, \Omega_d}{\sqrt{\pi} (d+2)} n_0 \mathbbm{d}^{d-1} \sqrt{\frac{T_H(t)}{m}},
\end{align}
that is employed to make the space variables dimensionless (i.e.~$\tilde{x}_i = x_i/\ell$, where tilde denotes the dimensionless space variable), and a dimensionless time
\begin{align}
\tilde{t} := \int_0^t \nu_H(t^\prime) \, \mathrm{d}t^\prime
\end{align}
that measures time as the number of effective collisions per particle \citep{BRM1998, BP2004, GarzoSantos2007}. 
The resulting system of PDEs having time-independent coefficients reads
\begingroup
\allowdisplaybreaks
\begin{align}
\label{massBalPertDimless}
\frac{\partial \tilde{n}}{\partial \tilde{t}} + \frac{\partial \tilde{v}_i}{\partial \tilde{x}_i} &= 0,
\\
\label{momentBalPertDimless}
\frac{\partial \tilde{v}_i}{\partial \tilde{t}} + \frac{\partial \tilde{\sigma}_{ij}}{\partial \tilde{x}_j} + \frac{\partial \tilde{n}}{\partial \tilde{x}_i} + \frac{\partial \tilde{T}}{\partial \tilde{x}_i} - \frac{1}{2}\zeta_0^\ast \tilde{v}_i &= 0,
\\
\label{energyBalPertDimless}
\frac{\partial \tilde{T}}{\partial \tilde{t}}+ \frac{2}{d} \left(\frac{\partial \tilde{q}_i}{\partial \tilde{x}_i} + \frac{\partial \tilde{v}_i}{\partial \tilde{x}_i} \right) + \zeta_0^\ast \left(\tilde{n} + \frac{1}{2} \tilde{T}\right) &= 0,
\\
\label{stressBalPertDimless}
\frac{\partial \tilde{\sigma}_{ij}}{\partial \tilde{t}} + \frac{\partial \tilde{m}_{ijk}}{\partial \tilde{x}_k} + \frac{4}{d+2} \frac{\partial \tilde{q}_{\langle i}}{\partial \tilde{x}_{j \rangle}} + 2 \frac{\partial \tilde{v}_{\langle i}}{\partial \tilde{x}_{j \rangle}} 
+ \xi_\sigma \tilde{\sigma}_{ij}
&= 0,
\\
\label{HFBalPertDimless}
\frac{\partial \tilde{q}_i}{\partial \tilde{t}} + \frac{1}{2}\frac{\partial \tilde{R}_{ij}}{\partial \tilde{x}_j} 
+ \frac{\partial \tilde{\sigma}_{ij}}{\partial \tilde{x}_j} 
+ \frac{d+2}{2} \left[\frac{\partial \tilde{\Delta}}{\partial \tilde{x}_i} 
+a_2 \frac{\partial \tilde{n}}{\partial \tilde{x}_i} 
+ (1+2 a_2) \frac{\partial \tilde{T}}{\partial \tilde{x}_i}\right]
+ \xi_q \tilde{q}_i
&= 0,
\\
\label{mijkBalPertDimless}
\frac{\partial \tilde{m}_{ijk}}{\partial \tilde{t}} 
+ \frac{3}{d+4}\frac{\partial \tilde{R}_{\langle ij}}{\partial \tilde{x}_{k\rangle}}
+3\frac{\partial \tilde{\sigma}_{\langle ij}}{\partial \tilde{x}_{k\rangle}} 
+ \xi_m \tilde{m}_{ijk}
&=0,
\\
\label{DeltaBalPertDimless}
\frac{\partial \tilde{\Delta}}{\partial \tilde{t}} + \xi_1  \frac{\partial \tilde{q}_i}{\partial \tilde{x}_i} 
+ \frac{1}{d(d+2)} \frac{\partial \tilde{\varphi}_i}{\partial \tilde{x}_i} 
+ \nu_{\!\Delta}^\ast \, \tilde{\Delta}
&=0,
\\
\label{RijBalPertDimless}
\frac{\partial \tilde{R}_{ij}}{\partial \tilde{t}} 
+ \frac{2}{d+2}  \frac{\partial \tilde{\varphi}_{\langle i}}{\partial \tilde{x}_{j\rangle}}
+ \frac{4(d+4)}{d+2} \frac{\partial \tilde{q}_{\langle i}}{\partial \tilde{x}_{j\rangle}}
+2(d+4)a_2 \frac{\partial \tilde{v}_{\langle i}}{\partial \tilde{x}_{j\rangle}} 
+ 2\frac{\partial \tilde{m}_{ijk}}{\partial \tilde{x}_k} 
&
\nonumber\\
+ \xi_R \tilde{R}_{ij} 
-\nu_{R\sigma}^\ast
\tilde{\sigma}_{ij}
&= 0,
\\
\label{phiBalPertDimless}
\frac{\partial \tilde{\varphi}_i}{\partial \tilde{t}} 
+ 4\frac{\partial \tilde{R}_{ij}}{\partial \tilde{x}_j} 
 + \xi_2 \left(\frac{\partial \tilde{\Delta}}{\partial \tilde{x}_i} 
+4 a_2 \frac{\partial \tilde{T}}{\partial \tilde{x}_i} 
- a_2 \frac{\partial \tilde{\sigma}_{ij}}{\partial \tilde{x}_j}\right)
+\xi_\varphi \tilde{\varphi}_i
- \nu_{\varphi q}^\ast \tilde{q}_i 
&= 0,
\end{align}
\endgroup
where
\begin{align}
\label{xi}
\left.
\begin{gathered}
\xi_\sigma  = \nu_\sigma^\ast - \zeta_0^\ast,
\qquad
\xi_q = \nu_q^\ast - \frac{3}{2} \zeta_0^\ast,
\qquad
\xi_m = \nu_m^\ast - \frac{3}{2} \zeta_0^\ast,
\qquad
\xi_R = \nu_R^\ast - 2 \zeta_0^\ast,
\\
\xi_\varphi = \nu_\varphi^\ast - \frac{5}{2} \zeta_0^\ast,
\qquad
\xi_1 = \frac{8}{d(d+2)} \left(1 - \frac{d+2}{2} a_2\right),
\qquad
\xi_2 = (d+2)(d+4).
\end{gathered}
\right\}
\end{align}
System \eqref{massBalPertDimless}--\eqref{phiBalPertDimless}, admits a normal mode solution of the form 
\begin{align}
\label{normalmodesol}
\tilde{\bm{\Psi}} = \check{\bm{\Psi}} \exp {\big[ \mathbbm{i} ( \bm{k} \cdot \tilde{\bm{x}} - \omega \, \tilde{t}) \big]},
\end{align}
where $\tilde{\bm{\Psi}}= (\tilde{n}, \tilde{v}_i, \tilde{T}, \tilde{\sigma}_{ij}, \tilde{q}_i, \tilde{m}_{ijk}, \tilde{\Delta}, \tilde{R}_{ij}, \tilde{\varphi}_i)^{\mathsf{T}}$ is the vector containing all the dimensionless perturbations 
and $\check{\bm{\Psi}} = (\check{n}, \check{v}_i, \check{T}, \check{\sigma}_{ij}, \check{q}_i, \check{m}_{ijk}, \check{\Delta}, \check{R}_{ij}, \check{\varphi}_i)^{\mathsf{T}}$ the vector containing their corresponding complex amplitudes.
Furthermore, in the normal mode solution \eqref{normalmodesol},
$\mathbbm{i}$ is the imaginary unit, $\bm{k}$ the dimensionless wavevector of the disturbance and $\omega$ the dimensionless frequency of the associated wave. 
For temporal stability analysis, the wavevector $\bm{k}$ is assumed to be real and the frequency $\omega$ is assumed to be complex. 
The real part of the frequency, $\mathrm{Re}(\omega)$, measures the phase velocity $\bm{v}_{\mathrm{ph}}$ of the wave via $\bm{v}_{\mathrm{ph}} = \mathrm{Re}(\omega)\bm{k}/k^2$, where $k=|\bm{k}|$ is the wavenumber, 
and the imaginary part of the frequency, $\mathrm{Im}(\omega)$, controls the growth/decay of the disturbance in time and is referred to as the \emph{growth rate}. 
Form \eqref{normalmodesol} of the normal mode solution deduces that the disturbance will grow (or decay) in time if the growth rate is positive (or negative). Consequently, for stability of the system, the growth rate must be non-positive, i.e.~$\mathrm{Im}(\omega) \leq 0$. 


Assuming that the wavevector of the disturbance is parallel to the $x$-axis, i.e.~$\bm{k} = k \, \hat{\bm{x}}$, where 
$\hat{\bm{x}}$ is the unit vector in the $x$-direction, system~\eqref{massBalPertDimless}--\eqref{phiBalPertDimless}, 
using relations in appendix~\ref{app:tracefreeGradients},
can be decomposed into two independent eigenvalue problems, namely the longitudinal problem and the transverse problem, for the amplitude of the disturbance. 
It is worthwhile to note that in two dimensions, there is only one transverse direction along the $y$-axis while in three dimensions, there are two transverse directions along the $y$- and $z$-axes. 
Consequently, there is one transverse problem in two dimensions and two transverse problems in three dimensions. Nevertheless, the coefficient matrices associated with the both transverse problems in three dimensions are essentially the same; therefore it is sufficient to analyse only one transverse problem (let us say, that associated with the $y$-direction) in three dimensions. 
Thus, the longitudinal and transverse problems read
\begin{subequations}
\label{eigvalProbs}
\begin{align}
\mathscr{L} \begin{bmatrix}
        \check{n} \\
        \check{v}_x \\
        \check{T} \\
        \check{\sigma}_{xx} \\
        \check{q}_x \\
        \check{m}_{xxx} \\
        \check{\Delta} \\
        \check{R}_{xx} \\
        \check{\varphi}_x
        \end{bmatrix}
= 
\begin{bmatrix}
0 \\
0 \\
0 \\
0 \\
0 \\
0 \\
0 \\
0 \\
0 
\end{bmatrix}
\qquad\textrm{and}\qquad
\mathscr{T} 
\begin{bmatrix}
\check{v}_y \\
\check{\sigma}_{xy} \\
\check{q}_y \\
\check{m}_{xxy} \\
\check{R}_{xy} \\
\check{\varphi}_y
\end{bmatrix}
= 
\begin{bmatrix}
0 \\
0 \\
0 \\
0 \\
0 \\
0 
\end{bmatrix}
\tag{\theequation $a$,$b$}
\end{align}
\end{subequations}
respectively, where the matrices $\mathscr{L} \equiv \mathscr{L}(k,\omega, d, e)$ and $\mathscr{T} \equiv \mathscr{T}(k,\omega, d, e)$ are presented in appendix~\ref{app:matrices}.

%
For nontrivial solutions of the longitudinal and transverse problems \eqref{eigvalProbs}, the determinants of both matrices $\mathscr{L}$ and $\mathscr{T}$ must vanish, i.e.~$\det{(\mathscr{L})}=0$ and $\det{(\mathscr{T})}=0$. 
These conditions are the dispersion relations for the longitudinal and transverse systems and can, respectively, be written as%
\begin{subequations}
\label{dispRels}
\begin{align}
\omega^9 
+ \sum\limits_{r=1}^9 \mathbbm{a}_r \, \omega^{9-r} = 0
\qquad\textrm{and}\qquad
\omega^6 
+ \sum\limits_{s=1}^6 \mathbbm{b}_s \, \omega^{6-s} = 0,
\tag{\theequation $a$,$b$}
\end{align}
\end{subequations}
%
where the coefficients $\mathbbm{a}_r$ ($r=1,2,\dots,9$) and $\mathbbm{b}_s$ ($s=1,2,\dots,6$) are functions of the wavenumber $k$, the dimension $d$ and the coefficient of restitution $e$; although the explicit values of these coefficients are not given here for conciseness. 
The solutions of the longitudinal and transverse problems \eqref{eigvalProbs} for each root $\omega \equiv \omega(k)$ of the corresponding dispersion relations \eqref{dispRels} are referred to as the eigenmodes for the longitudinal and transverse problems \eqref{eigvalProbs}, respectively.
\subsection{Eigenmodes from the longitudinal and transverse systems}\label{Eigmodes}
\begin{figure}
\begin{center}
\includegraphics[width = 0.45\textwidth]{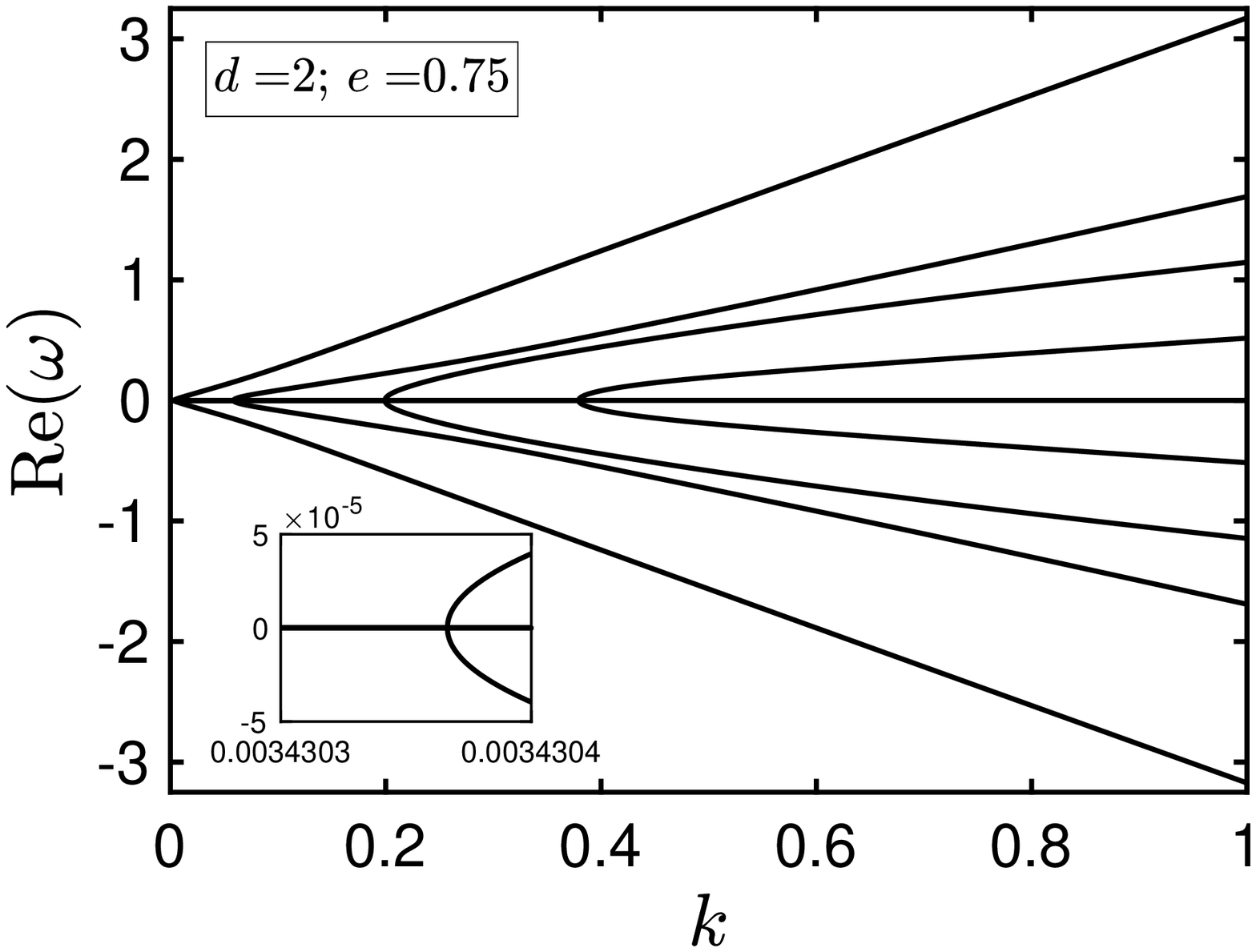}
\hfill
\includegraphics[width = 0.45\textwidth]{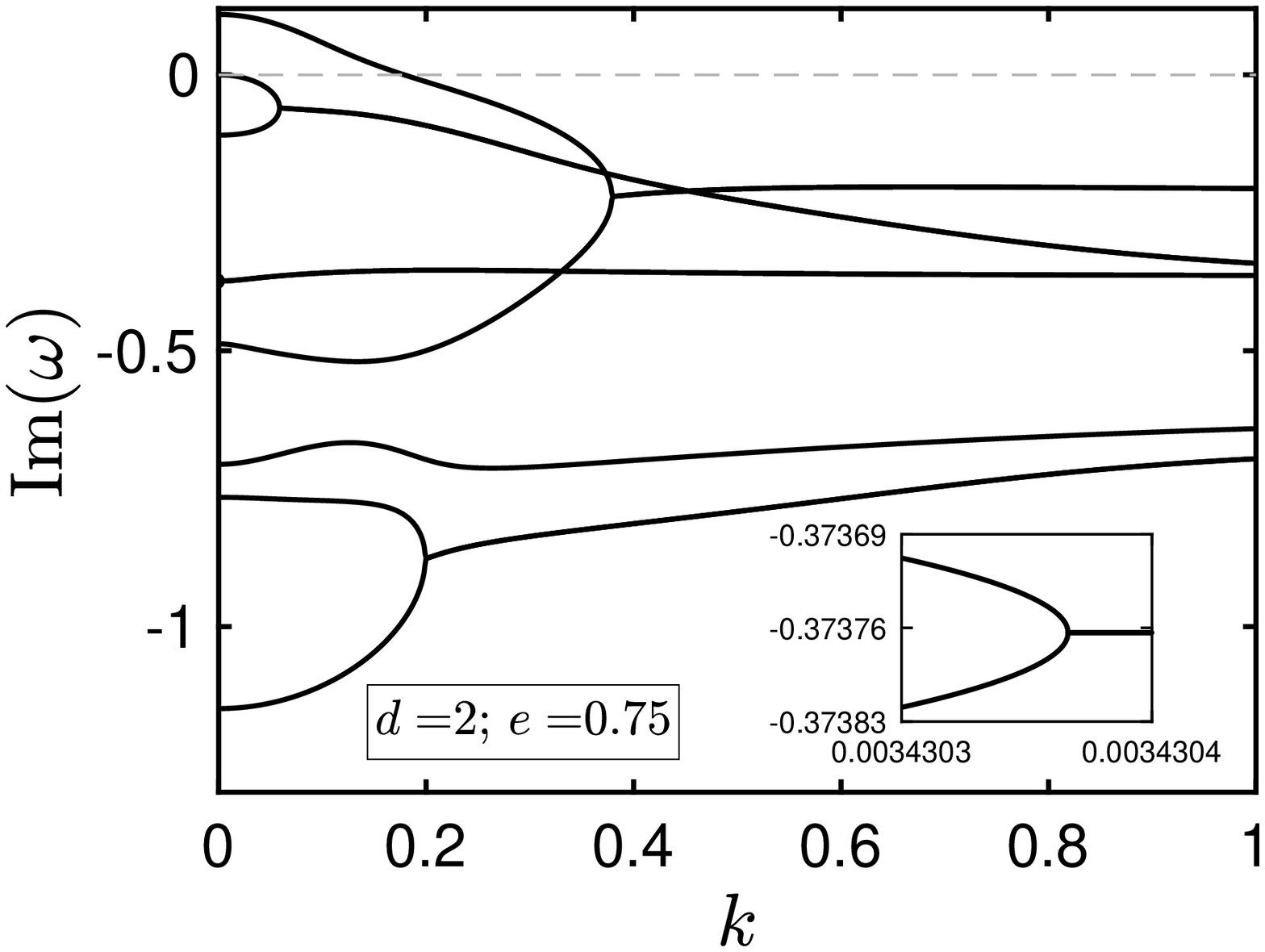}
\includegraphics[width = 0.45\textwidth]{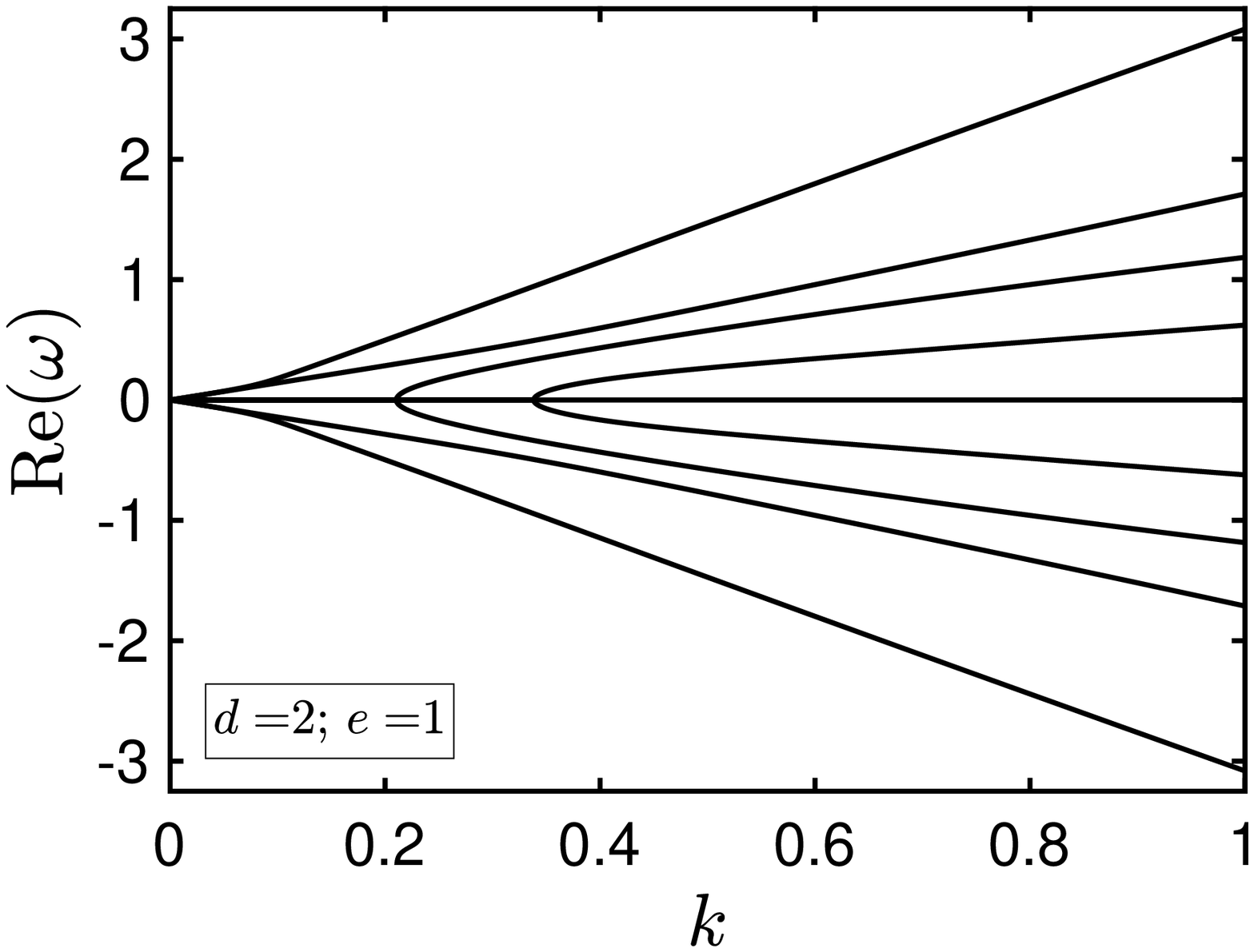}
\hfill
\includegraphics[width = 0.45\textwidth]{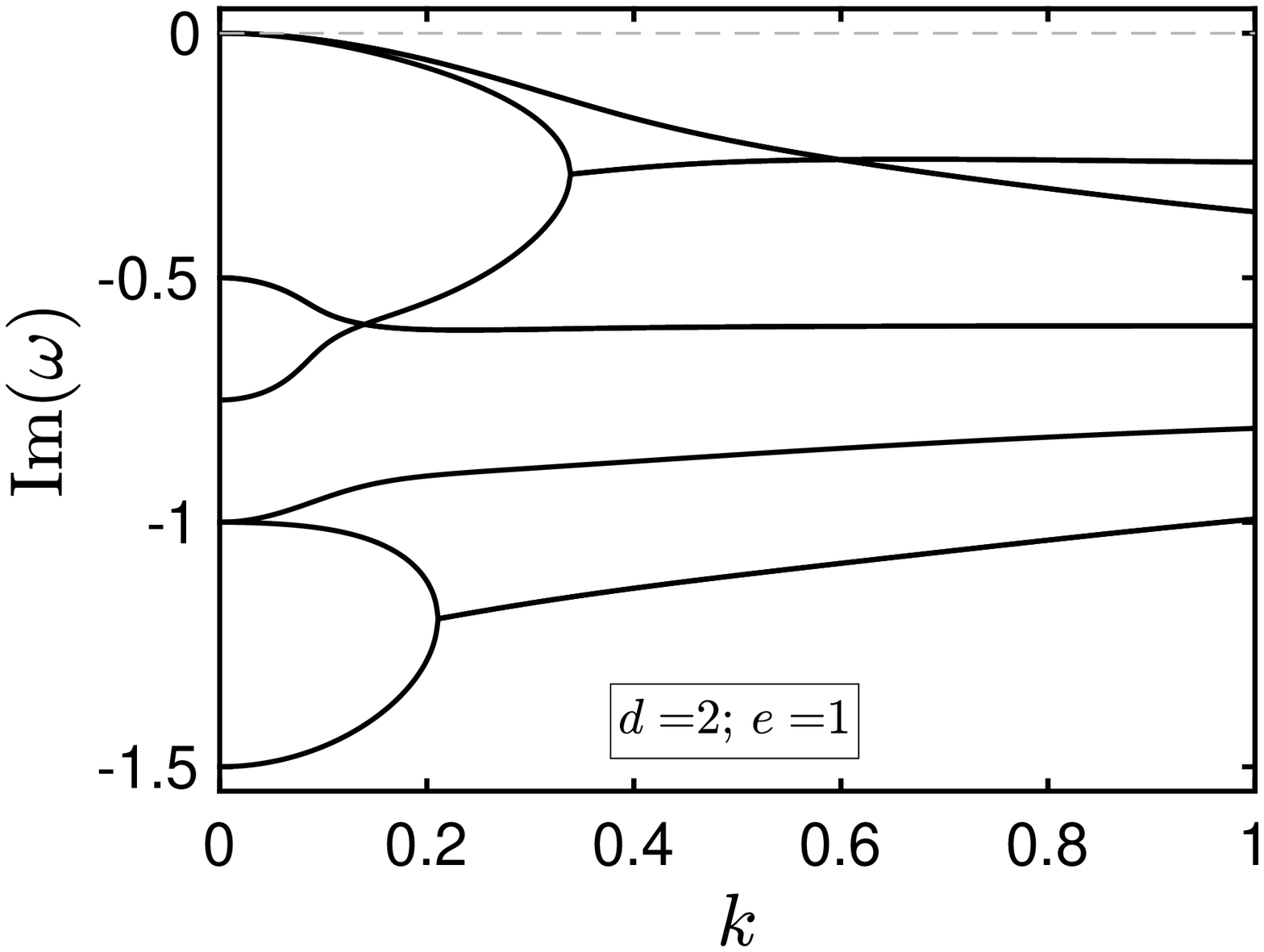}
\caption{The real and imaginary parts of the frequencies associated with the eigenmodes from the longitudinal system (\ref{eigvalProbs}$a$) obtained from the G29 equations for $d=2$. 
The top and bottom rows display the results for $e=0.75$ (inelastic case) and $e=1$ (elastic case), respectively.
}
\label{fig:long_d2}
\end{center}
\end{figure}

\begin{figure}
\begin{center}
\includegraphics[width = 0.45\textwidth]{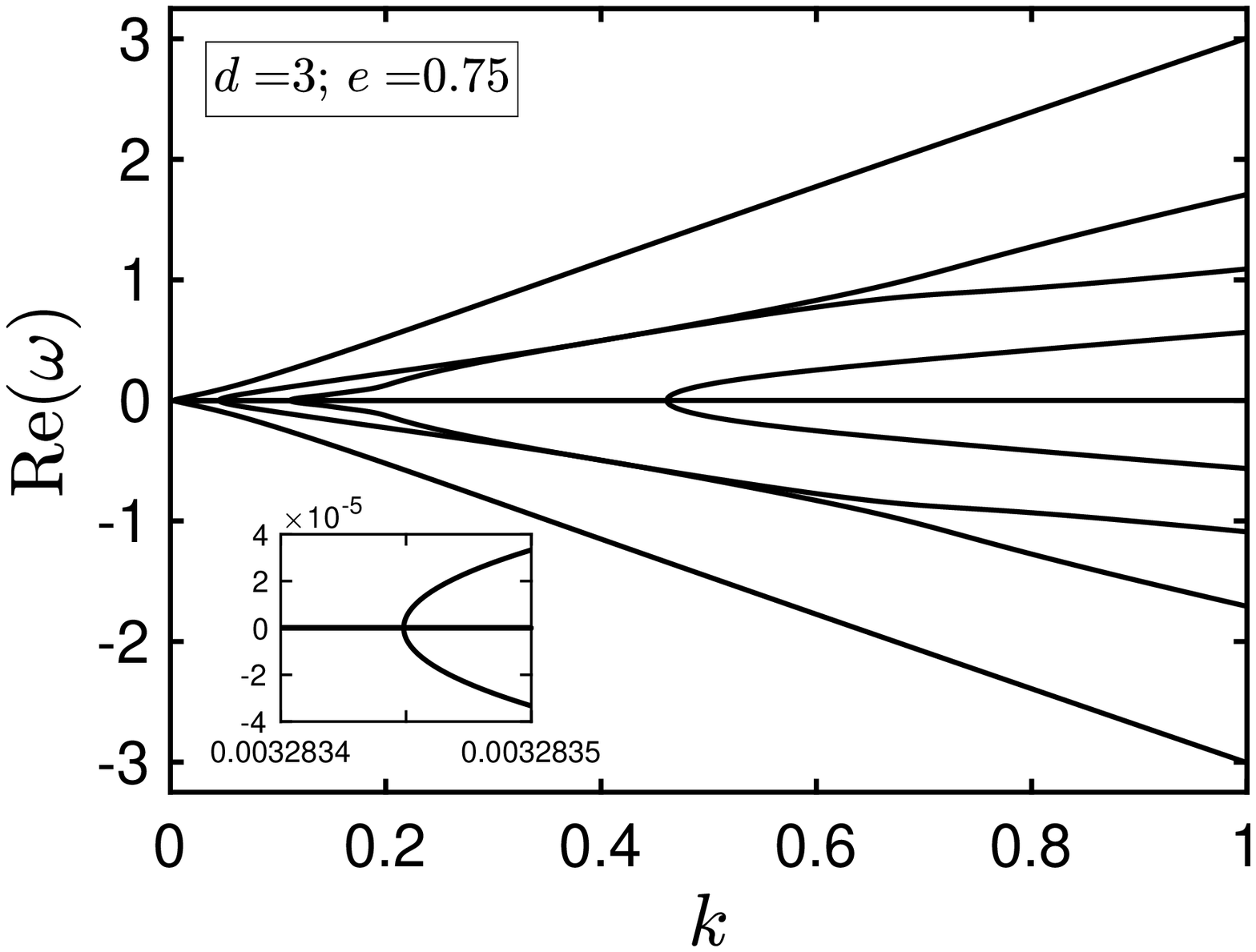}
\hfill
\includegraphics[width = 0.45\textwidth]{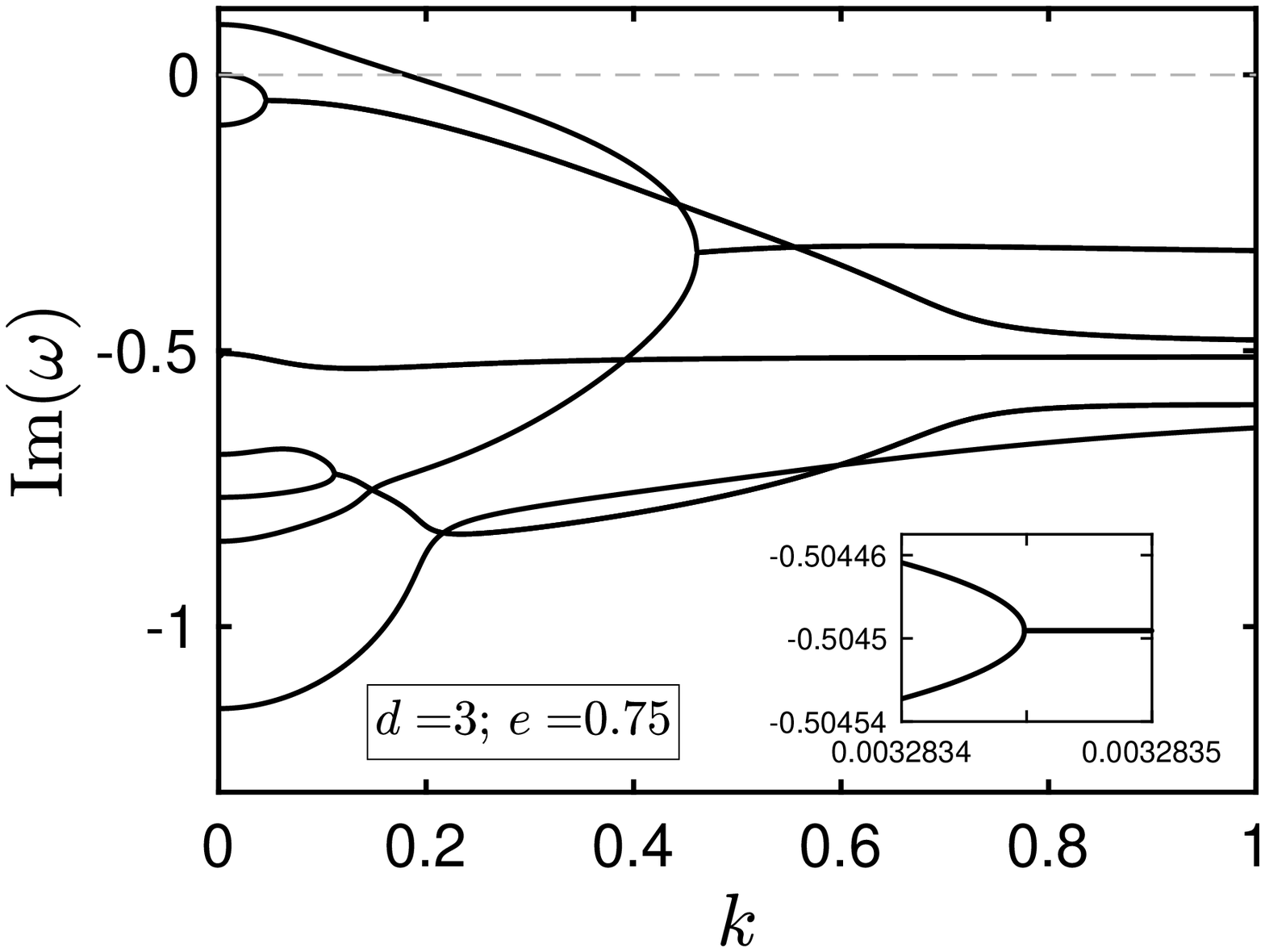}
\includegraphics[width = 0.45\textwidth]{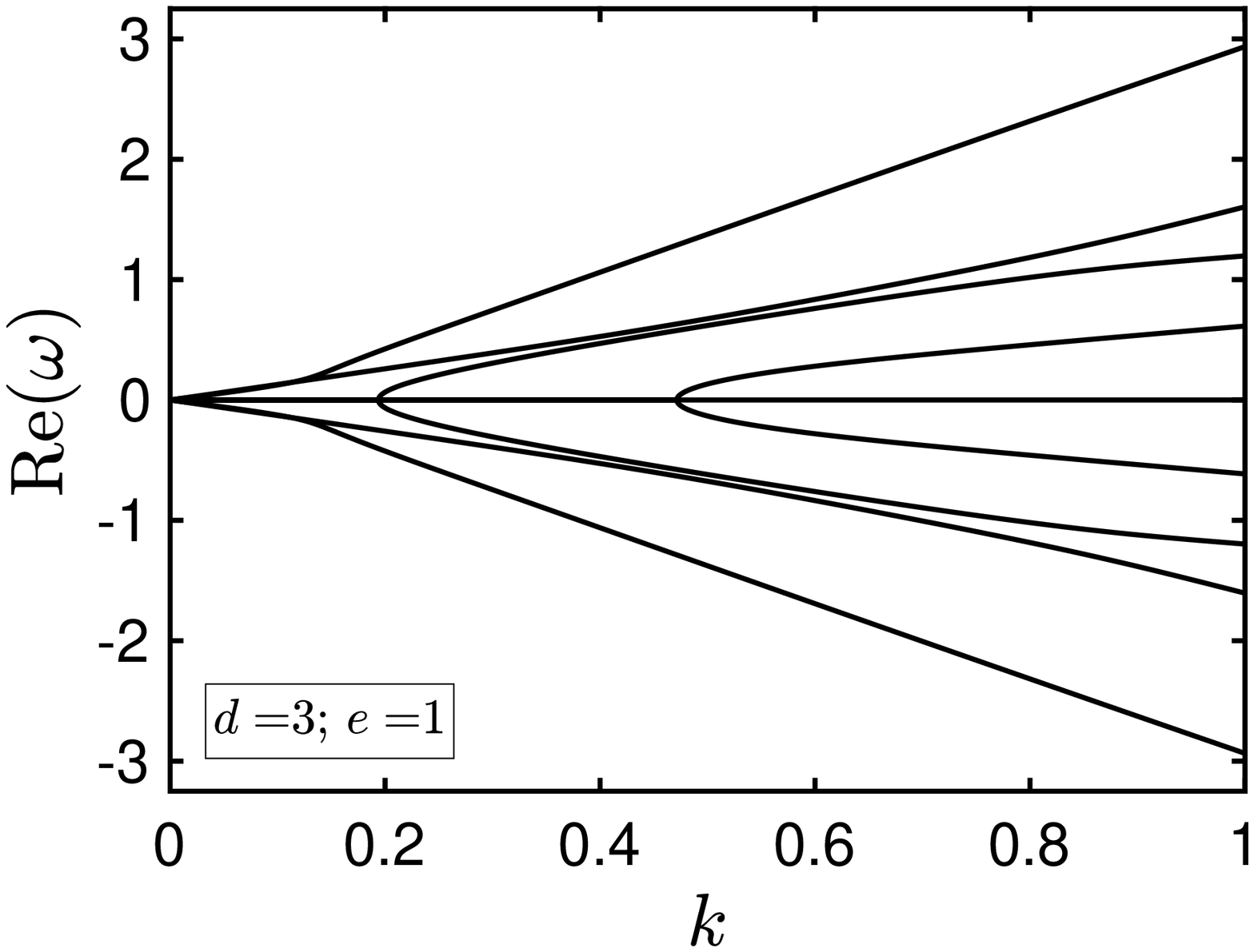}
\hfill
\includegraphics[width = 0.45\textwidth]{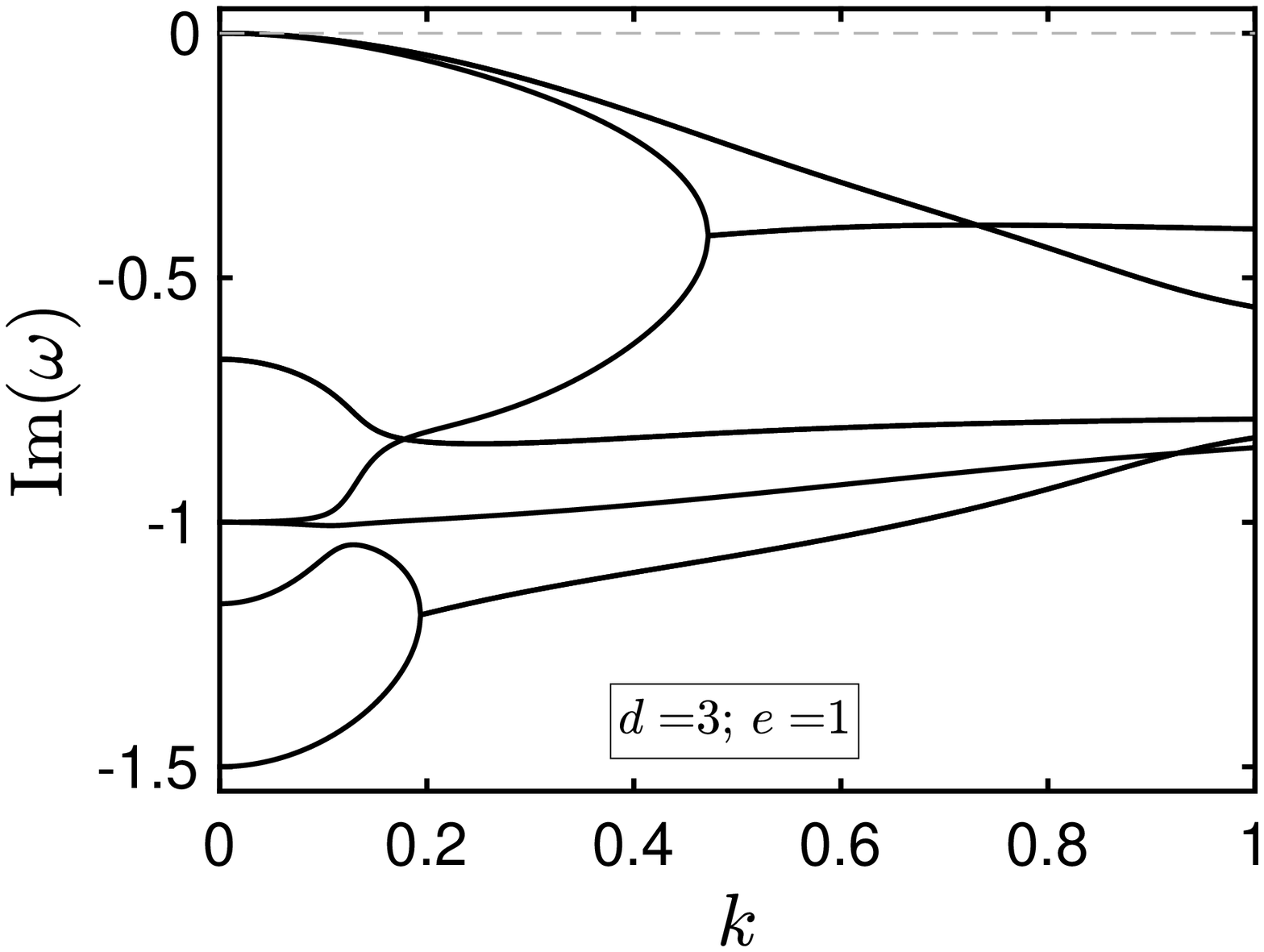}
\caption{The same as figure~\ref{fig:long_d2} but for $d=3$. 
}
\label{fig:long_d3}
\end{center}
\end{figure}

The nine roots of dispersion relation (\ref{dispRels}$a$) lead to nine eigenmodes for the longitudinal system (\ref{eigvalProbs}$a$) associated with the G29 equations while the six roots of dispersion relation (\ref{dispRels}$b$) yield six eigenmodes for the transverse system (\ref{eigvalProbs}$b$) associated with the G29 equations. 
The real part of the frequency ($\mathrm{Re}(\omega)$) associated with each eigenmode and its growth rate ($\mathrm{Im}(\omega)$) from both the longitudinal and transverse systems are illustrated in figures~\ref{fig:long_d2}--\ref{fig:tran_d3} for $d=2$ (figures~\ref{fig:long_d2} and \ref{fig:tran_d2}), $d=3$ (figures~\ref{fig:long_d3} and \ref{fig:tran_d3})---with figures~\ref{fig:long_d2} and \ref{fig:long_d3} being for the longitudinal system (\ref{eigvalProbs}$a$) and figures~\ref{fig:tran_d2} and \ref{fig:tran_d3} for the transverse system (\ref{eigvalProbs}$b$). 
The top and bottom rows in each figure depict the eigenmodes for the inelastic ($e=0.75$) and elastic ($e=1$) cases, respectively while the left and right columns in each figure delineate the real part of the frequency and growth rate, respectively. 
It is apparent from the left columns of the figures~\ref{fig:long_d2} and \ref{fig:long_d3} that four pairs out of the nine eigenmodes from the longitudinal system have nonzero $\mathrm{Re}(\omega)$, i.e.~the four pairs of associated eigenmodes are travelling whereas one eigenmode is purely imaginary, i.e.~it has $\mathrm{Re}(\omega)=0$ for all wavenumbers and hence always remains stationary. 
Similarly, it is clear from the left columns of figures~\ref{fig:tran_d2} and \ref{fig:tran_d3} that two pairs out of the six eigenmodes from the transverse system have nonzero $\mathrm{Re}(\omega)$, i.e.~the two pairs of associated eigenmodes are travelling, whereas two eigenmodes are purely imaginary and hence remain stationary for all wavenumbers. 
A travelling eigenmode is commonly referred to as a sound mode and a stationary eigenmode as a heat mode \citep{BP2004,Garzo2005}. 
\subsubsection{Longitudinal systems (figures~\ref{fig:long_d2} and \ref{fig:long_d3})}
For $e=0.75$ (top rows of figures~\ref{fig:long_d2} and \ref{fig:long_d3}), the first pair of sound modes originates at $k \approx 0.00343$ in the case of $d=2$ (at $k \approx 0.00328$ in the case of $d=3$) (see the insets on the left columns of figures~\ref{fig:long_d2} and \ref{fig:long_d3}), followed by a second pair of sound modes commencing at $k \approx 0.0587$ in the case of $d=2$ (at $k \approx 0.0453$ in the case of $d=3$) travelling slower than the first pair, followed by a third pair of sound modes starting at $k \approx 0.1992$ in the case of $d=2$ (at $k \approx 0.1114$ in the case of $d=3$) travelling even slower than the second pair (in general), followed by a fourth pair of sound modes commencing at $k \approx 0.3791$ in the case of $d=2$ (at $k \approx 0.461$ in the case of $d=3$) travelling even slower than the third pair. 
It should be noted, however, that below these wavenumbers, the respective eigenmodes are heat modes since their real parts are zero.
Each pair of the sound modes starts propagating in opposite directions at the aforesaid wavenumbers as the eigenvalues corresponding to each pair of the sound modes are a pair of complex conjugates.
Beyond the aforesaid wavenumbers, the imaginary parts of each (respective) pair of sound modes coincide due to the same reason.
This can be clearly seen in the top rows and right columns of figures~\ref{fig:long_d2} and \ref{fig:long_d3}, in which the imaginary parts of the first, second, third and fourth pairs of sound modes coincide beyond $k \approx 0.00343, 0.0587, 0.1992, 0.3791$, respectively,  in the case of $d=2$ (beyond $k \approx 0.00328, 0.0453, 0.1114, 0.461$ respectively,  in the case of $d=3$).
From the top rows and right columns of figures~\ref{fig:long_d2} and \ref{fig:long_d3}, it is evident that all the eigenmodes except one heat mode from the fourth pair (for which $\mathrm{Im}(\omega)$ coincide beyond $k \approx 0.3791$ in the case of $d=2$ and beyond $k \approx 0.461$ in the case of $d=3$) are stable as $\mathrm{Im}(\omega) \leq 0$ for them. 
The unstable heat mode remains unstable for wavenumbers $k < k_c$ but becomes stable for wavenumbers $k\geq k_c$, where $k_c$ is called the critical wavenumber, the wavenumber at which $\mathrm{Im}(\omega)$ flips its sign. 
For $e=0.75$, $k_c \approx 0.179$ in the case of $d=2$ and $k_c \approx 0.18$ in the case of $d=3$. 
The general behaviour of the eigenmodes of the longitudinal system for moderate to large values of $e$ is similar to those for $e=0.75$ (as shown in the top rows of figures~\ref{fig:long_d2} and \ref{fig:long_d3}), although $k_c \to 0$ as $e \to 1$. 
Thus, for moderately to nearly elastic granular gases, there exists a critical wavenumber $k_c$, below which the unstable heat mode renders the longitudinal system unstable. 
On the other hand, for sufficiently small values of $e$, the growth rates of some of the eigenmodes of the longitudinal system remain positive even for large wavenumbers and, hence, there does not exist a critical wavenumber in this case. 
This means that the longitudinal system remains always unstable for all coefficients of restitution below a certain value, which is not true \cite[see, e.g.,][]{GST2018}.
Let us refer to this value of the coefficient of restitution---below which a system remains always unstable---as the threshold coefficient of restitution $e_{\mathrm{th}}$. 
For the longitudinal system associated with the G29 equations, $e_{\mathrm{th}} \approx 0.56356$ in the case of $d=2$ and $e_{\mathrm{th}} \approx 0.40157$ in the case of $d=3$.

\begin{figure}
\begin{center}
\includegraphics[width = 0.45\textwidth]{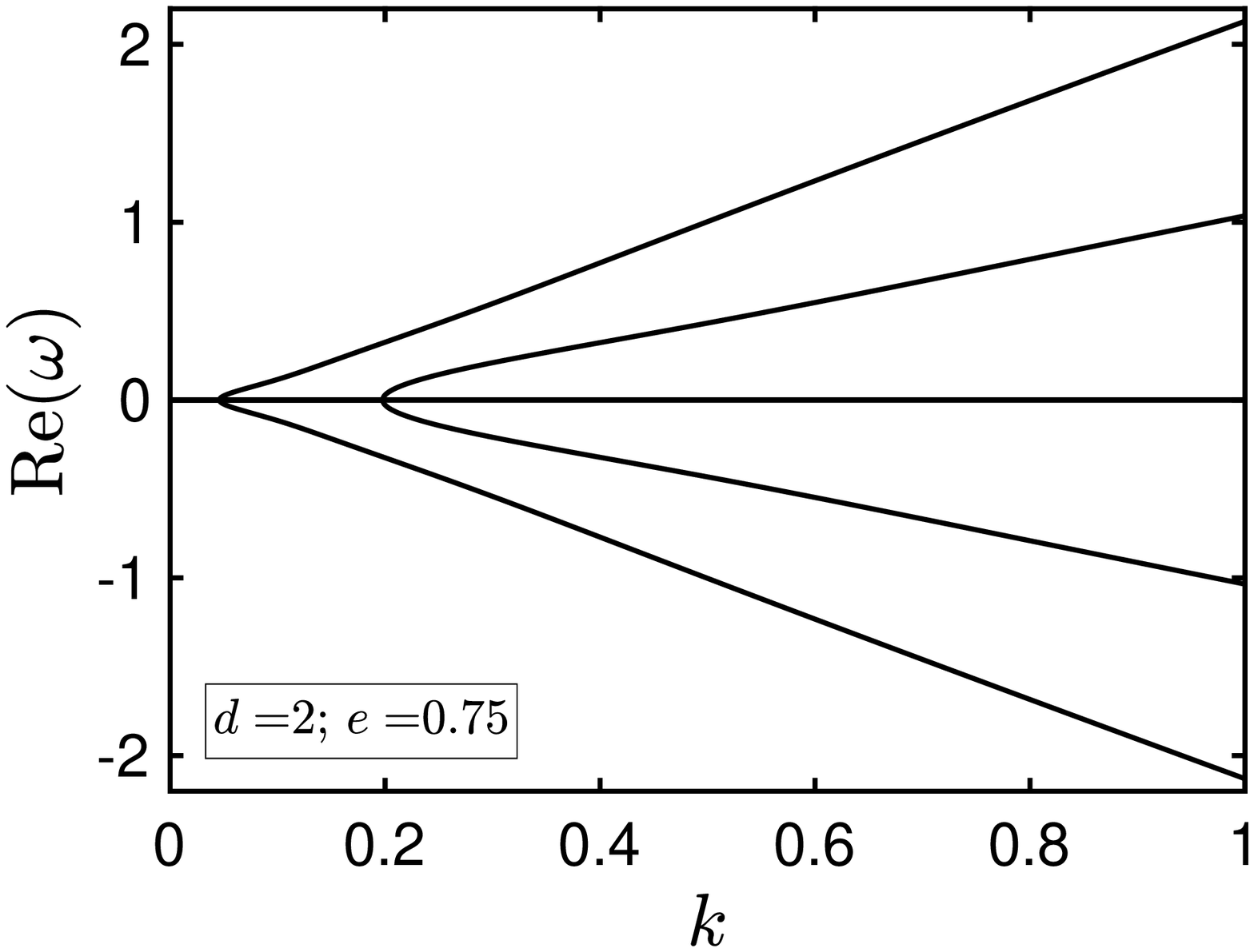}
\hfill
\includegraphics[width = 0.45\textwidth]{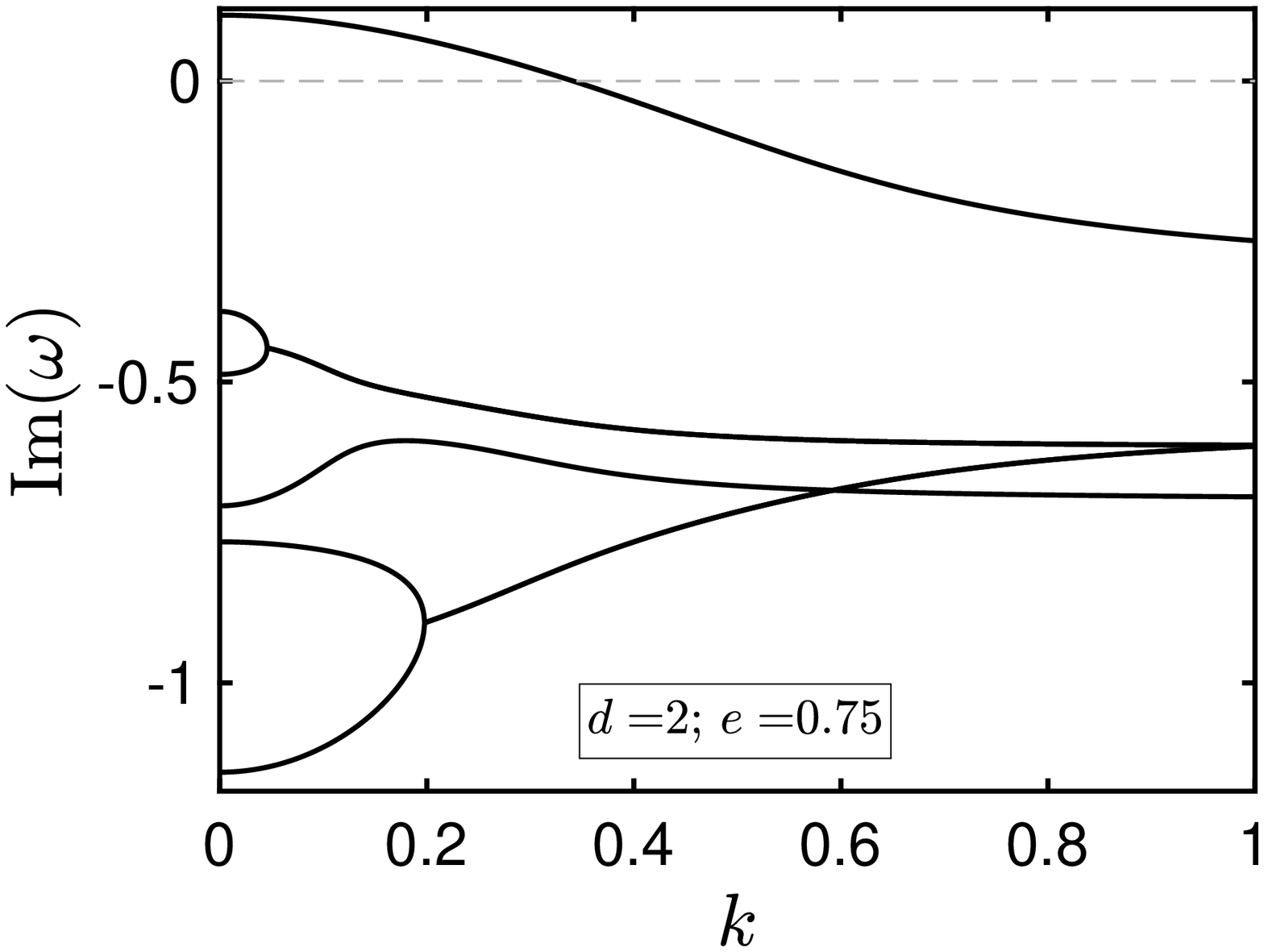}
\includegraphics[width = 0.45\textwidth]{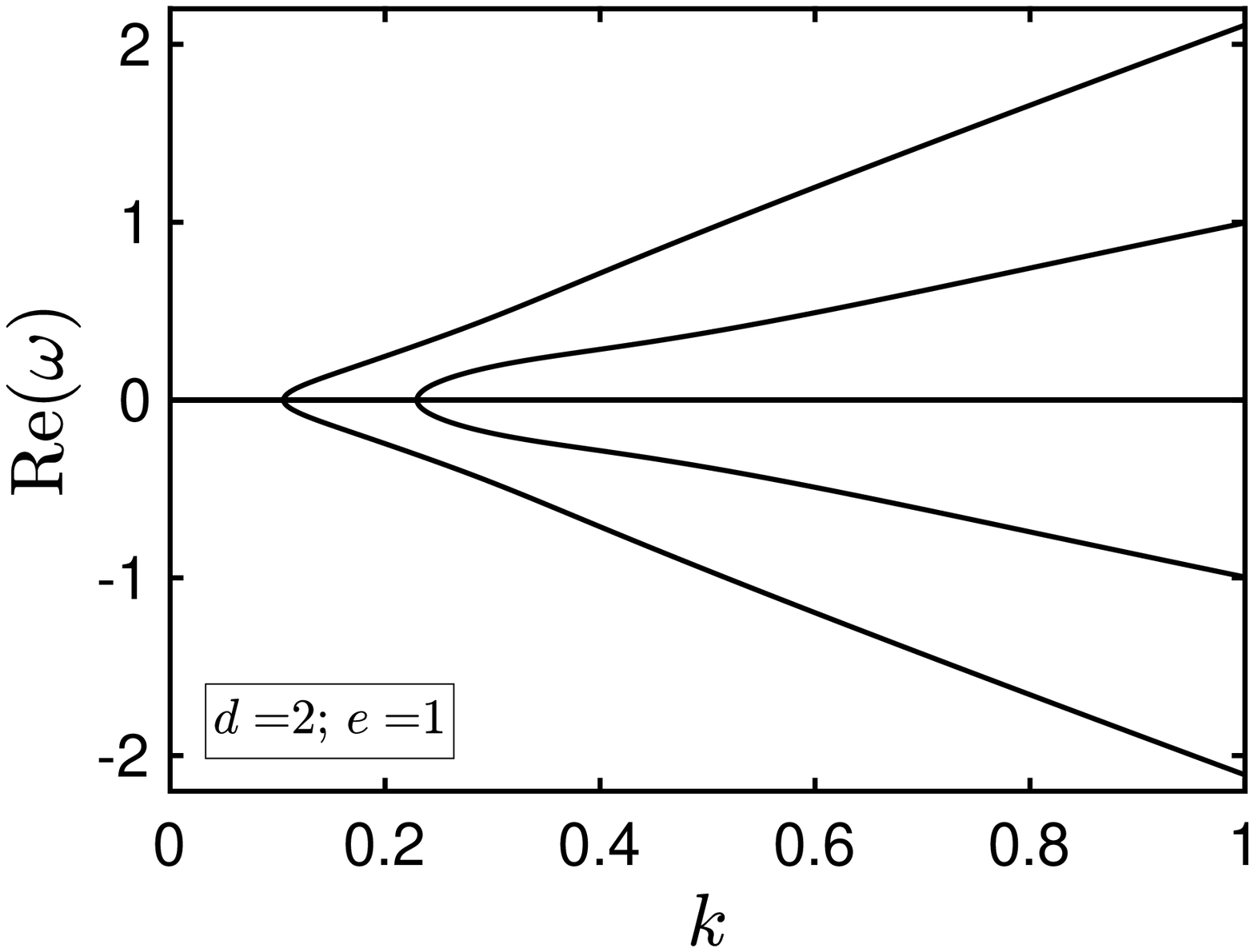}
\hfill
\includegraphics[width = 0.45\textwidth]{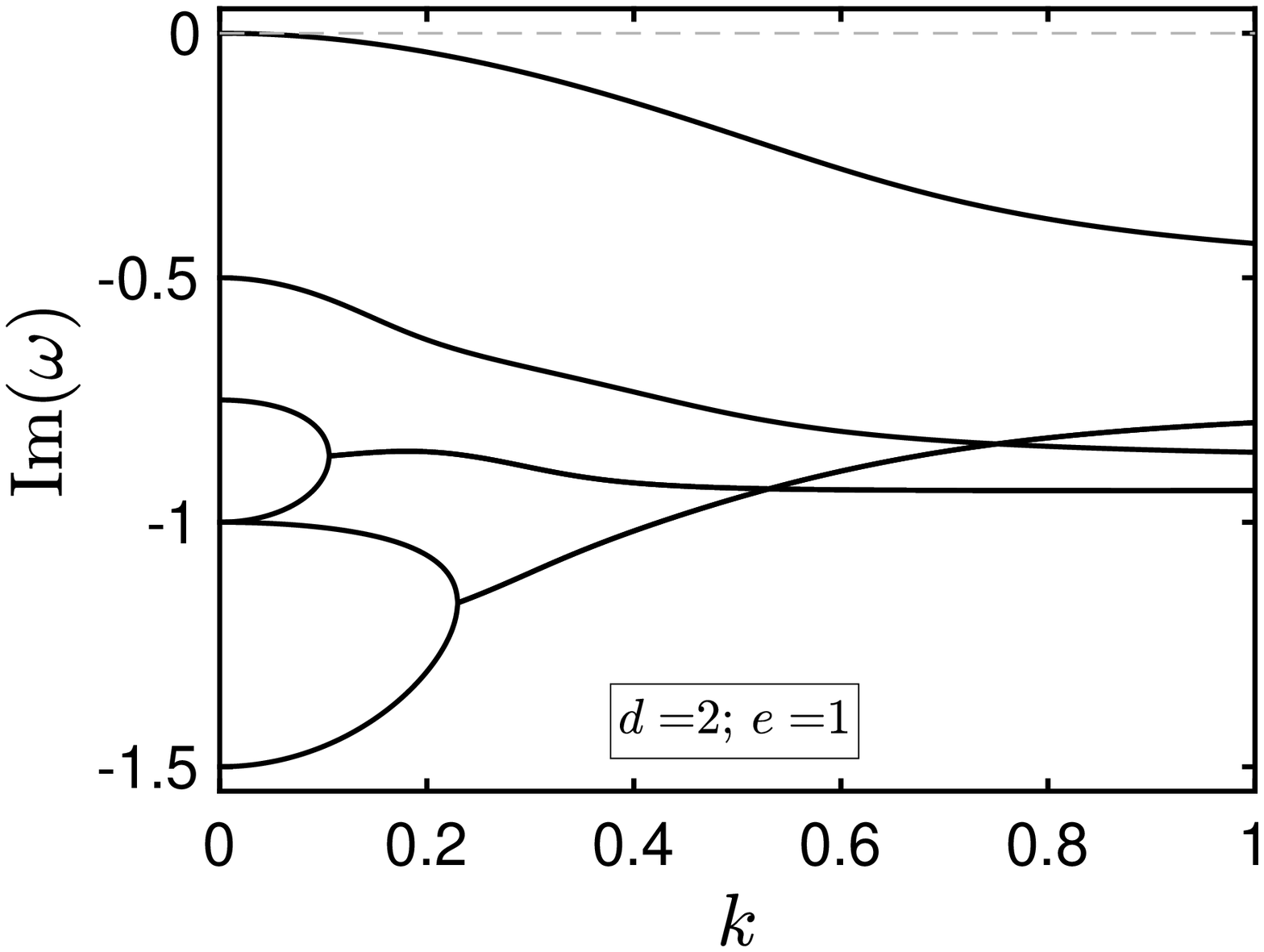}
\caption{The real and imaginary parts of the frequencies associated with the eigenmodes from the transverse system (\ref{eigvalProbs}$b$) obtained from the G29 equations for $d=2$. 
The top and bottom rows display the results for $e=0.75$ (inelastic case) and $e=1$ (elastic case), respectively.
}
\label{fig:tran_d2}
\end{center}
\end{figure}

\begin{figure}
\begin{center}
\includegraphics[width = 0.45\textwidth]{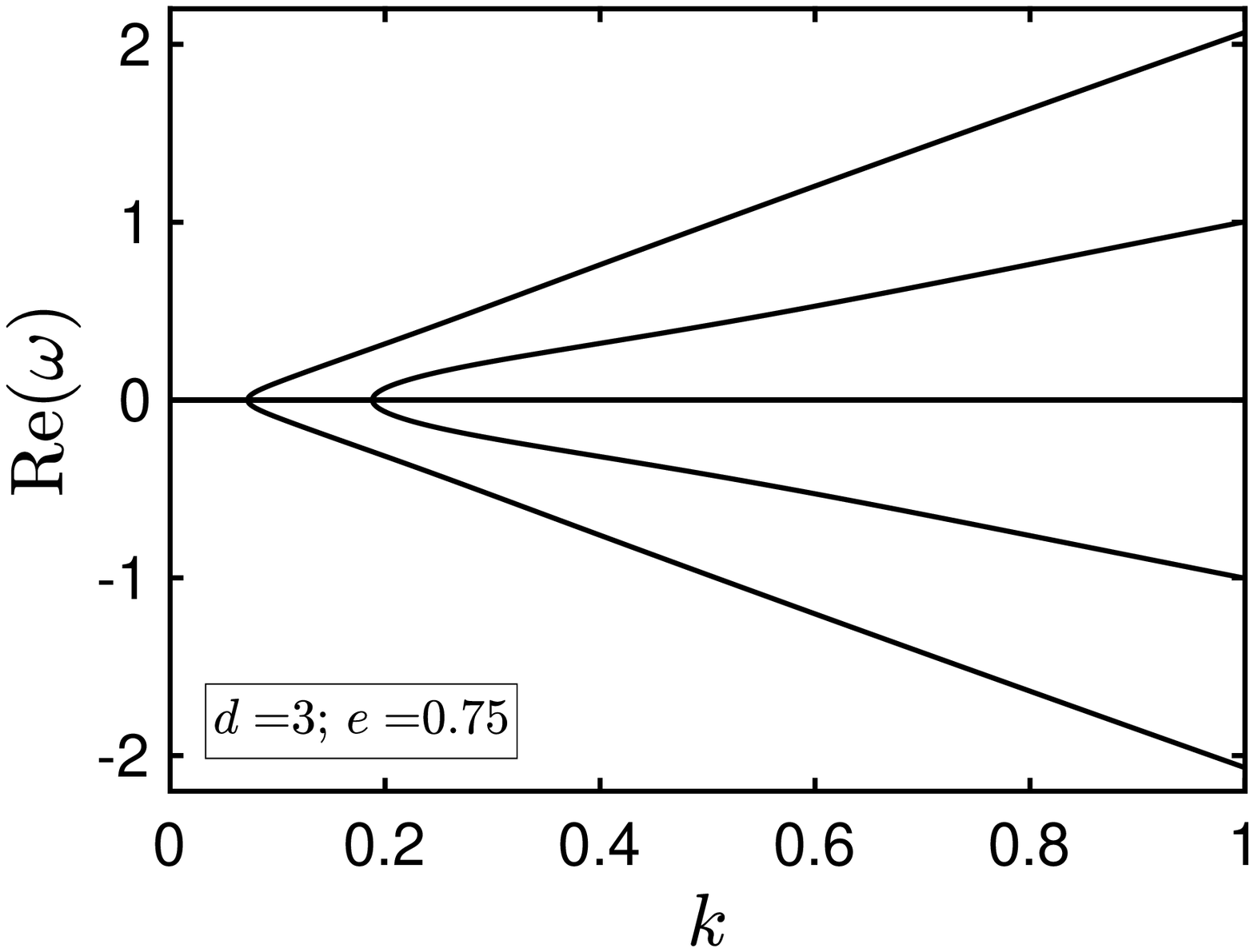}
\hfill
\includegraphics[width = 0.45\textwidth]{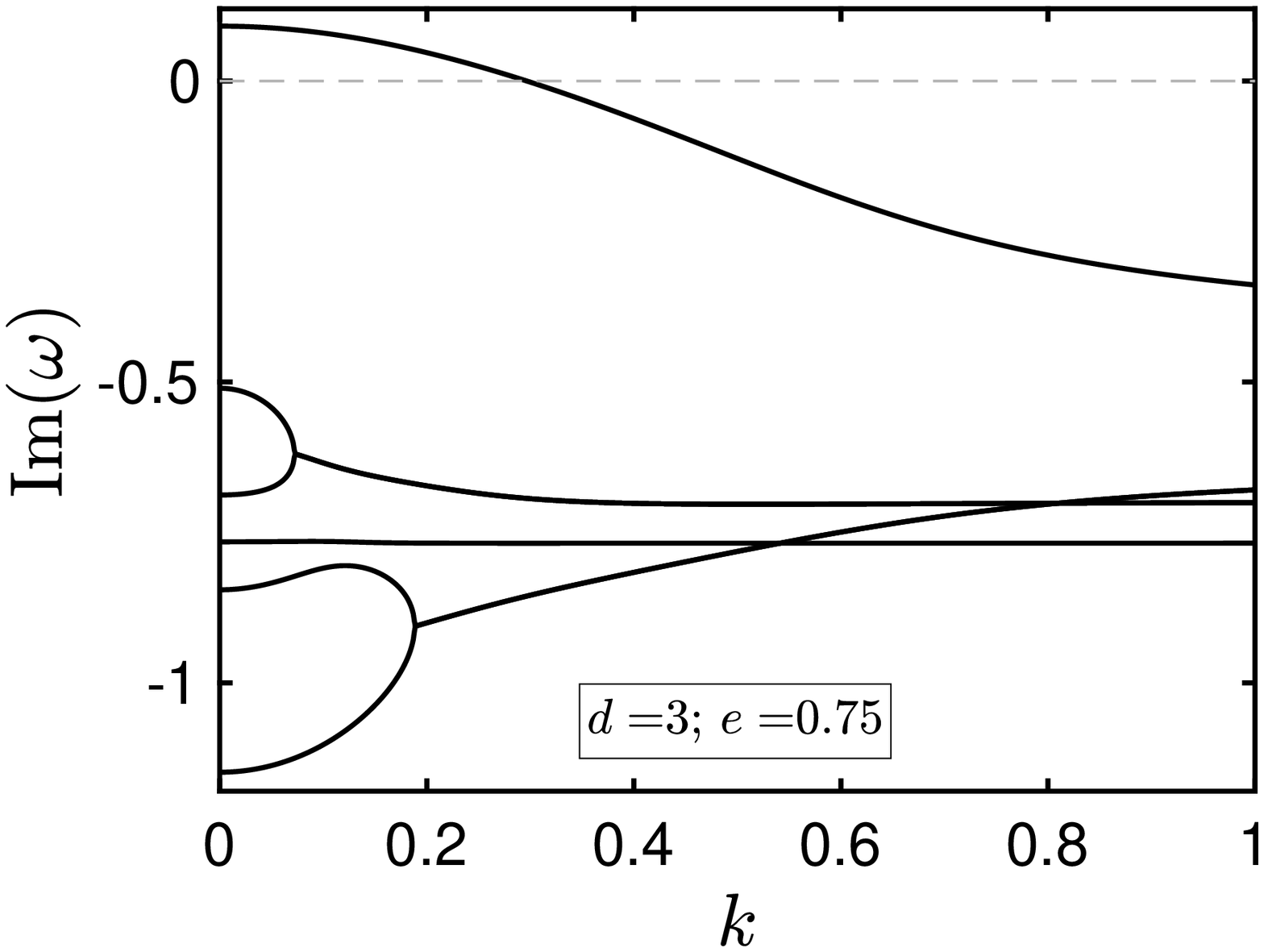}
\includegraphics[width = 0.45\textwidth]{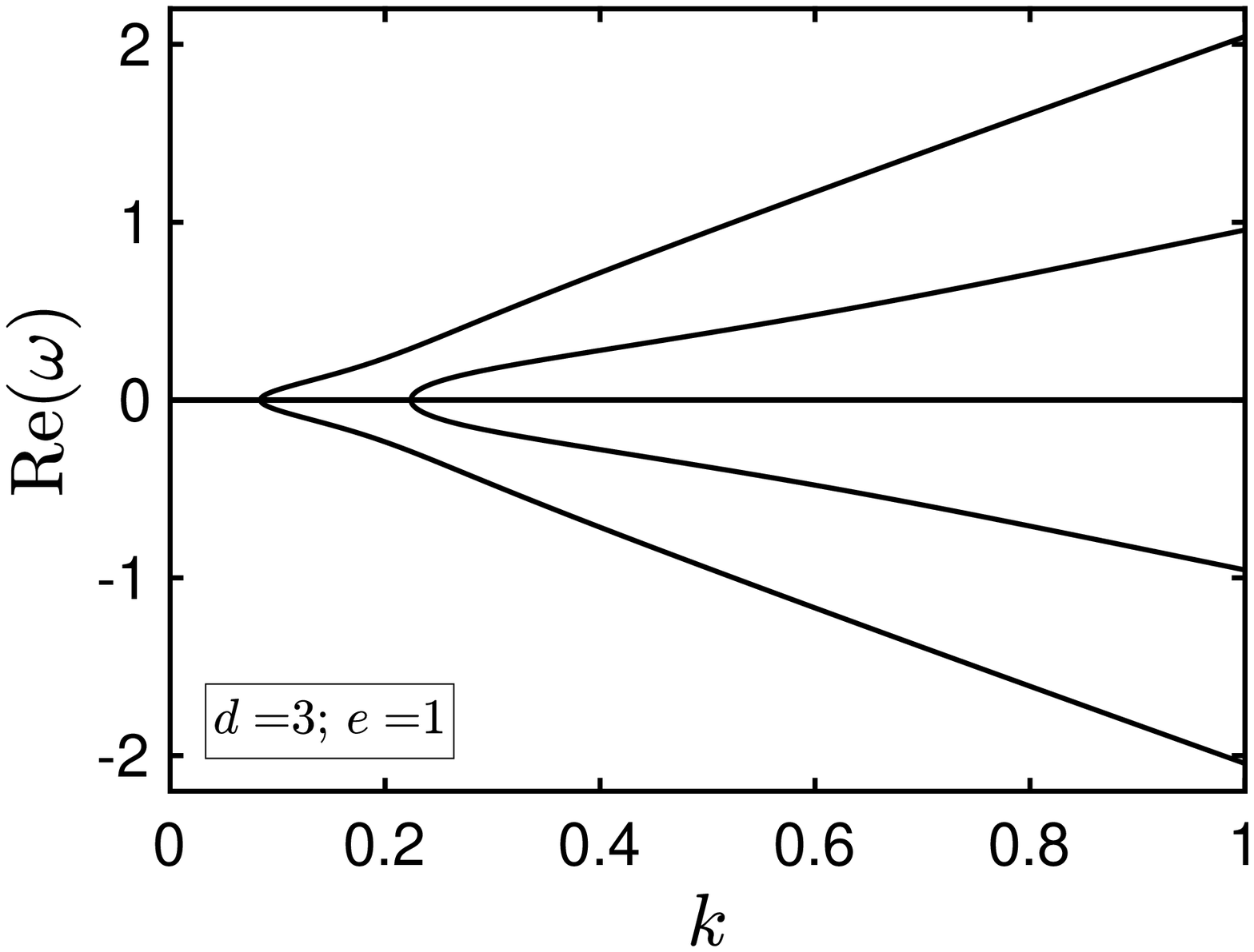}
\hfill
\includegraphics[width = 0.45\textwidth]{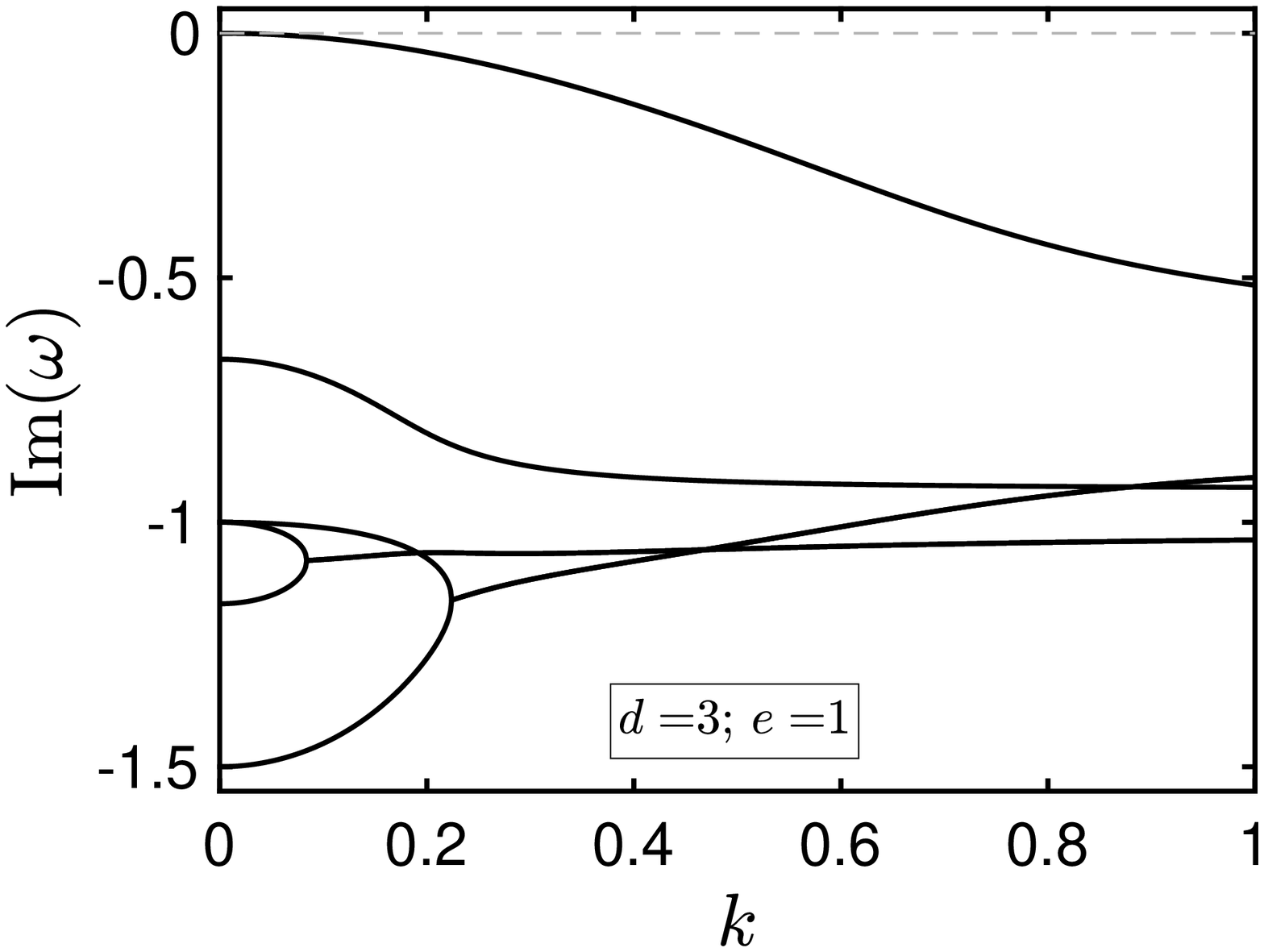}
\caption{The same as figure~\ref{fig:tran_d2} but for $d=3$.
}
\label{fig:tran_d3}
\end{center}
\end{figure}

For $e=1$ (bottom rows of figures~\ref{fig:long_d2} and \ref{fig:long_d3}), two pairs (one faster and the other slower) of sound modes start propagating in opposite directions already at $k=0$, followed by a third pair of sound modes appearing at $k \approx 0.2104$ in the case of $d=2$ (at $k \approx 0.1937$ in the case of $d=3$) travelling slower than both the first and second pairs, followed by an even slower fourth pair of sound modes commencing at $k \approx 0.3383$ in the case of $d=2$ (at $k \approx 0.4714$ in the case of $d=3$).
Accordingly, the imaginary parts of the frequencies for the two pairs of sound modes commencing at $k=0$ coincide for all wavenumbers, that for the third pair coincide beyond $k \approx 0.2104$ in the case of $d=2$ (beyond $k \approx 0.1937$ in the case of $d=3$) and that for the fourth pair coincide beyond $k \approx 0.3383$ in the case of $d=2$ (beyond $k \approx 0.4714$ in the case of $d=3$); see the bottom rows and right columns of the figures.
From the bottom rows and right columns of figures~\ref{fig:long_d2} and \ref{fig:long_d3}, it can be perceived that $\mathrm{Im}(\omega) \leq 0$ for all eigenmodes in this case. 
This means that the longitudinal system remains stable for all wavenumbers in the elastic case ($e=1$).
\subsubsection{Transverse system (figures~\ref{fig:tran_d2} and \ref{fig:tran_d3})} 


For $e=0.75$ (top rows of figures~\ref{fig:tran_d2} and \ref{fig:tran_d3}), the first pair of sound modes starts propagating in opposite directions at $k \approx 0.0458$ in the case of $d=2$ (at $k \approx 0.07216$ in the case of $d=3$), followed by a second pair of sound modes commencing at $k \approx 0.1977$ in the case of $d=2$ (at $k \approx 0.1882$ in the case of $d=3$) travelling slower than the first pair.
Beyond these wavenumbers, the imaginary parts of each (respective) pair of travelling sound modes merge due to the aforementioned reason,  
which is clearly reflected in the top rows and right columns of figures~\ref{fig:tran_d2} and \ref{fig:tran_d3}: the first pair coincides beyond $k\approx 0.0458$ in the case of $d=2$ (beyond $k\approx 0.07216$ in the case of $d=3$) and the second beyond $k \approx 0.1977$ in the case of $d=2$ (beyond $k \approx 0.1882$ in the case of $d=3$).
The remaining two out of six eigenmodes remain stationary for all wavenumbers, i.e.~these modes have no oscillations since their frequencies are purely imaginary. The non-oscillatory eigenmodes of the transverse system (\ref{eigvalProbs}$b$) are referred to as the shear modes \citep{BP2004, Garzo2005}. 
Clearly, there are two shear modes in the transverse system but one of them is unstable for wavenumbers $k < k_c$ as its growth rate is positive (see the top rows and right columns of figures~\ref{fig:tran_d2} and \ref{fig:tran_d3}), where $k_c$ is the critical wavenumber for the transverse system. 
For $e=0.75$, $k_c \approx 0.341$ in the case of $d=2$ and $k_c \approx 0.296$ in the case of $d=3$.
The general behaviour of the eigenmodes of the transverse system for moderate to large values of $e$ is also similar to that for $e=0.75$ (as shown in the top rows of figures~\ref{fig:tran_d2} and \ref{fig:tran_d3}) with $k_c \to 0$ as $e \to 1$. 
Thus, for moderately to nearly elastic granular gases, the unstable shear mode renders the transverse system unstable for wavenumbers $k<k_c$; nevertheless, the system becomes stable for all wavenumbers $k \geq k_c$.
However, analogously to the longitudinal system, the growth rates of some of the eigenmodes of the transverse system also remain positive for large wavenumbers  for all $e$ below a threshold coefficient of restitution $e_{\mathrm{th}}$, implying that there does not exist a $k_c$ for all $e<e_{\mathrm{th}}$ and that the transverse system remains always unstable for all $e<e_{\mathrm{th}}$, which is also not true \cite[see, e.g.,][]{GST2018}. 
For the transverse system associated with the G29 equations, $e_{\mathrm{th}} \approx 0.32349$ in the case of $d=2$ and $e_{\mathrm{th}} \approx 0.16867$ in the case of $d=3$. 

For $e=1$ (bottom rows of figures~\ref{fig:tran_d2} and \ref{fig:tran_d3}), the first pair of travelling sound modes starts propagating in opposite directions at $k \approx 0.1056$ in the case of $d=2$ (at $k \approx 0.08398$ in the case of $d=3$), followed by a second pair of sound modes commencing at $k \approx 0.2296$ in the case of $d=2$ (at $k \approx 0.22396$ in the case of $d=3$) travelling slower than the first pair.
Accordingly, the imaginary parts of frequencies for the first pair of travelling sound modes merge beyond $k \approx 0.1056$ in the case of $d=2$ (beyond $k \approx 0.08398$ in the case of $d=3$) and that for the second pair merge beyond $k \approx 0.2296$ in the case of $d=2$ (beyond $k \approx 0.22396$ in the case of $d=3$); see the bottom rows and right columns of the figures.
The remaining two eigenmodes are stable shear modes in the elastic case since the real parts of their associated frequencies are zero and imaginary parts (growth rates) are non-positive for all wavenumbers.
Consequently, the transverse system also remains stable for all wavenumbers in the elastic case ($e=1$).
\subsection{Comparison among various Grad moment theories}\label{CompTheories}
\begin{figure}
\begin{center}
\includegraphics[width=0.47\textwidth]{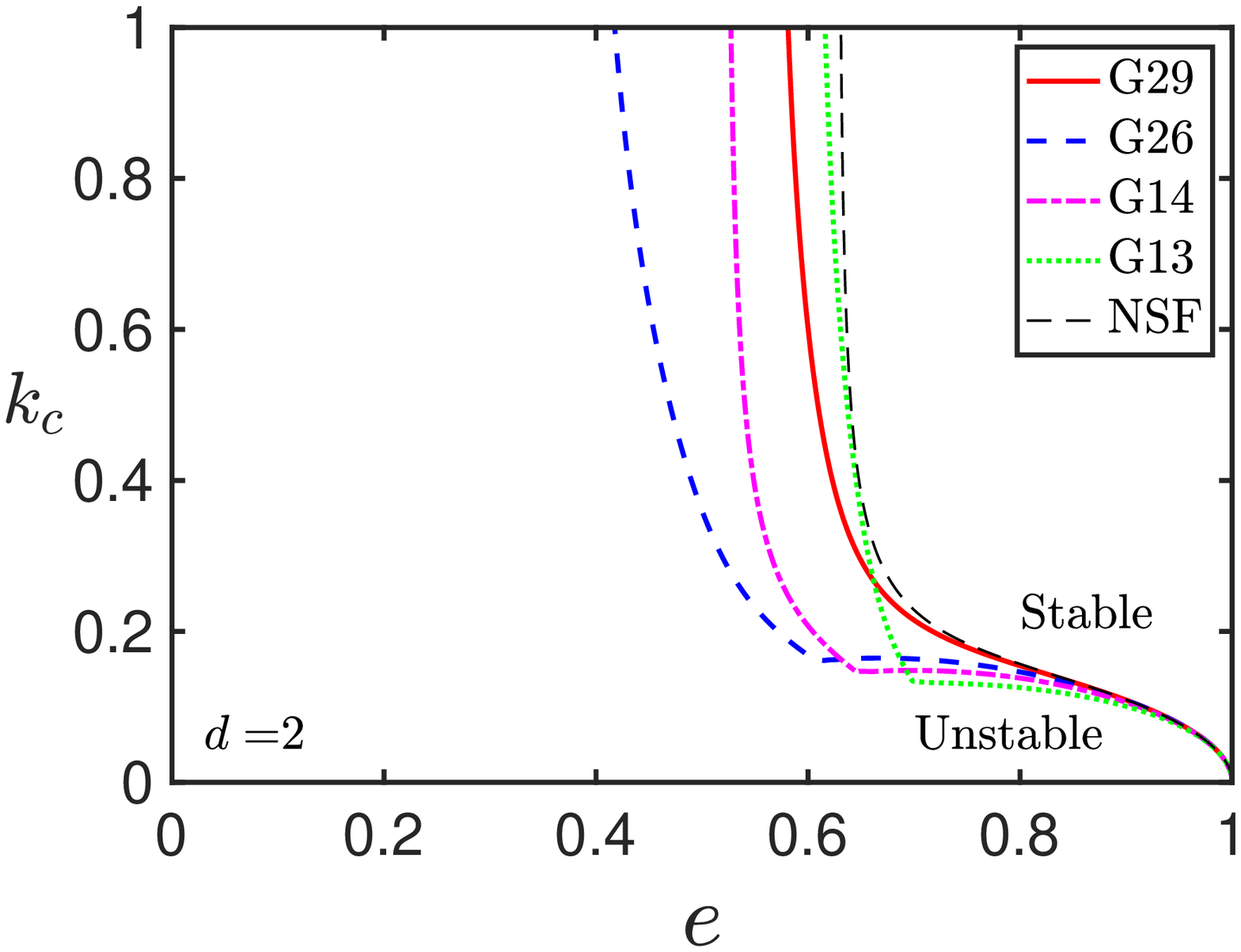}
\hfill
\includegraphics[width=0.47\textwidth]{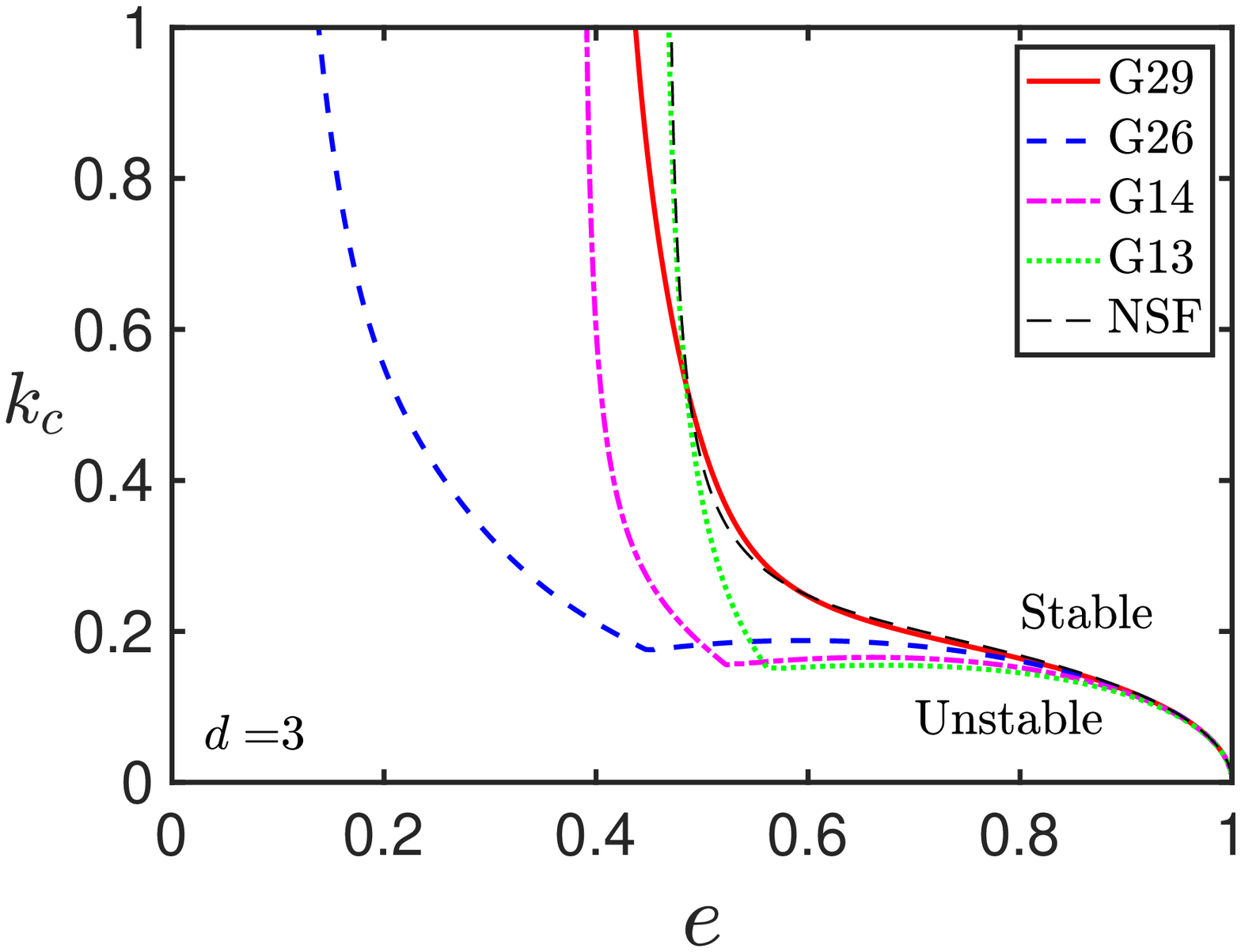}
\\
\includegraphics[width=0.47\textwidth]{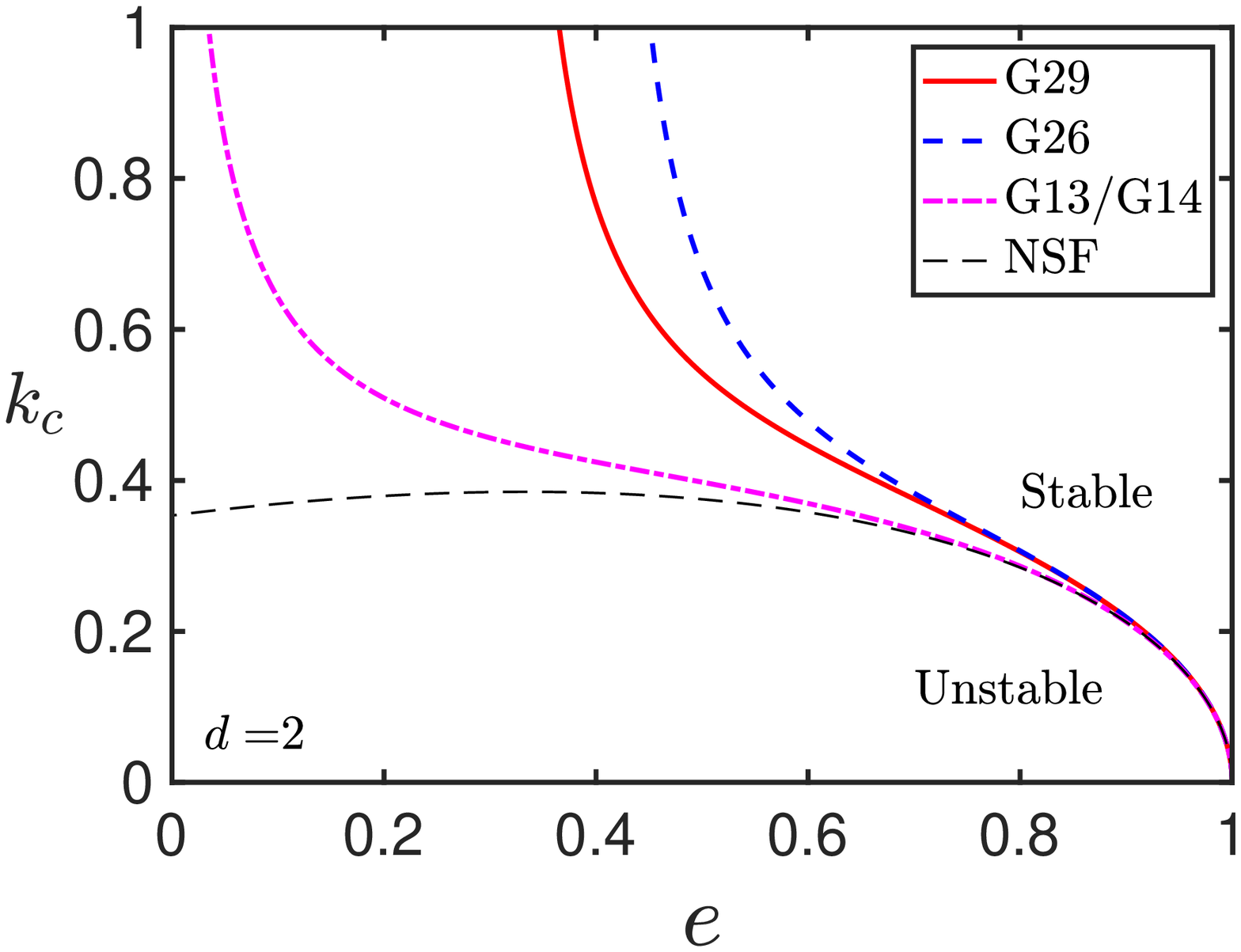}
\hfill
\includegraphics[width=0.47\textwidth]{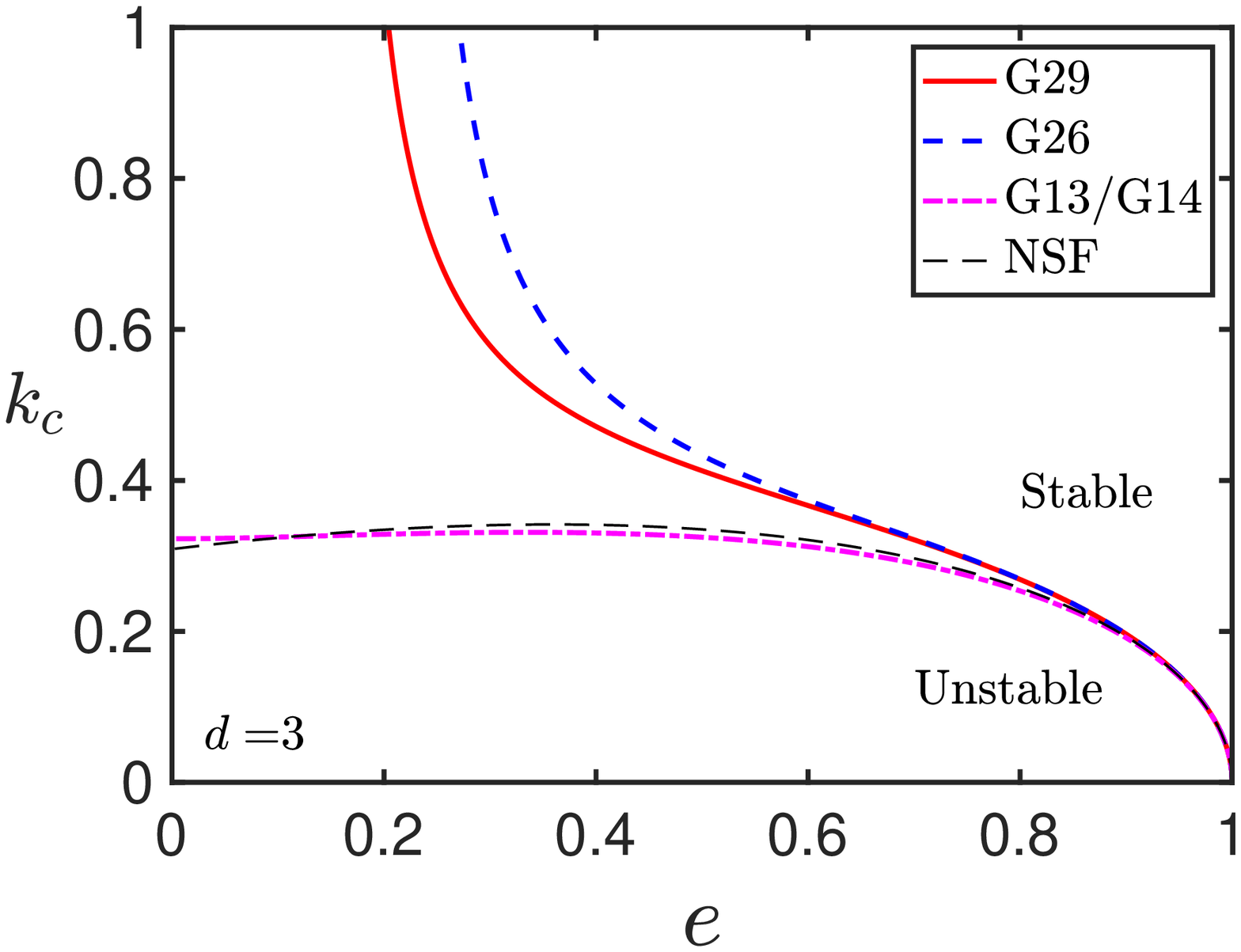}
\caption{Critical wavenumbers in the $(e,k_c)$-plane from the NSF and different Grad moment theories. 
The top and bottom rows exhibit the critical wavenumbers for the longitudinal and transverse systems, respectively, while the left and right columns display the results for $d=2$ and $d=3$, respectively.  
Each system is unstable (stable) below (above) its corresponding curve.
}
\label{fig:zero_contours_all}
\end{center}
\end{figure}

As discussed in \S\,\ref{Subsec:various}, a lower-level Grad moment system can be obtained from the G29 equations by discarding the appropriate field variables. 
Accordingly, the longitudinal and transverse problems associated with the G13, G14 and G26 systems can be obtained by eliminating the appropriate variables and corresponding rows and columns of the matrices $\mathscr{L}$ and $\mathscr{T}$ in \eqref{eigvalProbs}.  
For comparison purpose, I shall also include the results obtained from the linear stability analysis of the system of the NSF equations for IMM along with these Grad moment systems. 
The linear-dimensionless NSF equations in the perturbed field variables are \eqref{massBalPertDimless}--\eqref{energyBalPertDimless} with 
\begin{align}
\tilde{\sigma}_{ij} = - 2 \eta^\ast \frac{\partial \tilde{v}_{\langle i}}{\partial \tilde{x}_{j \rangle}} 
\quad\textrm{and}\quad
\tilde{q}_i = - \frac{d(d+2)}{2(d-1)} \left(\kappa^\ast \frac{\partial \tilde{T}}{\partial \tilde{x}_i}
+ \lambda^\ast \frac{\partial \tilde{n}}{\partial \tilde{x}_i} \right),
\end{align}
where the reduced transport coefficients $\eta^\ast$, $\kappa^\ast$ and $\lambda^\ast$ for IMM are given by \eqref{ReducedTransCoeff}. 
From the linear-dimensionless NSF equations, it is straightforward to obtain the longitudinal and transverse problems associated with the system of the NSF equations by following a similar procedure as above. 
The longitudinal and transverse problems associated with the system of the NSF equations read 
\begin{align}
\label{eigvalProbsNSF}
\mathscr{L}_{\mathrm{NSF}} 
\begin{bmatrix}
\check{n} \\ 
\check{v}_x \\  
\check{T}
\end{bmatrix}
= \begin{bmatrix}
0 \\ 
0 \\  
0
\end{bmatrix}
\quad\textrm{and}\quad
\left(-\omega - \mathbbm{i} \eta^\ast k^2 + \mathbbm{i} \dfrac{\zeta_0^\ast}{2}\right) \check{v}_y = 0,
\end{align}
respectively, where the matrix $\mathscr{L}_{\mathrm{NSF}} \equiv \mathscr{L}_{\mathrm{NSF}}(k,\omega, d, e)$ is also presented in appendix~\ref{app:matrices}. 
As far as the linear stability of the HCS is concerned, the NSF theory and all Grad moment theories---although not shown here explicitly for the NSF, G13, G14 and G26 equations---predict a similar behaviour for moderate to large values of the coefficient of restitution in the sense that one heat mode from the longitudinal system associated with each theory and one shear mode from the transverse system associated with each theory are unstable below some critical wavenumbers for granular gases while all the modes remain stable in the elastic case ($e=1$).
The stability of a (longitudinal or transverse) system is regulated by its least stable eigenmode. 
Therefore, to analyse the stability of the longitudinal and transverse systems associated with different moment theories, the critical wavenumbers for the least stable mode (unstable shear mode for the longitudinal system and unstable heat mode for the transverse system) from each moment theory are plotted in the $(e,k_c)$-plane in figure~\ref{fig:zero_contours_all}. 
The figure also includes the critical wavenumber profiles (shown by thin dashed black lines) for the least stable modes of the longitudinal and transverse systems associated with the NSF theory for IMM. 
The top and bottom rows of figure~\ref{fig:zero_contours_all} display the critical wavenumbers for the longitudinal and transverse systems, respectively, while the left and right columns exhibit the results for $d=2$ and $d=3$, respectively.
Since the critical wavenumber is that wavenumber where the growth rate ($\mathrm{Im}(\omega)$) changes its sign, the curves in figure~\ref{fig:zero_contours_all} are essentially the zero contours---in the $(e,k_c)$-plane---of the growth rate of the least stable mode 
in each system.
Each curve for the critical wavenumber corresponds to $\mathrm{Im}(\omega)=0$ and hence divides the $(e,k_c)$-plane into two parts demarcating the stable and unstable regions: the region below a curve is unstable as $\mathrm{Im}(\omega)>0$ in this region whereas that above this curve is stable as $\mathrm{Im}(\omega)<0$ in this region.
In general, the NSF and all the moment theories considered here predict qualitatively similar critical wavenumber profiles. In particular, the critical wavenumber is zero in the elastic ($e=1$) case since both the longitudinal and transverse systems are stable in this case, and it increases with increasing inelasticity, in general. For nearly elastic granular gases ($0.9 \leq e \leq 1$), the respective critical wavenumber profiles from the NSF and all moment theories coincide for both the longitudinal and transverse systems.

For the longitudinal system (top row in figure~\ref{fig:zero_contours_all}), the NSF and all the moment theories yield smooth critical wavenumber profiles for moderate to large values of the coefficient of restitution discerning the stability regions. 
However, each theory gives a threshold value of the coefficient of restitution below which the longitudinal system remains unstable (because for $e<e_{\mathrm{th}}$, the growth rates of some of the eigenmodes of the longitudinal system remain positive even for large wavenumbers), which is not true \citep[see, e.g.,][]{GST2018}. 
This simply means that the IMM model is not adequate for granular gases with moderate to large inelasticity.
The threshold values of the coefficient of restitution $e_{\mathrm{th}}$ for the longitudinal systems associated with the NSF and different Grad moment theories are given in table~\ref{tab:e_th}.
Furthermore, the G13, G14 and G26 theories lead to kinks at $e \approx 0.7,0.648,0.608$ in the case of $d=2$ (at $e \approx 0.562,0.522,0.446$ in the case of $d=3$), respectively, indicating that the region of applicability increases on increasing the number of moments.

For the transverse system (bottom row in figure~\ref{fig:zero_contours_all}), the critical wavenumbers for the G13 and G14 theories are exactly the same due to the fact that the scalar fourth moment $\check{\Delta}$ does not enter the transverse system associated with any moment theory (see (\ref{eigvalProbs}$b$)). 
Hence the critical wavenumbers from both the theories are depicted by a single dot-dashed magenta line.
Among the theories considered, the transverse systems associated with the NSF and G13 (or G14) theories give the critical wavenumbers for all coefficients of restitution whereas those associated with the G26 and G29 theories again lead to  threshold values of the coefficient of restitution below which the transverse systems associated with them remain unstable for all wavenumbers (due to the same reason as above). 
This again restricts the employability of the IMM model to moderately to nearly elastic granular gases.   
The threshold values of the coefficient of restitution $e_{\mathrm{th}}$ for the transverse systems associated with the G26 and G29 theories are also given in table~\ref{tab:e_th}. 
%
The critical wavenumbers from the G26 (shown by dashed blue line) and G29 (shown by solid red line) theories closely follow each other for $0.7 \lesssim e \leq 1$ in the case of $d=2$ and for $0.6 \lesssim e \leq 1$ in the case of $d=3$. 
Similarly, the critical wavenumbers from the NSF (shown by thin dashed black line) and G13 or G14 (shown by dot-dashed magenta line) theories closely follow each other for $0.7 \lesssim e \leq 1$ in the case of $d=2$ and for all values of $e$ in the case of $d=3$.

\begin{table}
\centering
\begin{tabular}{ccccccc}
\toprule
& & NSF & G13 & G14 & G26 & G29 
\\
\midrule
\multirow{2}{*}{Longitudinal} & $d=2$ & 0.62798 & 0.60211 & 0.52174 & 0.37473 & 0.56356
\\[1ex]
& $d=3$ & 0.46551 & 0.46033 & 0.38608 & 0.06256 & 0.40157
\\[0.5ex]
\midrule
\multirow{2}{*}{Transverse} & $d=2$ & -- & -- & -- & 0.41360 & 0.32349
\\[1ex]
& $d=3$ & -- & -- & -- & 0.23030 & 0.16867
\\
\bottomrule
\end{tabular}
\caption{\label{tab:e_th} Threshold values of the coefficient of restitution $e_{\mathrm{th}}$ below which the longitudinal and transverse systems for IMM remain always unstable.
}
\end{table}

From figures~\ref{fig:long_d2}--\ref{fig:zero_contours_all}, it is concluded that the instabilities of the longitudinal and transverse systems above---for moderately to nearly elastic granular gases---are confined to small wavenumbers (or long wavelengths), i.e.~these instabilities are long-wave instabilities. 
Thus, there is a minimum system size, referred to as the critical system size, such that the instabilities will not appear in a system having size smaller than the critical system size.
%
\subsection{Critical system size}
\begin{figure}
\begin{center}
\includegraphics[width=0.4\textwidth]{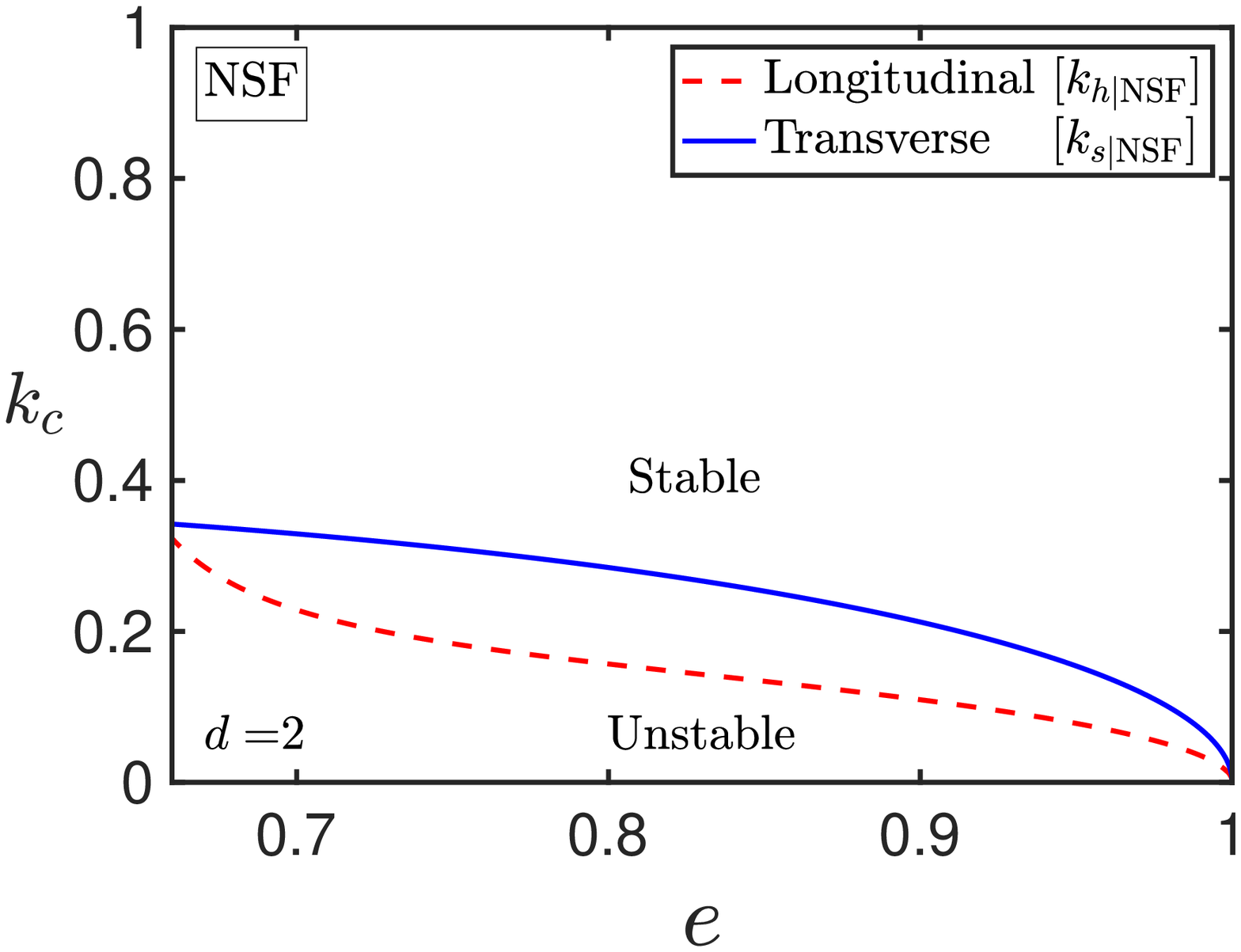}
\hspace*{10mm}
\includegraphics[width=0.4\textwidth]{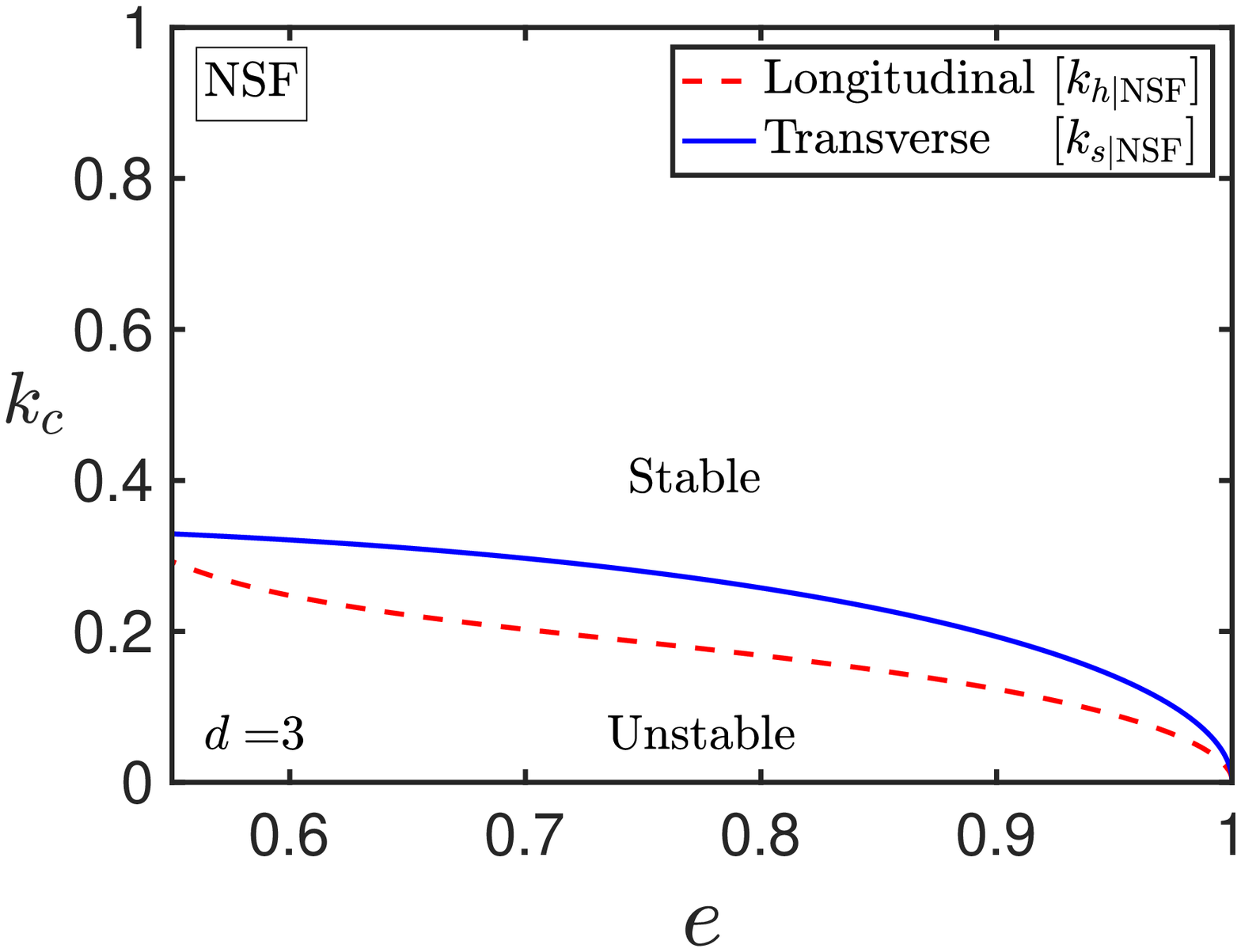}
\\
\includegraphics[width=0.4\textwidth]{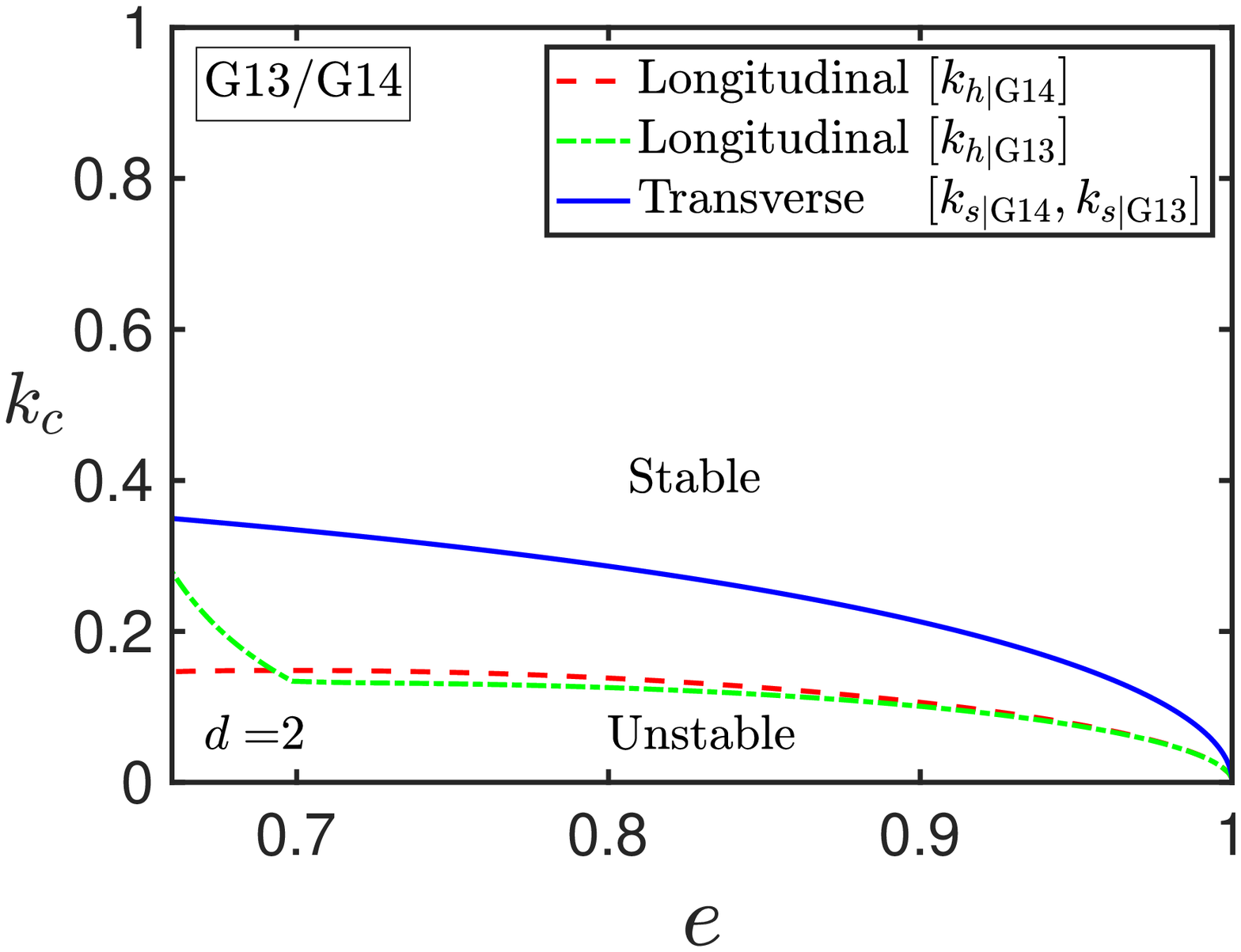}
\hspace*{10mm}
\includegraphics[width=0.4\textwidth]{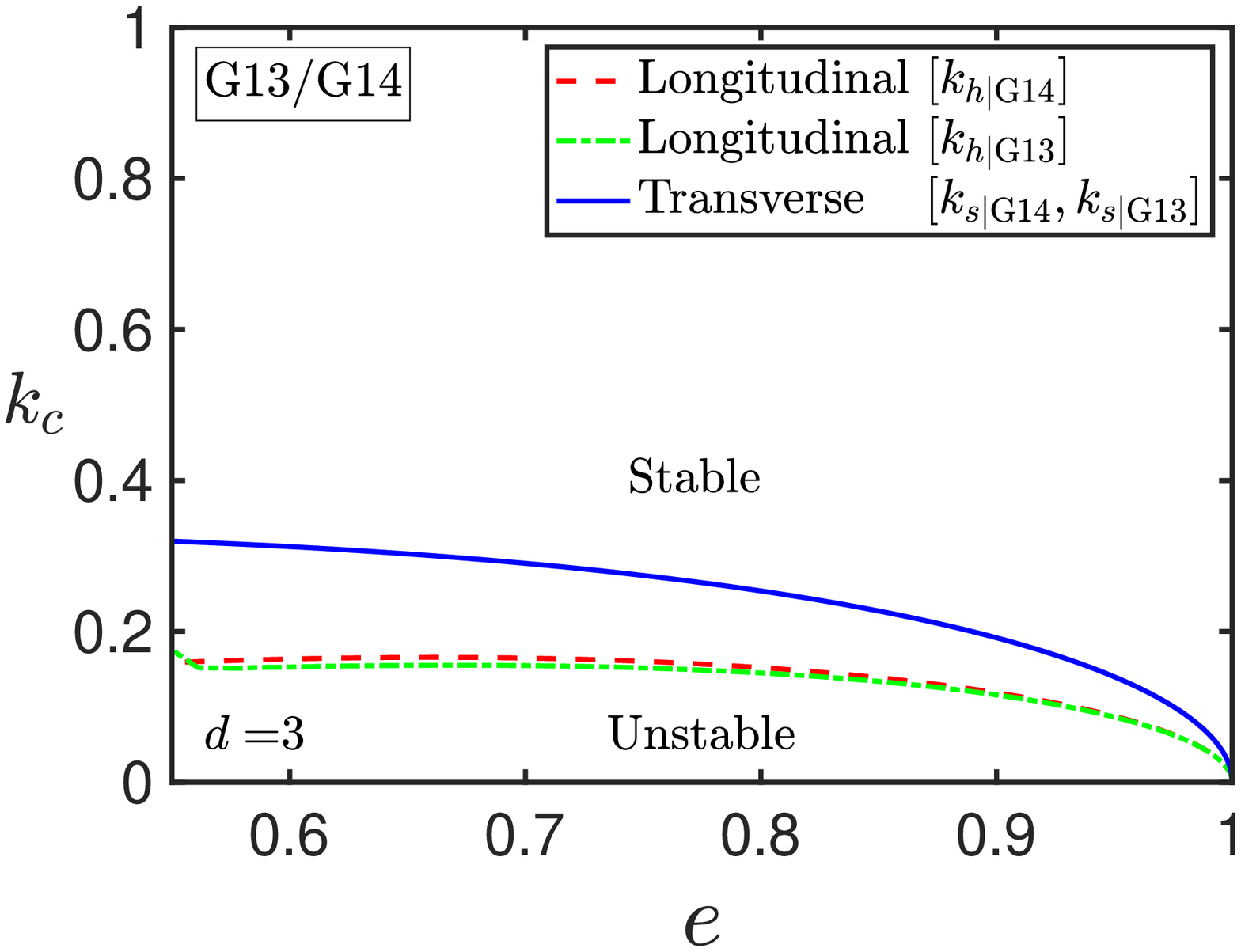}
\\
\includegraphics[width=0.4\textwidth]{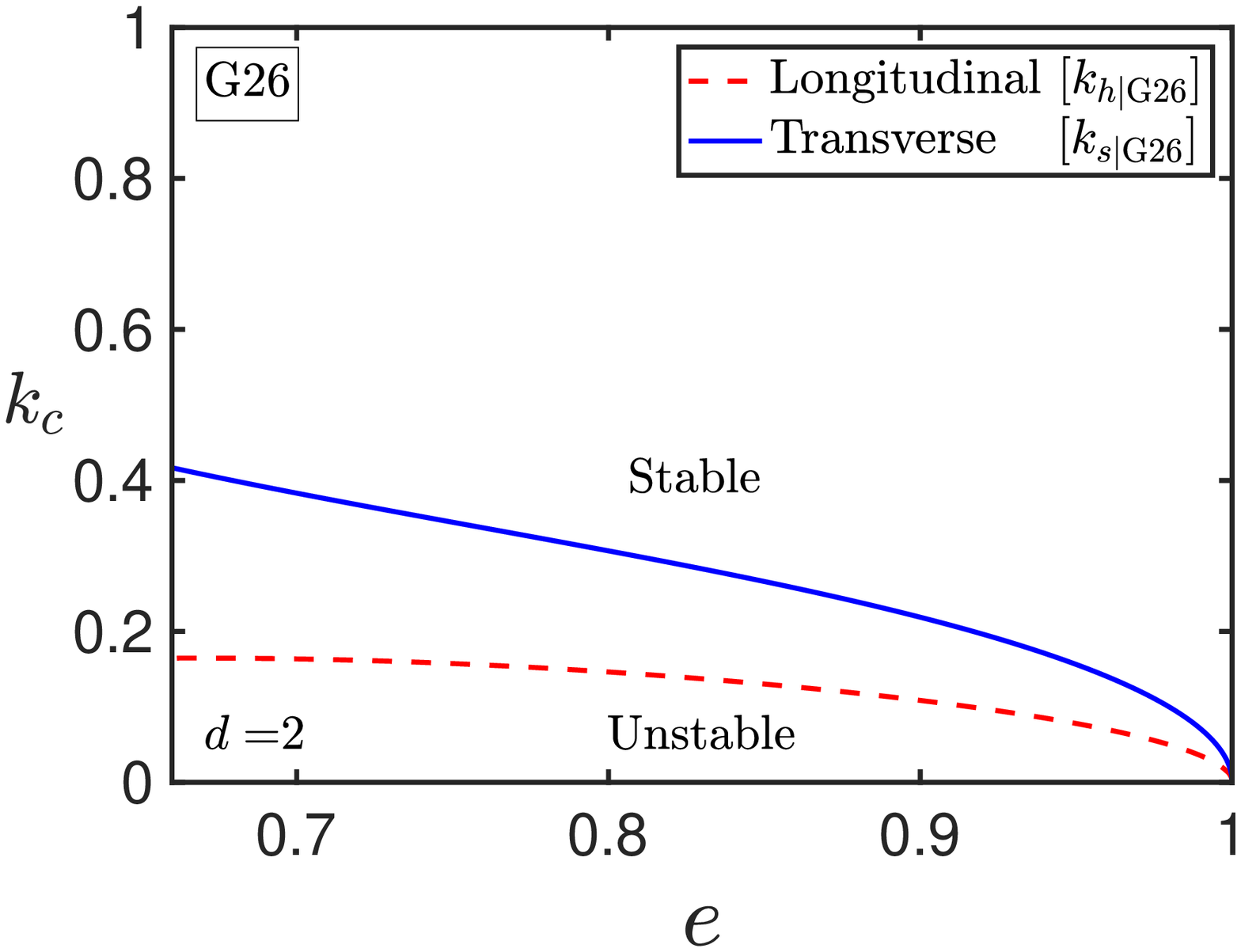}
\hspace*{10mm}
\includegraphics[width=0.4\textwidth]{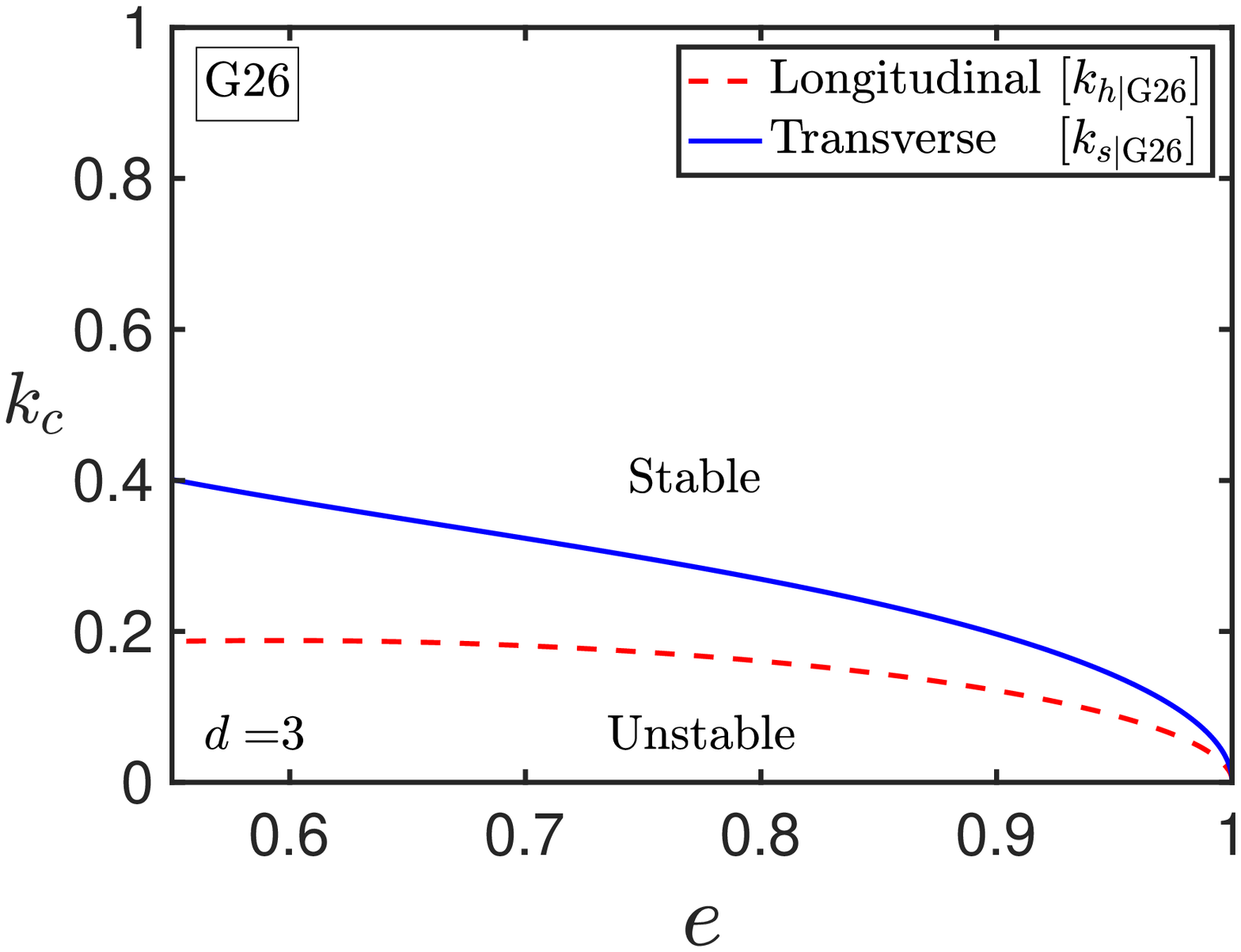}
\\
\includegraphics[width=0.4\textwidth]{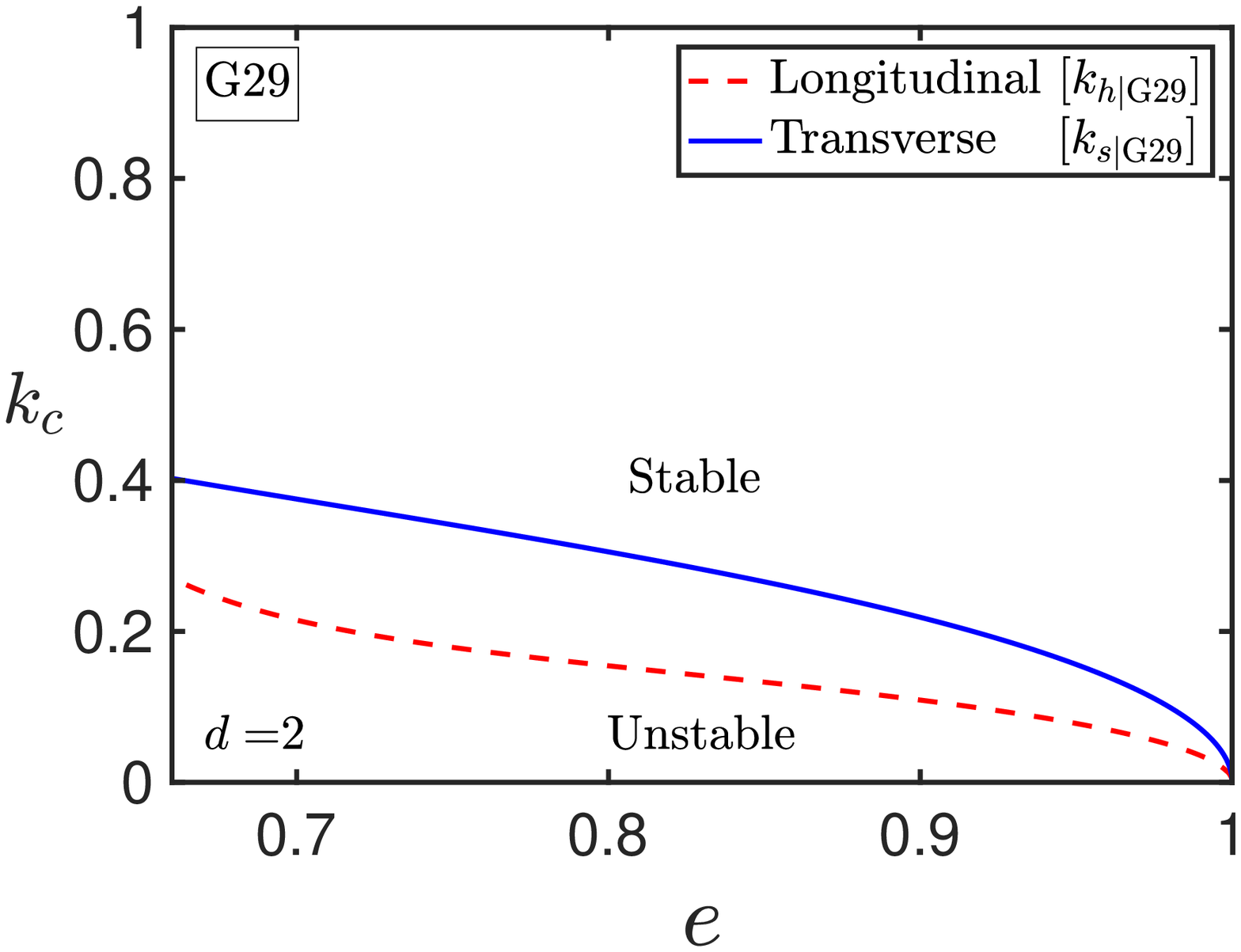}
\hspace*{10mm}
\includegraphics[width=0.4\textwidth]{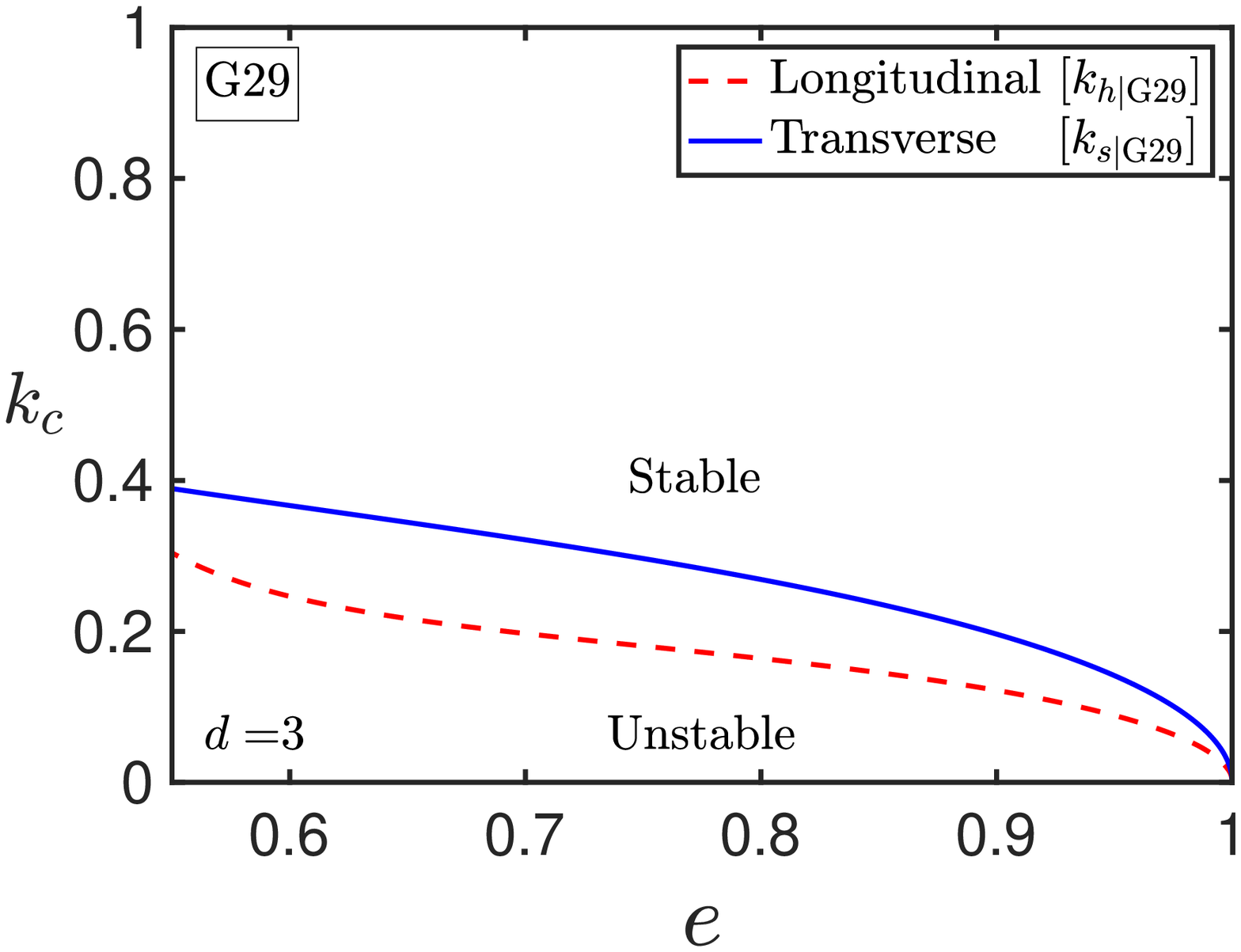}
\caption{Critical wavenumbers for the longitudinal and transverse systems---associated with the NSF and different Grad moment theories (G13, G14, G26, G29)---plotted together in the $(e,k_c)$-plane. 
The left and right columns display the results for $d=2$ and $d=3$, respectively.  
}
\label{fig:longtran}
\end{center}
\end{figure}

It is well-established---through the linear stability analysis of hydrodynamic models, through the DSMC method as well as through molecular dynamics (MD) simulations---that the HCS of a freely cooling granular gas is unstable but a minimum critical system size is necessary for the onset of instabilities \cite[see, e.g.,][]{BRM1998, BP2004, Garzo2005, GST2018}. 
Moreover, it is also known that during the instability phenomenon of the HCS, 
the instability of the unstable shear mode first engenders the formation of vortices in the system through linear effects \citep{BP2004, Garzo2005} and
subsequently clustering set in due to the instability of the shear mode through nonlinear effects \citep{BR-MC1999, Goldhirsch2003}.
In order to verify through the moment theories that it is the unstable shear mode which is responsible for the onset of instabilities in a freely cooling granular gas, the critical wavenumbers for the longitudinal and transverse systems are plotted together in the $(e,k_c)$-plane (but only for that range of $e$, which apparently leads to meaningful critical wavenumber profiles) in figure~\ref{fig:longtran}. 
Denoting the critical wavenumbers associated with the unstable heat and shear modes for any moment system by $k_h$ and $k_s$, respectively, it can be easily deduced that a heat (shear) mode of the longitudinal (transverse) system for wavenumbers $k>k_h$ ($k>k_s$) will always decay while that for wavenumbers $k<k_h$ ($k<k_s$) will grow exponentially. 
For the ranges of the coefficient of restitution shown in figure~\ref{fig:longtran}, $k_s \geq k_h$.
Since the wavelength, and hence the critical system size, is inversely proportional to the wavenumber, it is the unstable shear mode from the transverse system which becomes unstable first. 
From the above discussion, the critical system size can be determined with the knowledge of the critical wavenumber for the transverse system itself.
Indeed, it is possible to obtain the analytical expressions for the critical wavenumbers from the moment theories as follows.

%

It has been verified (although shown here only for the G29 system in the top rows of figures~\ref{fig:long_d2}--\ref{fig:tran_d3} in the case of $e=0.75$ for the sake of succinctness) that the real part of the frequency for the least stable eigenmode is either always zero (for longitudinal systems associated with the NSF, G14 and G26 theories and for transverse systems associated with all the theories) or is nonzero only for wavenumbers above the critical wavenumber (for longitudinal systems associated with the G13 and G29 theories). 
Therefore, 
for the least stable mode, $\omega = 0$ at critical wavenumber. 
Consequently, the critical wavenumbers for the longitudinal and transverse systems associated with a moment theory can also be determined by substituting $\omega = 0$ in their dispersion relations and solving the resulting equations for $k$. 
For instance, the critical wavenumbers for the longitudinal and transverse systems associated with the G29 theory can also be determined by inserting $\omega = 0$ in \eqref{dispRels} and solving the resulting equations, namely $\mathbbm{a}_9 = 0$ and $\mathbbm{b}_6 = 0$. 
Hence the roots of $\mathbbm{a}_9 = 0$ and $\mathbbm{b}_6 = 0$, respectively, yield the critical wavenumbers for the longitudinal and transverse systems associated with the G29 equations. 
%
It is worthwhile to note that the coefficients $\mathbbm{a}_9$ and $\mathbbm{b}_6$ are (even-degree) polynomials of degree eight and four in $k$, respectively. Similarly, the coefficients of $\omega^0$ in the dispersion relations for the longitudinal (transverse) systems associated with the NSF, G13, G14 and G26 are (even-degree) polynomials of degree four, four, four and six (two, two, two and four) in $k$, respectively. Consequently, each of them leads to more than one value of the critical wavenumber. 
Nevertheless, only one of them in each case is meaningful (positive) and that is the analytical expression for the corresponding critical wavenumber.
After some algebra, the explicit expressions for $k_h$ and $k_s$ for each of the NSF and Grad moment systems, in a compact form, can be written as
\begingroup
\allowdisplaybreaks
\begin{align}
\label{criticalkh_NSF} 
k_{h|\mathrm{NSF}} & = 
\sqrt{\frac{d-1}{2(d+2)}}\sqrt{\frac{\zeta_0^\ast }{\kappa^\ast-\lambda^\ast}},
\\
\label{criticalks_NSF} 
k_{s|\mathrm{NSF}} & =\sqrt{\frac{\zeta_0^\ast }{2\eta^\ast}},
\\
\label{criticalkh_G13} 
k_{h|\mathrm{G13}} &= 
\sqrt{\frac{d (d+2)}{2} } 
\sqrt{\frac{\zeta_0^\ast\, \xi_\sigma \, \xi_q}{\zeta_0^\ast \xi_3 + (d+2)^2 (1+a_2) \xi_\sigma}},
\\
\label{criticalkh_G14} 
k_{h|\mathrm{G14}} &=\sqrt{\frac{d (d+2)}{2} } 
\sqrt{\frac{\zeta_0^\ast \, \xi_\sigma \, \xi_q \, \nu_\Delta^\ast}{\xi_4}},
\\
\label{criticalks_G14} 
k_{s|\mathrm{G13}} &= k_{s|\mathrm{G14}} = \sqrt{\frac{d+2}{2} } 
\sqrt{\frac{\zeta_0^\ast\, \xi_\sigma \, \xi_q}{(d+2) \xi_q - \zeta_0^\ast}},
\\
\label{criticalkh_G26} 
k_{h|\mathrm{G26}} &=\sqrt{\frac{\sqrt{\vartheta_{12}^2 + \vartheta_{11} \vartheta_{13}}-\vartheta_{12}}{\vartheta_{11}}},
\\
\label{criticalks_G26} 
k_{s|\mathrm{G26}} &= \sqrt{\frac{\sqrt{\vartheta_{22}^2 + \vartheta_{21} \vartheta_{23}}-\vartheta_{22}}{\vartheta_{21}}},
\\
\label{criticalkh_G29} 
k_{h|\mathrm{G29}} &=\sqrt{\frac{1}{\vartheta_{31}}\left(\frac{\vartheta_{33}}{\vartheta_{34}} + \vartheta_{34}-\vartheta_{32}\right)},
\\
\label{criticalks_G29} 
k_{s|\mathrm{G29}} &= \sqrt{\frac{\sqrt{\vartheta_{42}^2 + \vartheta_{41} \vartheta_{43}}-\vartheta_{42}}{\vartheta_{41}}},
\end{align}
\endgroup
where the coefficients appearing in these expressions are relegated to appendix~\ref{app:coeffwavenumber} for better readability; and ``$|\mathrm{NSF}$'' and ``$|\mathrm{G\dots}$'' in subscripts denote the NSF and Grad moment systems which the critical wavenumbers belong to. The plots of $k_h$ and $k_s$ from the analytical expressions agree completely with those in figure~\ref{fig:longtran} (at least for the depicted ranges of the coefficient of restitution; for instance, $k_{h|\mathrm{G29}}$ becomes complex for smaller values of $e$).

The critical system size $L_c$ is estimated by $\tilde{L}_c = 2\pi / \max{\{k_h,k_s\}}$ \citep{Garzo2005, GST2018}, 
where $\tilde{L}_c := L_c / \ell$ is the dimensionless critical system size and $\ell$ is the length scale defined in \eqref{ell}. 
From figure~\ref{fig:longtran}, $k_s>k_h$ for the NSF and all the moment theories considered in this work (for the shown ranges of $e$ in the figure); therefore the critical system size $L_c$ is given by
\begin{align}
\label{criticalLength}
L_c = \frac{2\pi}{k_s} \times \ell = \frac{2\pi}{k_s} \times \frac{d+2}{4\sqrt{2}} \frac{\Gamma\left(\frac{d}{2}\right)}{\Gamma\left(\frac{d+1}{2}\right)} \ell_0,
\end{align}
where $\ell_0 = \Gamma\left(\frac{d+1}{2}\right) / \left(\sqrt{2} \, \pi^{(d-1)/2} n_0 \mathbbm{d}^{d-1}\right)$ is the mean free path of a dilute hard-sphere gas.

\begin{figure}
\centering
\includegraphics[width = 0.47\textwidth]{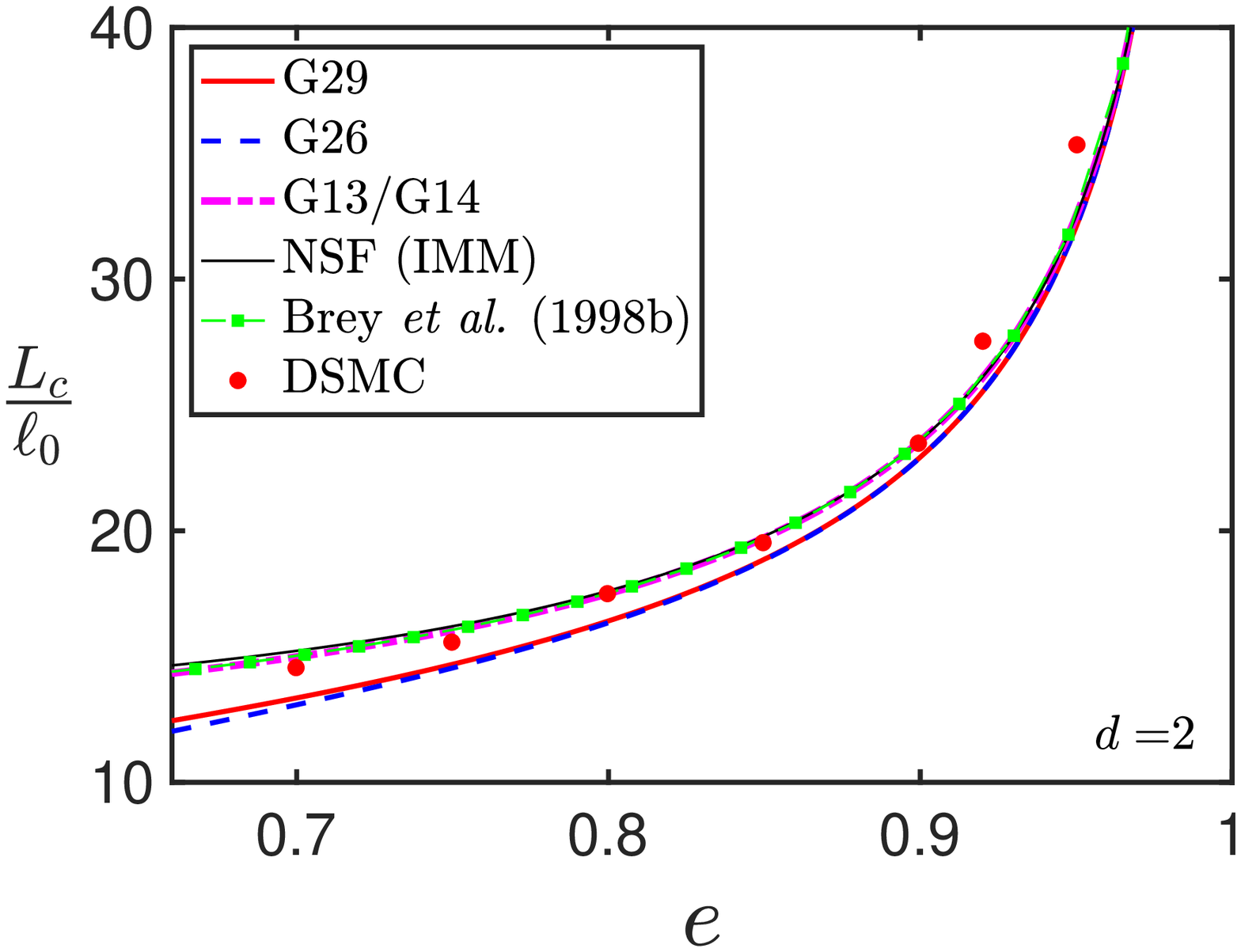}
\hfill
\includegraphics[width = 0.47\textwidth]{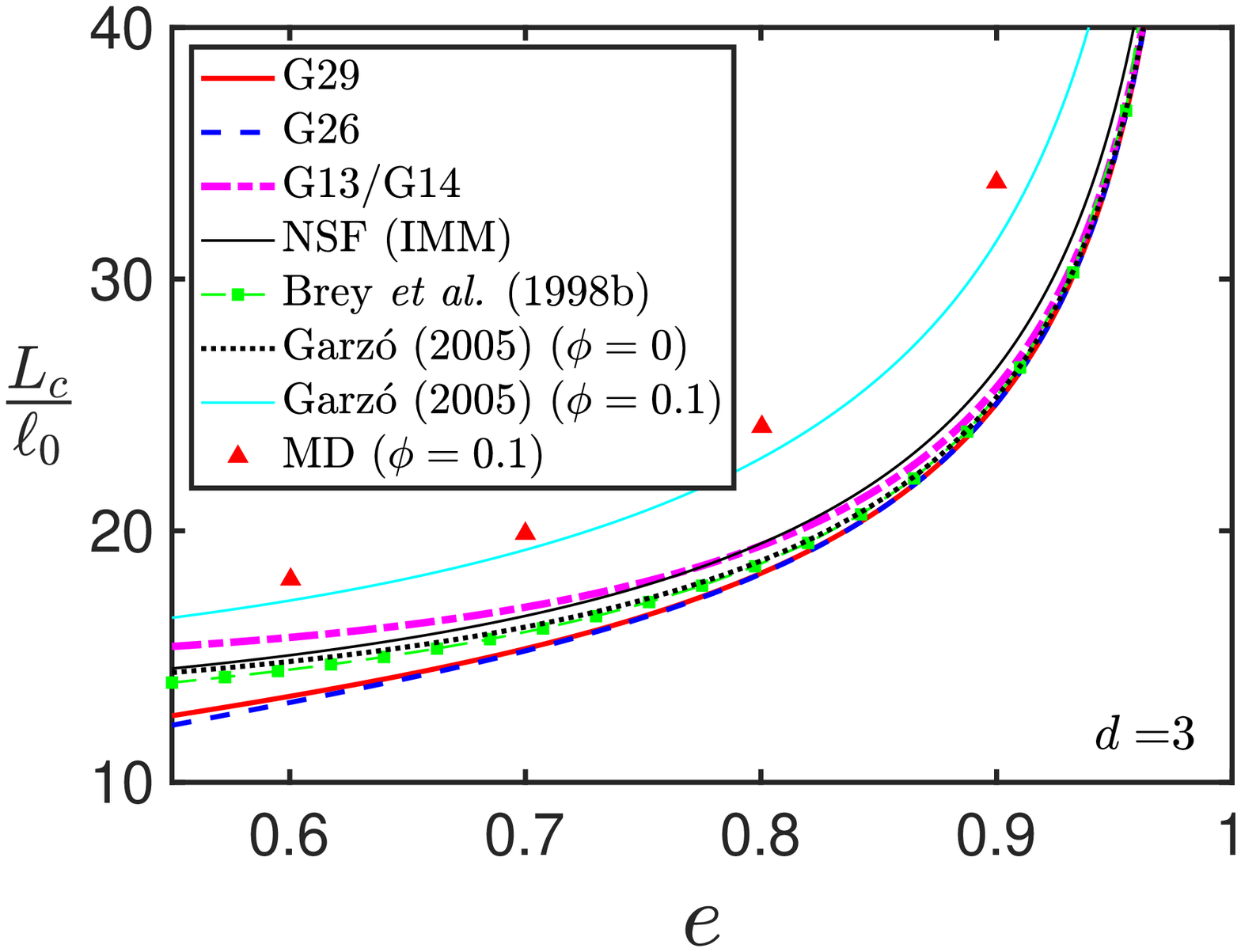}
\caption{Critical system size in units of the mean free path $\ell_0$ plotted over the coefficient of restitution $e$ for (left) $d=2$ and (right) $d=3$. 
The three-dimensional MD simulation results of \cite{MDCPH2011} (depicted by triangles) in the right panel are at solid fraction $\phi=0.1$ while the dilute limit refers to $\phi \to 0$.
The cyan line on the right panel represents the critical system size at solid fraction $\phi=0.1$ computed from the theoretical expression of \cite{Garzo2005}, and is included only to show the good agreement between theoretical results of \cite{Garzo2005} and MD simulations results of \cite{MDCPH2011}.
}
\label{fig:critical_length}
\end{figure}

The critical system size in units of the mean free path ($L_c/\ell_0$) is illustrated in figure~\ref{fig:critical_length} as a function of the coefficient of restitution $e$
for $d=2$ (left panel) and $d=3$ (right panel). 
The dot-dashed (magenta), dashed (blue) and solid (red) lines in the figure depict the critical system size from the G13 (or G14), G26 and G29 theories, respectively, computed with formula \eqref{criticalLength} using the analytical expressions for $k_s$ given in \eqref{criticalks_G14}, \eqref{criticalks_G26} and \eqref{criticalks_G29}, respectively. 
For comparison purpose, the figure also includes thin solid black lines depicting the critical system size from the NSF theory for IMM computed with formula \eqref{criticalLength} using the analytical expressions for $k_s$ given in \eqref{criticalks_NSF}. 
The dashed (green) lines with symbols delineate the critical system size computed from the theoretical expression given in \cite{BRM1998}, which was obtained via the linear stability analysis of a kinetic model for granular gases of IHS due to \cite{BDS1997}. 
It should be noted that the $l_0$ used in \cite{BRM1998} relates to the mean free path $\ell_0$ used in the present work via 
\begin{align*}
l_0 = \frac{2\sqrt{2}}{C} \frac{\pi^{\frac{d}{2}-1}}{\Gamma\left(\frac{d+1}{2}\right)} \ell_0
\quad\textrm{with}\quad
C \simeq \begin{cases}
         2 &\quad\textrm{for}\quad d=2,
         \\
         \frac{16}{5} &\quad\textrm{for}\quad d=3.
         \end{cases}
\end{align*}
The dotted (black) line on the right panel of figure~\ref{fig:critical_length} denotes the critical system size determined from the theoretical expression obtained by the linear stability analysis of the granular NSF equations for IHS in \cite{Garzo2005}. 
The (red) circles on the left panel of figure~\ref{fig:critical_length} are the results from two-dimensional DSMC simulations carried out by \citet{BRM1998}. 
The triangles on the right panel of figure~\ref{fig:critical_length} delineate the results from MD simulations of IHS carried out by \cite{MDCPH2011} at solid fraction $\phi=0.1$ and are included only for qualitative comparison, since the present work deals with dilute granular gases for which $\phi \to 0$. 
Note that the critical system size in \cite{MDCPH2011} is scaled with the diameter of a particle while that in the present work is scaled with the mean free path. Therefore the MD simulations results of \cite{MDCPH2011} have been multiplied by a factor $6\sqrt{2} \,\phi\, \chi(\phi)$ while displaying them in figure~\ref{fig:critical_length}. 
Here $\chi(\phi) = (2-\phi)/[2(1-\phi)^3]$ is the pair correlation function. 
The right panel of figure~\ref{fig:critical_length} also illustrates a solid cyan line, which depicts the critical system size for $\phi=0.1$ computed with the theoretical expressions derived in \cite{Garzo2005}, and is included just to show the good agreement between the theoretical results of \cite{Garzo2005} and MD simulations results of \cite{MDCPH2011}. 

Figure~\ref{fig:critical_length} reveals that the critical system size from all the theories and simulations decreases with increasing inelasticity. 
For disk flows ($d=2$, left panel of figure~\ref{fig:critical_length}), the critical system size from the NSF theory for IMM (thin solid black line) agrees well with that from the theoretical expression in \cite{BRM1998}. However, 
the critical system size from the G13 or G14 theory (dashed magenta line) seems to be slightly better than that from the NSF theory and agrees perfectly with that from the theoretical expression in \cite{BRM1998} (dashed green line with symbols), and also agrees reasonably well with the DSMC results of \cite{BRM1998} (red circles)---for $0.65\leq e \leq 1$. 
Nevertheless, the G26 and G29 theories somewhat underpredict the critical system size for all coefficients of restitution $e \gtrsim 0.65$, although their predictions are also close to the other theories for $e \gtrsim 0.9$.

For sphere flows ($d=3$, right panel of figure~\ref{fig:critical_length}), I could not find any simulation data for the dilute limit, i.e.~for solid fraction $\phi \to 0$. Therefore the data from MD simulations carried out by \citet{MDCPH2011} for solid fraction $\phi=0.1$ are included for comparison. 
It is important to note that the results from MD simulations by \citet{MDCPH2011} are in good agreement with those from theoretical expression of \cite{Garzo2005} not only for $\phi = 0.1$ but also for $\phi = 0.4$ \citep[see][figure 9]{MDCPH2011}. Therefore, in the dilute limit ($\phi \to 0$), the results from the theoretical expression of \cite{Garzo2005}, shown by the dotted black line in figure~\ref{fig:critical_length}, can be treated as a benchmark. 
Additionally, in this limit, the results from the theoretical expressions of \cite{Garzo2005} and \cite{BRM1998} are also in good agreement with each other. 
Clearly, the critical system size from the NSF theory for IMM is again in reasonably good agreement with that from \cite{Garzo2005}. 
Moreover, the critical system sizes from all the moment theories are also in good agreement with that from \cite{Garzo2005} for $e \gtrsim 0.85$, but deviate slightly from the results of \cite{Garzo2005} for $0.55 \lesssim e \lesssim 0.85$, where the G26 and G29 theories again underpredict the critical system size while the G13 or G14 theory overpredicts it. 
Between the G26 and G29 theories, the latter seems to perform slightly better at moderate values of the coefficient of restitution for both $d=2$ and $d=3$.

From figure~\ref{fig:critical_length}, it is apparent that the critical system size obtained from the NSF and Grad moment theories for IMM is in qualitatively good agreement with that obtained from the NSF-level theories and simulations for IHD/IHS, although it is also noticeable from the figure that some lower-order Grad moment theories (e.g.~the G13 or G14 theory) perform better than some higher-order Grad moment theories (e.g.~the G26 theory). 
A possible reason for this could be the choice of the effective collision frequency $\mathring{\nu}$ in the IMM model (see (\ref{effeciveFreq}$a$)) that was chosen in such a way that the cooling rates for IHS and IMM remain exactly the same while the collisional production terms in  the other moment equations for IMM  follow accordingly based on this choice of the effective collision frequency.
Consequently, with this choice of the effective collision frequency, even the NSF equations for IMM seem to perform better than some higher-order moment models.
Therefore it would be interesting to explore other possible choices for the effective collision frequency in the future in such a way that the results from the Boltzmann equations for IHS and IMM agree in an optimal way so that a higher-order moment model would perform better than a lower-order moment model in the case of IMM, similarly to moment models for IHS.
\section{Conclusion}
\label{Sec:Conclusion}
Grad moment equations---consisting of up to 29 moments---for a $d$-dimensional dilute granular gas composed of IMM have been derived from the Boltzmann equation for IMM via the Grad moment method.
A strategy for computing the collisional production terms associated with these moment equations in an automated way has been presented. 
Although the Maxwell interaction potential had been devised in such a way that the explicit form of the distribution function is not required to be known for determining the collisional production terms, and therefore the collisional production terms for IMM can, in principle, be evaluated using pen and paper, yet the complexity increases with an increase in the number of moments. Thus the presented strategy for computing the collisional production terms associated with the moment equations would really be useful when considering even more moments.  

The transport coefficients in the NSF laws for dilute granular gases of IMM have been determined by following a procedure due to \cite{Garzo2013} and it has been shown that the G13 equations for IMM are sufficient to derive the first-order (i.e.~the NSF-level) transport coefficients and that the higher moments do not play any role in determining the NSF transport coefficients for IMM since the stress and heat flux balance equations at this order do not have any dependence on the higher moments.
The higher-order moment equations will be required only for computing the transport coefficients beyond the NSF level.  
However, since the present work already provides some higher-order Grad moment equations, it would be interesting  to compute the transport coefficients beyond the NSF level  in the future by relating the Grad moment equations to the Burnett equations for IMM.
Although the Grad moment theories for IMM presented in this work seem to overestimate all the transport coefficients, the NSF-level transport coefficients for IMM obtained with the Grad moment theories and with the CE expansion \citep{Santos2003} are in complete agreement. 

The HCS of a freely cooling granular gas has then been investigated and it has been found that the decay of the granular temperature in the HCS obeys Haff's law but, in contrast to the case of IHS, does not depend on the higher moments since the fourth cumulant does not enter the energy balance equation for IMM. 
Yet, Haff's laws for IMM and IHS have been found to be in good agreement with each other.
Furthermore, the other higher moments have been found to relax much faster than the granular temperature.%

As an application of the derived moment models, a linear stability analysis has been performed to scrutinise the stability of the HCS due to small perturbations.
By decomposing each moment system into the longitudinal and transverse systems, it has been shown that a heat mode from the longitudinal system and a shear mode from the transverse system associated with each moment system are unstable for (moderately to nearly elastic) granular gases and that the unstable shear mode from the transverse system initiates instability in a homogeneously cooling granular gas. 
To assess the linear stability results, the critical system size for the onset of instability is investigated, and it has been found that the Grad moment theories for IMM yield a reasonably good estimate of the critical system size for granular gases with moderate to large coefficients of restitution. 

It is important to note that the only assumption on the coefficient of restitution in this work is that it is a constant. 
Therefore the present work should, in principle, be applicable to granular gases with any degree of inelasticity, which is not the case unfortunately. 
Nevertheless, this should not be thought of as a problem with moment models presented here, rather it is a problem with the IMM model itself; for instance, the IMM model yields negative values for the coefficient of thermal conductivity below a certain value of the coefficient of restitution \citep{Santos2003,GarzoSantos2011}.

It is anticipated that the Grad moment systems for dilute granular gases of IMM, similarly to those for monatomic gases, will also suffer from the loss of hyperbolicity. 
Therefore the hyperbolicity of these systems needs to be investigated in the future, which will also be useful in developing suitable numerical methods for solving them. 
Moreover, to overcome the undesirable consequences of the loss of hyperbolicity, a regularisation of Grad moment equations might also be necessary. 
The usefulness of the derived Grad moment systems is substantially limited by the unavailability of boundary conditions. 
Hence the development of boundary conditions complementing these Grad moment systems should be an immediate follow-up to the present work. 
Notwithstanding, the Grad moment systems presented in this work will be useful in describing granular processes involving large spatial gradients and are expected to pave the way to further developments including the developments of regularised moment models, required boundary conditions, efficient numerical frameworks including the MFS.
\section*{Acknowledgments}
The author appreciates the anonymous referees for their informative suggestions, which have helped to improve the paper.
The author gratefully acknowledges the financial supports through the ``MATRICS" project MTR/2017/000693 funded by the SERB, India and through the Commonwealth Rutherford Fellowship availed at the University of Warwick, UK. 
The author is thankful to Prof.~Vicente Garz{\'o} and Prof.~Andr{\'e}s Santos for some helpful suggestions on this work during the RGD31 symposium in Glasgow, to Dr.~James Sprittles, Prof.~Duncan Lockerby, Dr.~Anirudh Singh Rana and Dr.~Priyanka Shukla for some fruitful discussions, and to Prof.~Manuel Torrilhon for implementing in Mathematica the Einstein summation, which has been used in this work. 
\section*{Declaration of interests} The author reports no conflict of interest.
\appendix
\section{Symmetric and traceless part of a tensor}
\label{app:traceless}
Needless to say, the symmetric and traceless part of a tensor of rank zero or one is that tensor itself. 
The symmetric and traceless part of a rank two tensor $A_{ij}$ in $d$ dimensions is given by
\begin{align}
\label{tracefree2tensor}
A_{\langle ij \rangle} = A_{(ij)} - \frac{1}{d}A_{kk} \delta_{ij},
\end{align}
where the round brackets around the indices in \eqref{tracefree2tensor} and in what follows always denote the symmetric part of the corresponding tensor. 
Here, $A_{(ij)}=(A_{ij}+A_{ji})/2$ is the symmetric part of $A_{ij}$.
\newline
The symmetric and traceless part of a rank three tensor $A_{ijk}$ in $d$ dimensions is given by
\begin{align}
\label{tracefree3tensor}
A_{\langle ijk \rangle} = A_{(ijk)} - \frac{1}{d+2} \Big[ A_{(ill)} \delta_{jk}+ A_{(ljl)} \delta_{ik}+ A_{(llk)} \delta_{ij}\Big],
\end{align}
where $A_{(ijk)}=(A_{ijk}+A_{ikj}+A_{jik}+A_{jki}+A_{kji}+A_{kij})/6$ is the symmetric part of $A_{ijk}$.
\newline
The symmetric and traceless part of a rank four tensor $A_{ijkl}$ in $d$ dimensions is given by
\begin{align}
A_{\langle ijkl \rangle} &= A_{(ijkl)} - \frac{1}{d+4} \Big[ A_{(ijss)} \delta_{kl} + A_{(isks)} \delta_{jl} + A_{(issl)} \delta_{jk} + A_{(sjks)} \delta_{il} + A_{(sjsl)} \delta_{ik} 
\nonumber\\
&\quad + A_{(sskl)} \delta_{ij}\Big]
+\frac{1}{(d+2)(d+4)} A_{(rrss)} \Big[\delta_{ij} \delta_{kl} +\delta_{ik} \delta_{jl} + \delta_{il} \delta_{jk}\Big],
\end{align}
where $A_{(ijkl)}=(A_{ijkl}+A_{ijlk}+A_{ikjl}+A_{iklj}+A_{iljk}+A_{ilkj} + A_{jikl} + A_{jilk} + A_{jkil} + A_{jkli} + A_{jlik} + A_{jlki} + A_{kijl} + A_{kilj} + A_{kjil} + A_{kjli} + A_{klij} + A_{klji} + A_{lijk} + A_{likj} + A_{ljik} + A_{ljki} + A_{lkij} + A_{lkji})/24$ is the symmetric part of $A_{ijkl}$.
\newline
The symmetric and traceless part of a 
rank $n$ tensor $A_{i_1 i_2 \dots i_n}$ in $d$ dimensions is given by
\begin{align}
A_{\langle i_1 i_2 \dots i_n \rangle} &= A_{(i_1 i_2 \dots i_n)} + \beta_{n,1} \Big[ A_{(rr i_3 i_4 \dots i_n)} \delta_{i_1 i_2} + \textrm{all permutations} \Big]
\nonumber\\
&\quad + \beta_{n,2} \Big[A_{(rr ss i_5 i_6 \dots i_n)} \delta_{i_1 i_2}\delta_{i_3 i_4} + \textrm{all permutations} \Big] + \dots,
\end{align}
where the coefficients $\beta_{n,k}$ are given by
\begin{align}
\beta_{n,k} = \frac{(-1)^k}{\prod\limits_{j=0}^{k-1}\big(d+2n-2j-4\big)}
\end{align}
and the symmetric part of $A_{i_1 i_2 \dots i_n}$ is given by 
\begin{align}
A_{(i_1 i_2 \dots i_n)} = \frac{A_{i_1 i_2 i_3 i_4\dots i_n} + A_{i_2 i_1 i_3 i_4\dots i_n} + \textrm{all permutations}}{n!}.
\end{align}

%
\section{Computation of the collisional production terms}\label{app:CompProd}
For an arbitrary function $\psi(t,\bm{x},\bm{c})$, the collisional production term (or collisional moment) associated with it---on using the symmetry properties of the Boltzmann collision operator---reads \citep{GarzoSantos2007}
\begin{align}
\label{Jpsi} 
\int\psi(\bm{c}) \, J[\bm{c}|f,f] \, \mathrm{d}\bm{c}
 = \frac{\mathring{\nu}}{n\,\Omega_d} \iiint \Big[ \psi(t,\bm{x},\bm{c}^{\prime}) - \psi(t,\bm{x},\bm{c}) \Big]
f(\bm{c}) f(\bm{c}_1)\,
\mathrm{d}\hat{\bm{k}} \, \mathrm{d}\bm{c} \, \mathrm{d}\bm{c}_1,
\end{align} 
where the velocity with single prime denotes the post-collisional velocity in a direct collision that transforms the pre-collision velocities $\bm{c}$ and $\bm{c}_1$ of the colliding molecules to the post-collisional velocities $\bm{c}^{\prime}$ and $\bm{c}_1^{\prime}$ via the relations \citep{BP2004}
\begin{subequations}
\label{VelTran}
\begin{align}
\bm{c}^{\prime}=\bm{c}-\frac{1+e}{2}(\hat{\bm{k}}\cdot\bm{g})\hat{\bm{k}}
\qquad \textrm{and} \qquad
\bm{c}_1^{\prime}&=\bm{c}_1+\frac{1+e}{2}(\hat{\bm{k}}\cdot\bm{g})\hat{\bm{k}}.
\tag{\theequation $a$,$b$}
\end{align}
\end{subequations}
%
Typically, $\psi$ is of the tensorial form: $\psi=m\,C^{2a}C_{\langle i_1}C_{i_2}\cdots C_{i_n \rangle}$ and hence the general form of the collisional production term is 
\begin{align}
\label{prodtermGen}
\mathcal{P}_{i_1\cdots i_n}^a
&=\frac{m \, \mathring{\nu}}{n\,\Omega_d} \iint
\left[ \int \left\{\left(C^{\prime}\right)^{2a}C_{\langle i_1}^{\prime}C_{i_2}^{\prime}\cdots C_{i_n \rangle}^{\prime}-C^{2a}C_{\langle i_1}C_{i_2}\cdots C_{i_n \rangle}\right\}
\mathrm{d}\hat{\bm{k}}
\right] 
\nonumber\\
&\quad \times
f(\bm{c}) f(\bm{c}_1)
\, \mathrm{d}\bm{c} \, \mathrm{d}\bm{c}_1.
\end{align}
However, since the squared velocities in \eqref{prodtermGen} can be easily expressed in index notation using the Einstein summation convention, for instance $C^2 = C_{i_0} C_{i_0}$, and the indices in each term of
\eqref{prodtermGen} can be adjusted accordingly, it is convenient to first compute 
\begin{align}
\label{prodtermSimp}
P_{i_1\cdots i_n}=\frac{m \,\mathring{\nu}}{n\,\Omega_d} \iint
\left[ \int \left( C_{i_1}^{\prime}C_{i_2}^{\prime}\cdots C_{i_n}^{\prime} - C_{i_1}C_{i_2}\cdots C_{i_n}\right) 
\mathrm{d}\hat{\bm{k}}
\right] 
f(\bm{c}) f(\bm{c}_1)
\, \mathrm{d}\bm{c} \, \mathrm{d}\bm{c}_1
\end{align}
instead of computing \eqref{prodtermGen} directly.

To compute the right-hand side of \eqref{prodtermSimp}, the post-collisional peculiar velocities (marked with primes) in \eqref{prodtermSimp} are replaced with the pre-collisional peculiar velocities by exploiting the definition of the peculiar velocity and relation (\ref{VelTran}$a$). 
This changes the product of the post-collisional peculiar velocities in \eqref{prodtermSimp} to
\begin{align}
\label{ProdPostCollVel}
C_{i_1}^{\prime}C_{i_2}^{\prime}\cdots C_{i_n}^{\prime}
=\sum\limits_{j=0}^{n}(-w_0)^j\binom{n}{j} \hat{k}_{(i_1} \hat{k}_{i_2}\cdots \hat{k}_{i_j} C_{i_{j+1}}C_{i_{j+2}} \cdots C_{i_{n-1}} C_{i_n)} (\hat{\bm{k}}\cdot\bm{g})^j,
\end{align}
where $w_0 = (1+e)/2$ and the round brackets around the indices again denote the symmetric part of the corresponding tensor (see appendix~\ref{app:traceless} for its definition). 
Substituting \eqref{ProdPostCollVel} into \eqref{prodtermSimp}, one obtains
\begin{align}
\label{Intermediate_Prodterm}
P_{i_1\ldots i_n}=\frac{m \, \mathring{\nu}}{n\,\Omega_d} \sum\limits_{j=1}^{n}(-w_0)^j\binom{n}{j}\iint\! g^j I_{(i_1\ldots i_j} C_{i_{j+1}}C_{i_{j+2}} \cdots C_{i_{n-1}} C_{i_n)} f(\bm{c}) f(\bm{c}_1)
\, \mathrm{d}\bm{c} \, \mathrm{d}\bm{c}_1,
\end{align}
where
\begin{align}
\label{SVI}
I_{i_1 i_2\dots i_n}=\int\!\hat{k}_{i_1}\hat{k}_{i_2}\cdots \hat{k}_{i_n}(\hat{\bm{k}}\cdot\hat{\bm{g}})^n \,
\mathrm{d} \hat{\bm{k}}
\end{align}
is termed as the scattering vector integral with $\hat{\bm{g}} = \bm{g} / g$. 
The structure of the integrand in \eqref{SVI} suggests that $I_{i_1 i_2\dots i_n}$ will have the form
\begin{align}
\label{SVIintermsofdeltas}
I_{{i_1} i_2\dots{i_n}} = 
\sum\limits_{\beta=0}^{\left\lfloor \frac{n}{2} \right\rfloor} a_\beta^{(n)}\,\delta_{(i_1 i_2}\delta_{i_3 i_4}\cdots\delta_{i_{2\beta-1} i_{2\beta}}\frac{g_{i_{2\beta+1}}}{g}\frac{g_{i_{2\beta+2}}}{g}\cdots\frac{g_{i_n)}}{g},
\end{align}
where the unknown coefficients $a_\beta^{(n)}$ depend only on the dimension $d$, and are computed separately in \S\,\ref{Subsec:Coeffajn}. 
Insertion of \eqref{SVIintermsofdeltas} into \eqref{Intermediate_Prodterm} transpires the tensorial structure of $P_{i_1\ldots i_n}$:
\begin{align}
\label{startpointinMathematica}
P_{i_1\ldots i_n} &=\frac{m \, \mathring{\nu}}{n\,\Omega_d} \sum\limits_{j=1}^{n} \sum\limits_{\beta=0}^{\left\lfloor \frac{j}{2} \right\rfloor}(-w_0)^j\binom{n}{j} \delta_{(i_1 i_2} \delta_{i_3 i_4} \cdots\delta_{i_{2\beta-1} i_{2\beta}}\nonumber\\
&\quad\times\iint\!a_\beta^{(j)}\,g^{2\beta} g_{i_{2\beta+1}} g_{i_{2\beta+2}}\cdots g_{i_j} C_{i_{j+1}} C_{i_{j+2}}\cdots C_{i_n)} f(\bm{c}) f(\bm{c}_1)
\, \mathrm{d}\bm{c} \, \mathrm{d}\bm{c}_1.
\end{align}
Expression \eqref{startpointinMathematica}, without the prefactor $m \, \mathring{\nu}/ (n\,\Omega_d)$, is entered as the starting point in {\textsc{Mathematica}}\textsuperscript{\textregistered} script.

For general interaction potentials, specific forms of the distribution functions $f(\bm{c})$ and $f(\bm{c}_1)$ must be provided in order to compute $P_{i_1\ldots i_n}$ further. Nevertheless, for IMM, specific forms of $f(\bm{c})$ and $f(\bm{c}_1)$ are not required since the coefficients $a_\beta^{(j)}$ for IMM do not depend on the relative velocity $\bm{g}$. 
Indeed, for IMM, now all the components and the magnitude of the relative velocity $\bm{g}$ are replaced in terms of the peculiar velocities $\bm{C}$ and $\bm{C}_1$ by using the relation $\bm{g}=\bm{C}-\bm{C}_1$. 
It may be noted that the exponent of the magnitude of $\bm{g}$ is even in \eqref{startpointinMathematica}, which makes it easier to replace $g^{2\beta}$ in \eqref{startpointinMathematica} using the relation $g^2=C^2+C_1^2-2C_kC_{1k}$. 
At this step, some vanishing integrals, such as
\begin{align*}
\iint\! g_i g_j g_k f(\bm{c}) f(\bm{c}_1)
\, \mathrm{d}\bm{c} \, \mathrm{d}\bm{c}_1=0 
\quad\textrm{and}\quad
\iint\! g^2 g_i f(\bm{c}) f(\bm{c}_1)
\, \mathrm{d}\bm{c} \, \mathrm{d}\bm{c}_1=0,
\end{align*} 
are automatically taken care of in the {\textsc{Mathematica}}\textsuperscript{\textregistered} script.
Now, the double integrals in each term under the summation in \eqref{startpointinMathematica} can be written as a product of two independent integrals, one over $\bm{c}$ and the other over $\bm{c}_1$, and each of these integrals can be expressed in terms of the considered moments. Note that the present work deals with the traceless moments and all the terms should be expressed as traceless moments. 
This is not very straightforward for tensors of rank more than three. Nevertheless, this step has also been incorporated in the {\textsc{Mathematica}}\textsuperscript{\textregistered} script to express all the results in terms of traceless tensors. Finally, taking the traceless part of each term in the result, one obtains the required collisional production term.
All the collisional production terms obtained in this work agree with those obtained in \cite{GarzoSantos2007} till fourth order, which validates the code. In principle, this {\textsc{Mathematica}}\textsuperscript{\textregistered} script would be able to compute the collisional production terms for moments of any order. 
Nonetheless, as the code is not optimised, it takes a significantly long computation time in computing the collisional production terms associated with more than sixth-order moments.
\subsection{Computation of the coefficients \texorpdfstring{$a_\beta^{(n)}$}{ }}
\label{Subsec:Coeffajn}
The unknown coefficients $a_\beta^{(n)}$ follow by appropriately contracting the two forms of $I_{i_1\ldots i_n}$ in \eqref{SVI} and \eqref{SVIintermsofdeltas} with combinations of $\hat{g}_i = g_i/g$ and with combinations of Kronecker deltas, successively. This results into linear systems of algebraic equations, which yield the coefficients $a_\beta^{(n)}$ as functions of the scalar integrals given by
\citep{vNE1998,GarzoSantos2007}
\begin{align}
\label{B_r}
B_r=\int\! (\hat{\bm{k}}\cdot\hat{\bm{g}})^{2r}\,
\mathrm{d}\hat{\bm{k}}
=\frac{\Omega_d}{\sqrt{\pi}}\frac{\Gamma\left(\frac{d}{2}\right)\Gamma\left(r+\frac{1}{2}\right)}{\Gamma\left(r+\frac{d}{2}\right)} ,
\quad\textrm{where}\quad r\in\mathbb{N}.
\end{align}
From \eqref{SVI} and \eqref{SVIintermsofdeltas}, one has
\begin{align*}
I_i =\int\!\hat{k}_i (\hat{\bm{k}}\cdot\hat{\bm{g}})\,
\mathrm{d}\hat{\bm{k}} = a_0^{(1)}\frac{g_i}{g}.
\end{align*}
Contracting the above equation with $\hat{g}_i$ and using the fact that $\hat{k}_i \hat{g}_i = \hat{\bm{k}}\cdot\hat{\bm{g}}$, it readily follows that
\begin{align}
\label{a01}
a_0^{(1)}=B_1.
\end{align}
Again, from \eqref{SVI} and \eqref{SVIintermsofdeltas}, one has
\begin{align*}
I_{ij} =\int\!\hat{k}_i \hat{k}_j (\hat{\bm{k}}\cdot\hat{\bm{g}})^2\,
\mathrm{d}\hat{\bm{k}} =a_0^{(2)}\frac{g_i g_j}{g^{\,2}}+a_1^{(2)}\delta_{ij}.
\end{align*}
Contracting the above equation with $\hat{g}_i\hat{g}_j$ and with $\delta_{ij}$ successively, one obtains
\begin{align*}
B_2=a_0^{(2)}+a_1^{(2)}\quad\textrm{and}\quad B_1=a_0^{(2)}+d\, a_1^{(2)},
\end{align*}
and, thus
\begin{align}
\label{a02a12}
a_0^{(2)}=\frac{d \, B_2-B_1}{d-1}\quad\textrm{and}\quad 
a_1^{(2)}=\frac{B_1-B_2}{d-1}.
\end{align}
The next integrals are treated analogously and one, eventually, finds
\begin{align}
\label{a03a13}
a_0^{(3)}=\frac{(d+2) B_3 - 3 B_2}{d-1}\quad\textrm{and}\quad a_1^{(3)}=\frac{3(B_2-B_3)}{d-1},
\end{align}
\begin{gather}
\label{a04a14a24}
a_0^{(4)}=\frac{(d+2)[(d+4)B_4 - 6 B_3] + 3 B_2}{d^2-1}, \quad a_1^{(4)}=-\frac{6[(d+2)B_4 - (d+3) B_3 + B_2]}{d^2-1}\nonumber\\
\textrm{and}\quad 
a_2^{(4)}=\frac{3(B_4 - 2 B_3 + B_2)}{d^2-1},
\end{gather}
and so on; see {\textsc{Mathematica}}\textsuperscript{\textregistered} file ``kintegrals.nb'' provided as supplementary material.
\section{Coefficients in the collisional production terms}
\label{app:coeffprodterms}
The coefficients in the collisional production terms \eqref{P1}--\eqref{Pi2} associated with the G29 equations for IMM are as follows.
\begingroup
\allowdisplaybreaks
\begin{align}
\label{zeta0ast}
\zeta_0^\ast &= \frac{d+2}{4d}(1-e^2),
\\
\label{nusigma}
\nu_\sigma^\ast&=\frac{(1+e)(d+1-e)}{2d},
\\
\label{nuq}
\nu_q^\ast&=\frac{(1+e)[5d+4-(d+8)e]}{8d},
\\
\label{num}
\nu_m^\ast &=\frac{3}{2}\nu_\sigma^\ast=\frac{3(1+e)(d+1-e)}{4d},
\\
\label{nuR}
\nu_R^\ast&=\frac{(1+e)[7d^2+31d+18-(d^2+14d+34)e+3 (d+2)e^2 -6e^3]}{8d(d+4)},
\\
\label{nuphi}
\nu_\varphi^\ast&= \frac{(1+e) [32 d^2+129 d+64-(8 d^2+81 d+136) e+3(9d+16) e^2-3(d+24) e^3]}{32 d (d+4)},
\\
\alpha_0&=\frac{(1-e^2)(d+2)(4 d+5+3 e^2)}{8},
\\
\alpha_1&=\frac{(1+e)(d+2)[3(4d+3)-(4 d+17) e+3 e^2-3 e^3]}{16},
\\
\alpha_2&=\frac{(1+e) [3 d^2+13 d+10-(d^2+8 d+10) e+3(d+2)e^2-6 e^3]}{4d},
\\
\alpha_3&=\frac{(1+e) [14 d^2+57 d+34-3(d+6)(2d+3) e+15(d+2) e^2-3(d+14) e^3]}{4d},
\\
\varsigma_0
&=\frac{(1+e)^2 (1+6e-3e^2)}{8d^2(d+2)},
\\
\varsigma_1
&= - \frac{(1+e)^2 [d-2-3(d+4)e+6e^2]}{4d(d+4)},
\\
\varsigma_2
&= - \frac{(1+e)^2 [5 d-4-6(d+4) e-3(d-4) e^2]}{4 d (d+2)},
\\
\varsigma_3
&= \frac{(1+e)^2 [d+16+6(d+4) e-3(d+8) e^2]}{8 d (d+4)}.
\end{align}
\endgroup
\section{The G29 distribution function}
\label{app:G29}
The computation of the G29 distribution function is comparatively easier with the dimensionless variables. Let us introduce the dimensionless variables (denoted with bars) as follows:
\begin{align}
\label{dimlessvarfordisfun}
\left.
\begin{gathered}
\bar{\bm{C}} = \frac{\bm{C}}{\sqrt{\theta}}, \qquad
\bar{\bm{C}_1} = \frac{\bm{C}_1}{\sqrt{\theta}},
\qquad
\bar{\bm{g}} = \frac{\bm{g}}{\sqrt{\theta}}, 
\\
\bar{f}\equiv \bar{f}(t,\bm{x},\bm{C}) = \frac{\theta^{d/2}}{n}f(t,\bm{x},\bm{c}),
\qquad
\bar{f}_M = \frac{\theta^{d/2}}{n} f_M = \frac{1}{(2\pi)^{d/2}} \mathrm{e}^{-\bar{C}^2/2},
\\
\bar{u}_{{i_1}{i_2}\dots{i_r}}^a = \frac{u_{{i_1}{i_2}\dots{i_r}}^a}{\rho\theta^{a+\frac{r}{2}}},
\\
\bar{\sigma}_{ij} = \frac{\sigma_{ij}}{\rho\theta},
\quad
\bar{q}_i = \frac{q_i}{\rho\theta^{3/2}},
\quad
\bar{m}_{ijk} = \frac{m_{ijk}}{\rho\theta^{3/2}},
\quad
\bar{R}_{ij} = \frac{R_{ij}}{\rho\theta^2},
\quad
\bar{\varphi}_i = \frac{\varphi_i}{\rho\theta^{5/2}}
.
\end{gathered}
\right\}
\end{align}
In the dimensionless variables, the definitions of the 29 moments can be recast as
\begin{align}
\label{dimlessdef}
\left.
\begin{gathered}
1 = \int\! \bar{f}\,\mathrm{d}\bar{\bm{C}},
\quad 
0 = \int\! \bar{C}_i  \bar{f}\,\mathrm{d}\bar{\bm{C}},
\quad 
d = \int\! \bar{C}^2  \bar{f}\,\mathrm{d}\bar{\bm{C}},
\\
\bar{\sigma}_{ij} = \int\! \bar{C}_{\langle i}  \bar{C}_{j \rangle}\bar{f}\,\mathrm{d}\bar{\bm{C}},
\quad
\bar{q}_i = \frac{1}{2}\int\! \bar{C}^2 \bar{C}_i  \bar{f}\,\mathrm{d}\bar{\bm{C}},
\quad
\bar{m}_{ijk} = \int\! \bar{C}_{\langle i} \bar{C}_j \bar{C}_{k \rangle}\bar{f}\,\mathrm{d}\bar{\bm{C}},
\\
d(d+2)(1+\Delta) = \bar{u}^2 = \int\! \bar{C}^4  \bar{f}\,\mathrm{d}\bar{\bm{C}},
\\
\bar{R}_{ij} + (d+4)\bar{\sigma}_{ij} = \bar{u}_{ij}^1 = \int\! \bar{C}^2 \bar{C}_{\langle i}  \bar{C}_{j \rangle}\bar{f}\,\mathrm{d}\bar{\bm{C}}, 
\\
\bar{\varphi}_i + 4 (d+4) \bar{q}_i = \bar{u}_i^2 = \int\! \bar{C}^4 \bar{C}_i\bar{f}\,\mathrm{d}\bar{\bm{C}}.
\end{gathered}
\right\}
\end{align}
Let the G29 distribution function in the dimensionless form be given by
\begin{align}
\label{fgrad29dimless}
\bar{f}_{|\mathrm{G29}} =\bar{f}_M\Big(\lambda^0 &+\lambda_i^0\bar{C}_i+\lambda^1\bar{C}^2
+\lambda_{\langle ij\rangle}^0\bar{C}_i\bar{C}_j+\lambda_i^1\bar{C}^2\,\bar{C}_i
\nonumber\\
&+\lambda_{\langle ijk\rangle}^0\bar{C}_i\bar{C}_j\bar{C}_k
+\lambda^2\bar{C}^4 +\lambda_{\langle ij\rangle}^1\bar{C}^2 \bar{C}_i \bar{C}_j
+\lambda_i^2\bar{C}^4 \bar{C}_i 
\Big),
\end{align}
where the angle brackets again denote the symmetric-traceless tensors and $\lambda$'s are the unknown coefficients that are determined by replacing $\bar{f}$ with $\bar{f}_{|\mathrm{G29}}$ in definitions \eqref{dimlessdef}, and solving the resulting system of algebraic equations for $\lambda$'s.

The integrals over velocity space are typically evaluated by transforming the integral from a $d$-dimensional Cartesian coordinate system to a $d$-dimensional spherical coordinate system. A useful identity, which employs this transformation, for evaluating the integral of an \emph{even} function $h(C)$ in $C$ over the velocity space $\bm{C}$ is 
\begin{align}
\label{iden:sphtrans}
\int\!h(C)\,\mathrm{d}\bm{C} &= \int_{C=0}^{\infty} \int_{\theta_1=0}^\pi \int_{\theta_2=0}^\pi \dots \int_{\theta_{d-2}=0}^\pi \int_{\theta_{d-1}=0}^{2\pi} h(C)\,C^{d-1} \sin^{d-2} \theta_1 \sin^{d-3} \theta_2 \dots 
\nonumber\\
&\quad\times \sin^{2} \theta_{d-3} \sin \theta_{d-2}\,\mathrm{d}\theta_{d-1}\,\mathrm{d}\theta_{d-2}\dots \mathrm{d}\theta_{2}\,\mathrm{d}\theta_1 \,\mathrm{d}C
\nonumber\\
&=\frac{2 \pi^{d/2}}{\Gamma{(d/2)}}
\int_{C=0}^{\infty} h(C)\,C^{d-1} \,\mathrm{d}C,
\end{align}
where the following identities have been used: for $n \geq 0$,
\begin{align*}
\int_0^\pi\!\sin^n \theta \,\mathrm{d}\theta = \sqrt{\pi} \, \frac{\Gamma\left(\frac{n+1}{2}\right)}{\Gamma\left(\frac{n+2}{2}\right)}
\quad\textrm{and}\quad
\int_0^{2\pi}\!\sin^n \theta \,\mathrm{d}\theta = \big[1+(-1)^n\big]\sqrt{\pi} \, \frac{\Gamma\left(\frac{n+1}{2}\right)}{\Gamma\left(\frac{n+2}{2}\right)}.
\end{align*}
Note that a similar integral for an \emph{odd} function $h(C)$ in $C$ over the velocity space $\bm{C}$ vanishes. Furthermore, for an \emph{even} function $h(C)$ in $C$, it can be shown that 
\begin{align}
\label{int:CiCj}
\int\!C_i C_j  h(C)\,\mathrm{d}\bm{C} &= \frac{1}{d} \, \delta_{ij} \int\! C^2 h(C)\,\mathrm{d}\bm{C},
\\
\label{int:CiCjCkCl}
\int\!C_i C_j  C_k C_l \, h(C)\,\mathrm{d}\bm{C} &= \frac{1}{d(d+2)} \big(\delta_{ij} \delta_{kl} +\delta_{ik} \delta_{jl} +\delta_{il} \delta_{jk} \big) \int\! C^4 h(C)\,\mathrm{d}\bm{C},
\\
\label{int:CiCjCkClCrCs}
\int\!C_i C_j  C_k C_l C_r C_s  h(C)\,\mathrm{d}\bm{C} &= \frac{15}{d(d+2)(d+4)} \delta_{(ij} \delta_{kl} \delta_{rs)} \int\! C^6 h(C)\,\mathrm{d}\bm{C},
\end{align}
and, in general, 
\begin{align}
\label{int:Ci1toCin}
\int\!C_{i_1} C_{i_2} \dots C_{i_n}  h(C)\,\mathrm{d}\bm{C}
= 
\frac{n!}{2^{n/2} (\frac{n}{2})!} \frac{1}{\prod\limits_{j=0}^{\frac{n}{2}-1}(d+2j)}
\delta_{(i_1 i_2} \delta_{i_3 i_4} \dots \delta_{i_{n-1} i_n)} 
\int\! C^n h(C)\,\mathrm{d}\bm{C}
\end{align}
for an even $n$ while the integral vanishes for an odd $n$. As a consequence of \eqref{int:CiCj}--\eqref{int:CiCjCkClCrCs}, it is straightforward to show that
\begin{align}
\int\! C_{\langle i} C_{j \rangle}\,h(C)\,\mathrm{d}\bm{C} &= 0,
\\
\int\! C_i C_{\langle j} C_k C_{l \rangle}\,h(C)\,\mathrm{d}\bm{C} &= 0,
\\
\int\! C_{\langle i} C_j C_k C_{l \rangle}\,h(C)\,\mathrm{d}\bm{C} &= 0,
\\
\int\! C_{\langle i} C_j C_k C_{l \rangle} C_r C_s\,h(C)\,\mathrm{d}\bm{C} &= 0.
\label{iden:vanish}
\end{align}
Now, replacing $\bar{f}$ with $\bar{f}_{|\mathrm{G29}}$ in definitions \eqref{dimlessdef}, and using identities \eqref{iden:sphtrans}--\eqref{iden:vanish}, one obtains
\begin{align}
\left.
\begin{aligned}
1&=\lambda^0+d\lambda^1+d(d+2)\lambda^2,
\\
0&=\lambda_i^0+(d+2)\lambda_i^1+(d+2)(d+4)\lambda_i^2,
\\
1&=\lambda^0+(d+2)\lambda^1+(d+2)(d+4)\lambda^2,
\\
\bar{\sigma}_{ij} &= 2\lambda_{\langle ij\rangle}^0 + 2(d+4)\lambda_{\langle ij\rangle}^1,
\\
2 \bar{q}_i &=(d+2)\big[\lambda_i^0+(d+4)\lambda_i^1+(d+4)(d+6)\lambda_i^2 \big]
\\
\bar{m}_{ijk} &= 6 \lambda_{\langle ijk\rangle}^0,
\\
1+\Delta&=\lambda^0+(d+4)\lambda^1+(d+4)(d+6)\lambda^2,
\\
\bar{R}_{ij} + (d+4)\bar{\sigma}_{ij} &=2(d+4)\lambda_{\langle ij\rangle}^0 + 2(d+4)(d+6) \lambda_{\langle ij\rangle}^1,
\\
\bar{\varphi}_i + 4 (d+4) \bar{q}_i &= (d+2)(d+4) \big[ \lambda_i^0
+(d+6)\lambda_i^1
+(d+6)(d+8)\lambda_i^2\big].
\end{aligned}
\right\}
\end{align}
These equations yield
\begin{align}
\left.
\begin{gathered}
\lambda^0 = 1+ \frac{d(d+2)\Delta}{8}, 
\qquad 
\lambda^1 = -\frac{(d+2)\Delta}{4},
\qquad 
\lambda^2 = \frac{\Delta}{8},
\\
\lambda_i^0=\frac{\bar{\varphi}_i}{8} - \bar{q}_i,
\qquad
\lambda_i^1=\frac{\bar{q}_i}{d+2} - \frac{\bar{\varphi}_i}{4(d+2)},
\qquad
\lambda_i^2=\frac{\bar{\varphi}_i}{8(d+2)(d+4)},
\\
\lambda_{\langle ij\rangle}^0 =-\frac{\bar{R}_{ij} - 2  \bar{\sigma}_{ij}}{4},
\qquad
\lambda_{\langle ij\rangle}^1= \frac{\bar{R}_{ij}}{4(d+4)},
\qquad
\lambda_{\langle ijk \rangle}^0 = \frac{\bar{m}_{ijk}}{6}.
\end{gathered}
\right\}
\end{align}
Inserting these coefficients in \eqref{fgrad29dimless}, the dimensionless G29 distribution function reads
\begin{align}
\bar{f}_{|\mathrm{G29}} &=\bar{f}_M\bigg[1 +\frac{1}{2} \bar{\sigma}_{ij} \bar{C}_i\bar{C}_j + \bar{q}_i \bar{C}_i \left(\frac{1}{d+2} \bar{C}^2 - 1\right) +\frac{1}{6} \bar{m}_{ijk}\bar{C}_i\bar{C}_j\bar{C}_k  
\nonumber\\
&\quad + \frac{d(d+2)\Delta}{8} \left(1 - \frac{2}{d}\bar{C}^2 + \frac{1}{d(d+2)}\bar{C}^4\right)
+ \frac{1}{4} \bar{R}_{ij} \bar{C}_i \bar{C}_j 
\left(\frac{1}{d+4} \bar{C}^2 - 1 \right)
\nonumber\\
&\quad + \frac{1}{8} \bar{\varphi}_i \bar{C}_i \left(1 - \frac{2}{d+2}\bar{C}^2 + \frac{1}{(d+2)(d+4)}\bar{C}^4\right)
\bigg],
\end{align}
which on introducing the dimensions using \eqref{dimlessvarfordisfun} yields the G29 distribution function \eqref{G29disfun}.
\vspace*{-2mm}
\section{Explicit components of the traceless gradients}
\label{app:tracefreeGradients}
The explicit components of the traceless gradients in \eqref{massBalPertDimless}--\eqref{phiBalPertDimless} are computed as follows. 
Using \eqref{tracefree2tensor}, 
\begin{align*}
\frac{\partial \tilde{\Phi}_{\langle i}}{\partial \tilde{x}_{j\rangle}} 
= \frac{1}{2} \bigg(\frac{\partial \tilde{\Phi}_i}{\partial \tilde{x}_j} + \frac{\partial \tilde{\Phi}_j}{\partial \tilde{x}_i}\bigg)
- \frac{1}{d} \delta_{ij}\frac{\partial \tilde{\Phi}_k}{\partial \tilde{x}_k},
\end{align*}
where $\Phi\in\{v, q, \varphi\}$.
%
From the above equation, it follows that
\allowdisplaybreaks
\begin{align}
\frac{\partial \tilde{\Phi}_{\langle x}}{\partial \tilde{x}_{1\rangle}} 
&= \frac{\partial \tilde{\Phi}_x}{\partial \tilde{x}} 
- \frac{1}{d} \bigg(\frac{\partial \tilde{\Phi}_x}{\partial \tilde{x}}
+\frac{\partial \tilde{\Phi}_y}{\partial \tilde{y}}
+\frac{\partial \tilde{\Phi}_z}{\partial \tilde{z}}
\bigg)
= \frac{d-1}{d}\frac{\partial \tilde{\Phi}_x}{\partial \tilde{x}}
- \frac{1}{d} \frac{\partial \tilde{\Phi}_y}{\partial \tilde{y}}
- \frac{1}{d} \frac{\partial \tilde{\Phi}_z}{\partial \tilde{z}},
\\
\frac{\partial \tilde{\Phi}_{\langle x}}{\partial \tilde{x}_{2\rangle}} 
&= \frac{1}{2} \bigg( \frac{\partial \tilde{\Phi}_x}{\partial \tilde{y}}
+\frac{\partial \tilde{\Phi}_y}{\partial \tilde{x}} \bigg).
\end{align}
The other components, if needed, can be computed analogously.

For a symmetric-traceless rank two tensor $\tilde{\Phi}_{ij}$, definition  \eqref{tracefree3tensor} gives 
\begin{align*}
\frac{\partial \tilde{\Phi}_{\langle ij}}{\partial \tilde{x}_{k\rangle}} = \frac{1}{3} \bigg(\frac{\partial \tilde{\Phi}_{ij}}{\partial \tilde{x}_k} + 
\frac{\partial \tilde{\Phi}_{jk}}{\partial \tilde{x}_i} + 
\frac{\partial \tilde{\Phi}_{ik}}{\partial \tilde{x}_j} 
\bigg)
- \frac{2}{3(d+2)} \bigg( \frac{\partial \tilde{\Phi}_{il}}{\partial \tilde{x}_l} \delta_{jk}
+
\frac{\partial \tilde{\Phi}_{jl}}{\partial \tilde{x}_l} \delta_{ik}
+
\frac{\partial \tilde{\Phi}_{kl}}{\partial \tilde{x}_l} \delta_{ij}
\bigg),
\end{align*}
where $\Phi\in\{\sigma,R\}$ 
in the present work. 
From the above equation, it follows that
\begin{align}
\frac{\partial \tilde{\Phi}_{\langle xx}}{\partial \tilde{x}_{1\rangle}} &= \frac{\partial \tilde{\Phi}_{xx}}{\partial \tilde{x}}
- \frac{2}{d+2} \bigg( \frac{\partial \tilde{\Phi}_{xx}}{\partial \tilde{x}} 
+
\frac{\partial \tilde{\Phi}_{xy}}{\partial \tilde{y}} 
+
\frac{\partial \tilde{\Phi}_{xz}}{\partial \tilde{z}} 
\bigg)
\nonumber\\
&= \frac{d}{d+2} \frac{\partial \tilde{\Phi}_{xx}}{\partial \tilde{x}} 
- \frac{2}{d+2} \frac{\partial \tilde{\Phi}_{xy}}{\partial \tilde{y}} 
- \frac{2}{d+2} \frac{\partial \tilde{\Phi}_{xz}}{\partial \tilde{z}},
\\
\frac{\partial \tilde{\Phi}_{\langle xx}}{\partial \tilde{x}_{2\rangle}} 
&= \frac{1}{3} \bigg(\frac{\partial \tilde{\Phi}_{xx}}{\partial \tilde{y}} + 
2\frac{\partial \tilde{\Phi}_{xy}}{\partial \tilde{x}}
\bigg)
- \frac{2}{3(d+2)} \bigg( 
\frac{\partial \tilde{\Phi}_{xy}}{\partial \tilde{x}} 
+ \frac{\partial \tilde{\Phi}_{yy}}{\partial \tilde{y}} 
+ \frac{\partial \tilde{\Phi}_{yz}}{\partial \tilde{z}} 
\bigg)
\nonumber\\
&= \frac{2 (d+1)}{3(d+2)}  
\frac{\partial \tilde{\Phi}_{xy}}{\partial \tilde{x}} 
+ \frac{1}{3} \frac{\partial \tilde{\Phi}_{xx}}{\partial \tilde{y}} 
- \frac{2}{3(d+2)}  
\frac{\partial \tilde{\Phi}_{yy}}{\partial \tilde{y}} 
- \frac{2}{3(d+2)}  \frac{\partial \tilde{\Phi}_{yz}}{\partial \tilde{z}}.
\end{align}
The other components, if needed, can be computed analogously.
%
\section{Coefficient matrices in \eqref{eigvalProbs} and \eqref{eigvalProbsNSF}}
\label{app:matrices}
The matrix $\mathscr{L}$ in the longitudinal problem~(\ref{eigvalProbs}$a$) can be written in the form of block matrices for better readability as
\begin{align}
\label{matrixL}
\mathscr{L} =
\begin{bmatrix}
\mathscr{L}_{11} & \mathscr{L}_{12}
\\ 
\mathscr{L}_{21} & \mathscr{L}_{22}
\end{bmatrix}
\end{align}
with the block matrices being
\begingroup
\renewcommand{\arraystretch}{2.2}
\begin{align}
\mathscr{L}_{11} &=
\begin{bmatrix}
- \omega & k & 0 & 0 & 0 
\\
k & - \omega + \mathbbm{i} \dfrac{\zeta_0^\ast}{2} & k & k & 0 
\\
- \mathbbm{i} \zeta_0^\ast & \dfrac{2k}{d} & - \omega - \mathbbm{i} \dfrac{\zeta_0^\ast}{2} & 0 & \dfrac{2k}{d} 
\\
0 & \dfrac{2(d-1)k}{d} & 0 & - \omega - \mathbbm{i} \, \xi_\sigma & \dfrac{4(d-1)k}{d(d+2)} 
\\
\dfrac{(d+2) a_2 k}{2} & 0 &\dfrac{(d+2) (1+2 a_2) k}{2} & k & - \omega - \mathbbm{i} \,\xi_q
\end{bmatrix},
\end{align}
\begin{align}
\mathscr{L}_{12} &=
\begin{bmatrix}
0 & 0 & 0 & 0
\\
0 & 0 & 0 & 0
\\
0 & 0 & 0 & 0
\\
k & 0 & 0 & 0
\\
0 & \dfrac{(d+2) k}{2} & \dfrac{k}{2} & 0
\end{bmatrix},
\end{align}
\begin{align}
\mathscr{L}_{21} &=
\begin{bmatrix}
0 & 0 & 0 & \dfrac{3dk}{d+2} & 0
\\
0 & 0 & 0 & 0 & \xi_1 k
\\
0 & \dfrac{2(d+4)(d-1)a_2 k}{d} & 0 & \mathbbm{i}\, \nu_{R\sigma}^\ast & \dfrac{4(d+4)(d-1)k}{d(d+2)} 
\\
0 & 0 & 4 \xi_2 a_2 k & - \xi_2 a_2 k & \mathbbm{i} \,\nu_{\varphi q}^\ast
\end{bmatrix},
\end{align}
\begin{align}
\mathscr{L}_{22} &=
\begin{bmatrix}
- \omega - \mathbbm{i} \, \xi_m & 0 & \dfrac{3dk}{(d+2)(d+4)} & 0
\\
0 & - \omega - \mathbbm{i} \, \nu_\Delta^\ast & 0 & \dfrac{k}{d(d+2)}
\\
2k & 0 & - \omega - \mathbbm{i} \, \xi_R  & \dfrac{2(d-1)k}{d(d+2)}
\\
0 & \xi_2 k & 4k & - \omega - \mathbbm{i}  \, \xi_\varphi 
\end{bmatrix},
\end{align}
\endgroup
and the matrix $\mathscr{T}$  in the transverse problem~(\ref{eigvalProbs}$b$) reads
\begingroup
\renewcommand{\arraystretch}{2.2}
\begin{align}
\label{matrixT}
\mathscr{T} =
\begin{bmatrix}
\mathbbm{i} \dfrac{\zeta_0^\ast}{2} & k & 0 & 0 & 0 & 0
\\
k & 
- \mathbbm{i} \, \xi_\sigma & \dfrac{2 k}{d+2} & k & 0 & 0
\\
0 & k & 
- \mathbbm{i} \, \xi_q  & 0 & \dfrac{k}{2} & 0
\\
0 & \dfrac{2(d+1) k}{d+2} & 0 & 
-  \mathbbm{i} \, \xi_m & \dfrac{2(d+1) k}{(d+2)(d+4)}  & 0 
\\
(d+4) a_2 k & \mathbbm{i}\, \nu_{R\sigma}^\ast & \dfrac{2(d+4) k}{d+2} & 2 k & 
- \mathbbm{i} \, \xi_R & \dfrac{k}{d+2}
\\
0 & - \xi_2 a_2 k & \mathbbm{i} \,\nu_{\varphi q}^\ast & 0 & 4k & 
- \mathbbm{i} \, \xi_\varphi
\end{bmatrix}
- \omega \, I_6 
\end{align}
\endgroup
where $I_6$ is the identity matrix of dimensions $6 \times 6$. 
%

The matrix $\mathscr{L}_{\mathrm{NSF}}$ in \eqref{eigvalProbsNSF} reads
\begingroup
\renewcommand{\arraystretch}{2.2}
\begin{align}
\mathscr{L}_{\mathrm{NSF}} = 
\begin{bmatrix}
0 & k & 0
\\
k & - \mathbbm{i} \dfrac{2(d-1)}{d} \eta^\ast k^2 + \mathbbm{i} \dfrac{\zeta_0^\ast}{2} & k 
\\
- \mathbbm{i} \dfrac{d+2}{d-1} \lambda^\ast k^2 - \mathbbm{i} \zeta_0^\ast & \dfrac{2k}{d} & - \mathbbm{i} \dfrac{d+2}{d-1} \kappa^\ast k^2 - \mathbbm{i} \dfrac{\zeta_0^\ast}{2}
\end{bmatrix} - \omega \, I_3,
\end{align}
\endgroup
where $I_3$ is the identity matrix of dimensions $3 \times 3$.
%
\section{Coefficients in the analytical expressions of the critical wavenumbers}\label{app:coeffwavenumber}
Using the abbreviations given in \eqref{xi}, the coefficients appearing in the analytical expressions \eqref{criticalkh_G13}--\eqref{criticalks_G29} of the critical wavenumbers computed from various moment systems can be written as follows.
\begingroup
\allowdisplaybreaks
\begin{align}
\xi_3 &= (d-1) \big[2 (d+1) + 3 (d+2)a_2\big],
\\
\xi_4 &= \xi_3 \, \zeta_0^\ast \, \nu_\Delta^\ast 
+ (d+2) \xi_\sigma \Big[\big\{ (d+2) a_2 - 2\big\} \zeta_0^\ast + \xi_5 \Big],
\\
\vartheta_{11} &= - \frac{3}{4 (d+4)} \big[12 (d-1)(1+a_2) \zeta_0^\ast \, \nu_\Delta^\ast + \xi_6 \xi_7 \big],
\\
\vartheta_{12} &= \frac{\zeta_0^\ast \, \nu_\Delta^\ast \, \xi_m}{2d(d+2)} \big[\xi_3 \xi_R - (d-1) \xi_8 \big]
- \frac{\nu_\Delta^\ast \xi_7 \xi_9}{2}
- \frac{d+2}{8d} \xi_6 \, \xi_\sigma \, \xi_m \, \xi_R,
\\
\vartheta_{13} &= \xi_{10} \, \nu_\Delta^\ast \,\xi_R,
\\
\vartheta_{21} &= \frac{d+4}{d+2} (1-a_2) \xi_{11}
- \frac{d+1}{d+4} \zeta_0^\ast 
- \frac{2(d+1)}{d+4} \xi_q,
\\
\vartheta_{22} &= \frac{1}{4} \left[\big(\zeta_0^\ast \xi_R - \zeta_0^\ast \xi_\sigma + 2 \xi_q \xi_R\big) \xi_m
- \frac{\zeta_0^\ast \xi_7 \xi_{11}}{d+2}\right],
\\
\vartheta_{23} &= \xi_{10} \, \xi_R, 
\\
\vartheta_{31} &= \frac{9 (d-1) \xi_{12}}{2 d (d+2) (d+4)},
\\
\vartheta_{32} &= \xi_{13} - \xi_{14} - \xi_{15},
\\
\vartheta_{33} &= \vartheta_{32}^2 - \vartheta_{31} \xi_{16},  
\\
\vartheta_{34} &= \left(\sqrt{\xi_{17}^2 - \vartheta_{33}^3} - \xi_{17}\right)^{1/3},
\\
\vartheta_{41} &= \frac{\xi_{18}}{2(d+2)} \left(\xi_m - \frac{d+1}{d+4}\zeta_0^\ast\right)
- \frac{8+a_2 \xi_2}{2(d+2)^2}\zeta_0^\ast \xi_{11}
- \frac{(d+4)a_2-2}{d+2} \xi_{11} \xi_\varphi,
\\
\vartheta_{42} &= \frac{\big(\zeta_0^\ast + 2 \xi_q\big) \xi_m \xi_R \xi_\varphi}{4}
- \frac{\zeta_0^\ast\big(\xi_\sigma \xi_m \xi_{18} + 2 \xi_7 \xi_{11} \xi_\varphi\big)}{8(d+2)},
\\
\vartheta_{43} &= \xi_{10} \xi_R \xi_\varphi,
\end{align}
with
\begin{align}
\xi_5 &= (d+2) (1+a_2) \, \nu_\Delta^\ast,
\\
\xi_6 &= d \, \zeta_0^\ast \, \xi_1 - 4 (1+a_2) \, \nu_\Delta^\ast,
\\
\xi_7 &= 2 \xi_\sigma + (d+4) \xi_R + \nu_{R\sigma}^\ast,
\\
\xi_8 &= (d+2) \xi_\sigma - (d+4) \xi_R,
\\
\xi_9 &= \frac{\zeta_0^\ast}{d(d+2)} \left[(d-1) \xi_m + \frac{3 d^2 \xi_q}{2(d+4)}\right],
\\
\xi_{10} &= \frac{1}{2} \zeta_0^\ast \, \xi_\sigma \, \xi_q \, \xi_m,
\\
\xi_{11} &= \xi_m + \frac{2 (d+1)}{d+4} \xi_q,
\\
\xi_{12} &= (2 \zeta_0^\ast - \xi_6) \big[4(d+4) - 3 a_2 \xi_2\big] 
- (3 a_2 - 1) \xi_2 \left[3 \zeta_0^\ast (d \, \xi_1 - 2) + \frac{2 \xi_7}{d-1}\right],
\\
\xi_{13} &= \frac{8 (d-1)}{d^2 (d+2)} \xi_\sigma \xi_m \big(\xi_5 -a_2 \xi_2 \, \nu_\Delta^\ast\big)
+ \frac{2 \, \xi_9 \, \nu_\Delta^\ast}{d (d+2)} \big[4 \xi_3 - 9 (d-1) a_2 \xi_2 \big]- \frac{1}{d} (3 a_2 - 1) \xi_2 \xi_{19},
\\
\xi_{14} &= \frac{(d-1) \zeta_0^\ast \xi_m}{d^2 (d+2)}
\big(\xi_2 \xi_\sigma - 3 a_2 \xi_2 \xi_R - 4 \xi_8\big) 
+ \frac{1}{d} \left(4 - \frac{\xi_2}{d+2}\right)\left[\frac{d-1}{2} \zeta_0^\ast \xi_1 \xi_\sigma \xi_m - \xi_7 \xi_9 \right],
\\
\xi_{15} &= \frac{3}{4 (d+4)} 
\left[\frac{\zeta_0^\ast \xi_7 \xi_{18}}{d+2} + (d \, \xi_1 - 2) \zeta_0^\ast \xi_7 \xi_\varphi + (1+a_2) \nu_\Delta^\ast \left\{\frac{6 (d-1)}{d+2} \zeta_0^\ast \xi_{18} - 4 \xi_7 \xi_\varphi\right\}\right],
\\
\xi_{16} &= (\xi_5 \xi_{19} - \xi_7 \xi_9 \nu_\Delta^\ast) \xi_\varphi 
- \frac{1}{4} \zeta_0^\ast \xi_\sigma \xi_m \xi_R \left[(d+2) \xi_1 \xi_\varphi + \frac{\nu_{\varphi q}^\ast}{d}\right]
- \xi_{20},
\\
\xi_{17} &= \vartheta_{32}^3 - \frac{3}{2} \vartheta_{31} \big(\vartheta_{32} \xi_{16} + \xi_{10} \nu_\Delta^\ast \xi_R \xi_\varphi \vartheta_{31} \big),
\\
\xi_{18} &= 8 \xi_q + 2 (d+2) \xi_\varphi + \nu_{\varphi q}^\ast,
\\
\xi_{19} &= \frac{1}{d} \left[\xi_\sigma + \frac{3(d-1)}{d+2}\zeta_0^\ast\right] \xi_m \xi_R,
\\
\xi_{20} &= \frac{1}{2d(d+2)} \Big[2 \xi_2 \xi_{10} \xi_R + (d-1) \zeta_0^\ast \xi_{18} \xi_\sigma \xi_m \nu_\Delta^\ast\Big].
\end{align}
\endgroup

\bibliography{refer}
\bibliographystyle{jfm_doi}

\end{document}